\documentclass{aa}

\usepackage{graphicx}
\usepackage{natbib}
\usepackage{color}
\usepackage{upgreek}
\usepackage{soul}

\usepackage[colorlinks=true, linkcolor=blue, citecolor=blue, filecolor=blue, urlcolor=blue]{hyperref}

\bibpunct{(}{)}{;}{a}{}{,}

\usepackage{txfonts}
\usepackage{mathrsfs}
\begin{document}

\titlerunning{Magnetism of interacting binaries}
\authorrunning{Hubrig et al.}

\title{Magnetic field detections in massive systems at different stages of interaction}

   \author{
S.~Hubrig\inst{1}
\and 
 M.~Abdul-Masih\inst{2,3}
\and
S.~P.~J\"arvinen\inst{1}
\and
A.~Cikota\inst{4}    
\and
M.~Sch\"oller\inst{5}
\and
I.~Ilyin\inst{1}
\and
A.~Escorza\inst{2,3}
          }

          \institute{
Leibniz-Institut f\"ur Astrophysik Potsdam (AIP),
An der Sternwarte~16, 14482~Potsdam, Germany\\
 \email{shubrig@aip.de}
 \and
Instituto de Astrofisica de Canarias, C.~Via~Lactea, s/n, 38205~La~Laguna, Santa Cruz de~Tenerife, Spain
 \and
Universidad de La Laguna, Departamento de Astrofisica, Av.~Astrofisico Francisco Sanchez s/n, 38206~La~Laguna, Tenerife, Spain
\and
Gemini Observatory / NSF’s NOIRLab, Casilla~603, La~Serena, Chile
\and
European Southern Observatory, Karl-Schwarzschild-Str.~2, 85748~Garching, Germany\\
}

   \date{Received XXX 2025 / Accepted XXX 2025}

 
  \abstract
{
Despite the importance of magnetic fields in massive stars,
their origin is widely debated and still not well understood.
   }
   {
With the mounting evidence for the importance of studying magnetic fields in interacting massive binary and
multiple systems, it appears necessary to investigate the presence of magnetic fields in semi-detached systems with
ongoing mass transfer, and in contact systems where mass is actively being exchanged. 
}
   {
We present an analysis of 53 high-resolution HARPS\-pol spectropolarimetric
observations of a sample of 14 massive binary and multiple systems using the least-squares deconvolution technique.
The majority of the studied systems are classified as semi-detached or contact binaries.
}
{
Definite detections of the presence of a magnetic field are achieved in all studied systems apart from the rather faint
  system SV\,Cen, for which only a marginal detection was obtained.
The fact that the presence of magnetic fields is detected in all but one of the studied systems
strongly suggests that interaction between the system components plays a definite role in the generation of
magnetic fields in massive stars.
The measured mean longitudinal magnetic field strength for
all targets is of the order of a few hundred Gauss to a few kiloGauss.
The strongest longitudinal magnetic fields of 4 to 5\,kG are
discovered in the massive O-type triple system MY\,Ser
in both components of the contact binary. 
kiloGauss-order magnetic fields are also detected in two other systems, V1294\,Sco and V606\,Cen. 
It is possible that there is an implication of some system characteristics, such as multiplicity, the mass ratio
between the components, and a large fillout factor, on the measured magnetic field strength. 
Our results for the magnetic field measurements in interacting binaries present the first assessment of the occurrence
rate of magnetic fields in a representative sample of such systems.
}
   {}


   \maketitle


\section{Introduction}\label{sec:intro}

The important role of massive stars in the evolution of our Universe is widely recognised, but the origin of their
magnetism and its impact on the stellar evolution and the ultimate fate of massive stars as supernovae and compact objects
is still a matter of debate. The most popular scenario for the origin of the magnetic field involves a merging event,
mass transfer, or a common envelope (CE) evolution. Mass transfer or stellar merging rejuvenate the mass gaining star
(e.g.\ \citealt{Tout2008}; \citealt{Ferrario2009}; \citealt{Schneider2016}),
while the induced differential rotation is thought to be the key ingredient to generate a magnetic field
(e.g.\ \citealt{Wickr2014}). According to magnetohydrodynamical simulations of the merger of a binary by \citet{Schneider2019},
shear between the accretion stream and the stellar surface forms a modest magnetic field, which is then bolstered considerably
when the stellar cores merge.
In the context of this scenario, massive O stars are of special interest for studies of their magnetism
because more than 90\% of them are born and
live in binary or multiple systems \citep{{Offner2023}} and up to 70\% experience binary interaction during
their life (e.g.\ \citealt{Sana2012}).

Observational magnetic studies carried out in recent years
have revealed that systems in different evolutionary
states and different interaction phases indeed offer a way to disentangle different theoretically predicted magnetic
field generation channels and, consequently, to understand the magnetic field origin. 
To test the importance of
interaction processes for the generation of magnetic fields in massive stars, \citet{Hubrig2023}
performed an analysis of 61 high-resolution spectropolarimetric observations of 36 systems with O-type primaries.
The studied  sample included multiple systems with components at different evolutionary stages with wide and
tight orbits and different types of interaction.
The authors reported that out of the 36 systems, 22 exhibited in their LSD Stokes~$V$ profiles
definitely detected Zeeman features, 11 systems showed marginal evidence for the detection of a Zeeman feature,
and 3 systems did not show the presence of a magnetic field.

From a theoretical point of view, the role of magnetic fields in close binary evolution is not yet clear.
Usually, magnetic fields can be treated within magneto-hydrodynamical (MHD) simulations.
However, according to \citet{Campbell2018}, the implementation of MHD in binary stars  is
very challenging because the presence of magnetic fields makes the range of physical phenomena extremely rich
and complex. In addition to hydrodynamic flow instabilities, MHD instabilities are important. The role of magnetic
fields in the post-dynamical inspiral phase of CE evolution was recently discussed by
\citet{Gagnier2023}, who suggested that an initially weak magnetic field may undergo amplification by interacting
with spiral density waves and turbulence generated in the stellar envelope by the inspiraling companion. Also,
jet-like outflows appear in the MHD simulations carried out by \citet{Ondratschek2022} and \citet{Garc2021}.
Given the very challenging numerical implementation of MHD, only
observations of magnetic fields in binary and multiple systems at different stages of their interaction can
give the crucial information necessary to test theoretical predictions on the magnetic field origin.

With the mounting evidence for the importance of studying magnetic fields in interacting massive binary and
multiple systems, it appears necessary to investigate the presence and role of magnetic fields in
semi-detached systems with ongoing mass transfer, and in systems in the contact phase directly preceding
a merger event.
Binary interactions can fundamentally alter the evolutionary path of a
system, affecting the main sequence lifetime, the interior structures of the components, and the
positions on the Hertzsprung Russell diagram (e.g.\ \citealt{Marchant2024}).
When the primary star in massive binaries fills its Roche
lobe, mass transfer is initiated and the system enters a semi-detached configuration.
In Roche lobe overflow (RLOF) interactions, three different cases are
usually considered: case~A, if the RLOF episode occurs when the mass donor is on the core hydrogen-burning main sequence;
case~B, when the star is in the hydrogen shell burning phase; and case~C, when the star is in the helium shell
burning phase (e.g.\ \citealt{Kippenhahn1967,Vanbeveren1998}).
During the semi-detached phase, as mass is transferred from one component to the other, the mass receiver
can expand and fill its Roche lobe as well, leading to a contact phase.
The contact phase represents one of the most extreme forms of binary interactions throughout all branches
of massive binary evolution. While in a contact configuration, the components share a common surface through which mass
and energy are exchanged and experience a variety of interaction processes including but
not limited to tidal locking, mutual illumination, internal mixing,
and angular momentum exchange \citep{Wellstein2001,Mink2007}.
\citet{Henneco2024} identified five mechanisms that lead to contact and mergers: runaway mass transfer, mass loss through
the outer Lagrange point L2, expansion of the accretor, orbital decay because of tides, and non-conservative mass transfer.
The authors concluded that at least
40\% of mass-transferring binaries with initial primary-star masses of 5-20\,$M_{\odot}$ evolve into a contact
phase; $>$12\% and $>$19\% likely merge and evolve into a CE phase, respectively.

To address questions on the magnetic field origin in massive stars, we recently obtained
high-resolution ($R\approx110\,000$) HARPS\-pol (High Accuracy Radial velocity Planet Searcher
polarimeter; \citealt{Harps2008}) observations to explore the magnetic field incidence in a sample of
close, semi-detached, and contact binaries.
For a number of targets in our sample, the fillout factors, $f$, are known from the
physical properties of the components, including their radii, the radii of the inner and outer Roche lobes,
the surface potentials of the components, and the surface potentials of the inner and outer Roche lobes.
According to \citet{Mochnacki1972},
fillout factors describe the degree to which the equipotential surface corresponding to the photosphere fills out the
Lagrangian zero-velocity surfaces.
For detached components, which have photospheres lying inside their
inner Lagrangian surfaces, the fillout factor is $0 < f \le 1$, whereas for contact configurations
the photosphere lies between the
first and second Lagrangian surfaces corresponding to $1 \le f \le 2$.
Very few such systems have previously been analysed using spectropolarimetric observations.
The presence of rather strong longitudinal magnetic fields has been reported for the well-known contact binaries
MY\,Ser (=HD\,167971) and LY\,Aur (=HD\,35921) by \citet{Hubrig2023}, who used ESPaDOnS archival observations.
More recently, the presence of a magnetic field
was detected in the primary of the suspected contact ON3\,If*$+$O5.5\,V((f))
binary system NGC\,346 SSN\,7 ($P_{\rm orb}= 3.07$\,d)
located in the Small Magellanic Cloud using low-resolution
observations with the  ESO/VLT FOcal Reducer low dispersion Spectrograph
(FORS2; \citealt{Appenz1998}) in spectropolarimetric mode \citep{Hubrig2024}.

Notably, observational surveys of the presence of magnetic fields in contact systems are severely hampered
by the frequent occurrence of these systems in higher order multiple systems. According to \citet{Abdul-Masih2025},
of the confirmed contact systems, almost 70\% have at least one confirmed companion and about 25\% have
more than one additional companion. The majority of these detections come through eclipse timing variations (ETVs) due to
the light travel time effect, spectroscopic studies, and interferometry. \citet{Abdul-Masih2025} suggests that 
due to the compactness of the inner orbits in such systems, their multiplicity properties may be fundamentally
different from those of other O-type systems.
\citet{Pablo2018} mentioned several other obstacles in studies of contact systems.
Spectral lines of the components usually have significant overlap at most binary phases,
preventing accurate radial velocity measurements.
Further, in such systems surfaces are illuminated by two separate
energy sources, making the determination of the temperature distribution difficult.

In this work, we present our analysis of 53 high-resolution HARPS\-pol spectropolarimetric
observations acquired over the last two years for fourteen massive binary and multiple systems with components
at different evolutionary stages undergoing different types of interaction.
In Sect.~\ref{sec:obs} of this article,
 we detail our spectropolarimetric observations and describe the methodology of the data analysis.
The results of the magnetic field measurements for each target are presented in Sect.~\ref{sec:Bz}.
In Sect.~\ref{sec:disc}, we discuss the
implication of the reported magnetic field detections for future studies of massive stars.

\section{Observations and analysis}\label{sec:obs}

\begin{table*}
\caption{
Sample of the investigated multiple systems.
\label{tab:obsspt}
} 
\centering
\begin{tabular}{lrlclclc}
\hline
\hline
\multicolumn{1}{c}{Target} &
\multicolumn{1}{c}{$m_{\rm V}$} &
\multicolumn{1}{c}{Spectral} &
\multicolumn{1}{c}{Ref.} &
\multicolumn{1}{c}{Multiplicity} &
\multicolumn{1}{c}{Ref.} &
\multicolumn{1}{c}{Notes} &
\multicolumn{1}{c}{Ref.} \\
\multicolumn{1}{c}{} &
\multicolumn{1}{c}{} &
\multicolumn{1}{c}{Classification} &
\multicolumn{1}{c}{} &
\multicolumn{1}{c}{} &
\multicolumn{1}{c}{} &
\multicolumn{1}{c}{} &
\multicolumn{1}{c}{} \\
\hline
$\gamma$\,Equ  & 4.7 & A9\,Vp SrEu   & (1)  &  single &  &  standard     &  \\ 
29\,CMa  & 5.0 & O9\,Iabf+O9.7\,Ib   & (2)  &  SB2?   &  (2) & contact? & (20)  \\ 
V402\,Pup & 9.2 & O6\,V+O6\,V, O9.5\,V+O9.5\,V & (3) & quadruple & (3)  & contact & (3)  \\ 
TU\,Mus & 9.3 & O7\,V+O8\,V & (4) & quadruple? & (16)  & contact & (16) \\ 
SV\,Cen  &9.7 & B6.5III+B2V & (5)& SB2 & (5) &  semi-detached?     & (21) \\
V606\,Cen & 9.7& B0.5\,V+B3\,V & (6)& triple & (17) & contact  & (17) \\ 
V889\,Cen & 11.7 & O4f$^{+}$+O6-7:(f):& (7)& SB2  & (7) & contact & (7) \\
V918\,Sco & 5.5 & O7.5\,I(f)+ON9.7& (8) & SB2 & (18) & BSG, CWB, post-RLOF  & (18) \\ 
HR\,6187 & 5.7 & O3-3.5\,V+O5.5-6\,V+O6.5-7\,V& (9)& triple & (9) & CWB, merger? & (9) \\ 
$\mu^{01}$\,Sco &3.0 & B1.5\,V+B8-B3 & (10)   & SB2 & (10) & semi-detached & (10) \\ 
V1294\,Sco & 7.6 & O9.5IV(n)+B0V & (11)  & SB2& (11) & detached? & (22)\\ 
V701\,Sco  & 9.0  & B1-1.5\,V+ B1-1.5\,V & (12)  & triple?   &  (12) & contact    & (12) \\ 
MY\,Ser  & 7.5 & O7.5III+O9.5III+O9.5I & (13) & SB3  & (19) & PACWB, contact & (23) \\ 
V356\,Sgr  & 7.0 & A2I+B3V & (14) & SB2    & (14) &  contact?     & (14) \\ 
V337\,Aql & 8.9 &  B0+B3   & (15)  &  SB2   & (15) & semi-detached  & (15) \\ 
\hline
\end{tabular}
\tablefoot{
The first column gives the target name 
and the second column the visual magnitude.
In Columns~3 and 4 we present for each target the spectral classification and the corresponding
reference.
The multiplicity indicator and the corresponding reference are listed in columns 5 and 6.
Here, the entries SB2/3  correspond to spectroscopic binaries with double and triple line systems,
and '?' to a multiplicity classification uncertainty.
Columns~7 and 8 list notes concerning the interaction status of the system members
and the corresponding reference.
BSG stands for blue supergiants, CWB for colliding wind binaries,
and PACWB for particle-accelerating colliding-wind binaries exhibiting synchrotron radio emission.
References:
(1) \citet{Bychkov2016};
(2) \citet{Bagnuolo1994};
(3) \citet{Lorenzo2017};
(4) \citet{Penny2008};
(5) \citet{Drechsel1994};
(6) \citet{Lorenz1999};
(7) \citet{Raucq2017};
(8) \citet{Sota2014};
(9) \citet{Mahy2018};
(10) \citet{Antwerpen2010};
(11) \citet{Maiz2004};
(12) \citet{Qian2006};
(13) \citet{Becker2018};
(14) \citet{Rensbergen2011};
(15) \citet{Soydugan2014};
(16) \citet{Qian2007};
(17) \citet{Li2022};
(18) \citet{Thaller1998};
(19) \citet{Ibanoglu2013};
(20) \citet{Pablo2018};
(21) \citet{Rucinski1992};
(22) \citet{Rosu2022};
(23) \citet{Becker2013}.}
\end{table*}

Our spectropolarimetric observations were carried out on January~2 and 4 and on June~17 to 20 in 2024
and on April~12 to 16 2025 using HARPS\-pol attached to ESO's 3.6m telescope on La~Silla.
HARPSpol has a resolving power of
about 110\,000 and a wavelength coverage from 3780 to 6910\,\AA{},
with a small gap between 5259 and 5337\,\AA{}. The normalisation of the spectra to the continuum level was described
in detail by \citet{Hubrig2013}. 
One additional spectropolarimetric observation
of 29\,CMa with ESPaDOnS (the Echelle SpectroPolarimetric Device for Observations of
Stars) acquired on March~3 2013 was retrieved from the CFHT
(the Canada-France-Hawaii Telescope) science archive. ESPaDOnS covers the wavelength range from 3750 to 10\,500\,\AA{}
and the spectral resolution of the ESPaDOnS observations is about 65\,000.

With HARPSpol and  ESPaDOnS, we have access to measurements
of the mean longitudinal magnetic field $\left<B_{\rm z} \right>$, which is
the line-of-sight component of the magnetic field, weighted with
the line intensity and averaged over the visible hemisphere. The
longitudinal magnetic field is strongly dependent on the viewing
angle between the field orientation and the observer and is
modulated as the star rotates.
The assessment of the longitudinal magnetic field measurements 
is presented in our previous papers (e.g.\ \citealt{Hubrig2018}; \citealt{Jarvinen2020}). 
Similar to our previous studies, to increase the 
signal-to-noise ratio ($S/N$) by a multi-line approach, we employed the least-squares deconvolution (LSD) technique. 
The details of this technique, 
as well as how the LSD Stokes~$I$, Stokes~$V$, and diagnostic null spectra are calculated, were 
presented by \citet{Donati1997}.

Several complications existing in the analysis of the presence
of magnetic fields in multiple systems have recently been discussed in the work by \citet{Hubrig2023}.
The amplitudes of the Zeeman features (the features appearing in the Stokes~$V$ spectra of magnetic stars)
are much lower in multiple systems in comparison with their size in single stars. These features
also appear blended in composite spectra and can show severe shape distortions.
Also the shapes of blended spectral lines in the Stokes~$I$ spectra look different depending on the visibility of each
system component at different orbital phases. The number of spectral lines suitable for the LSD technique analysis
for massive stars is in addition  much lower in comparison to less massive stars. Furthermore, 
special care has to be taken to populate the LSD line masks for each system because the composite spectra
of multiple systems usually show very different spectral signatures corresponding to the different spectral
classification of the individual components.

Therefore, we searched in the Stokes~$V$ spectra of O and B-type systems for
both Zeeman features with typical and atypical shapes. To evaluate whether
the detected features are spurious or definite detections, we followed
the generally adopted procedure to use the false alarm probability
(FAP), based on reduced $\chi^{2}$ test statistics \citep{Donati1992}:
the presence of the Zeeman feature is considered as a definite detection (DD)
if ${\rm FAP} \leq 10^{-5}$,
as a marginal detection (MD) if  $10^{-5}<{\rm FAP}\leq 10^{-3}$,
and as a non-detection (ND) if ${\rm FAP}>10^{-3}$.
The list of targets, their visual magnitudes, spectral classifications, multiplicity information,
and interaction status, together with the related literature sources, are presented in Table~\ref{tab:obsspt}.

\section{Results of our LSD analysis for individual targets }\label{sec:Bz}

\begin{table*}
\caption{
Logbook of observations and results of magnetic field 
measurements for the studied systems.
\label{tab:obsall}
}
\centering
\begin{tabular}{lccrccc r@{$\pm$}l c}
\hline\hline \noalign{\smallskip}
\multicolumn{1}{c}{Target} &
\multicolumn{1}{c}{Date} &
\multicolumn{1}{c}{MJD} &
\multicolumn{1}{c}{$S/N$} &
\multicolumn{1}{c}{Line} &
\multicolumn{1}{c}{FAP}&
\multicolumn{1}{c}{Det.} &
\multicolumn{2}{c}{$\left< B_{\rm z} \right>$} &
\multicolumn{1}{c}{Remark} \\
\multicolumn{1}{c}{} &
\multicolumn{1}{c}{} &
\multicolumn{1}{c}{} &
\multicolumn{1}{c}{} &
\multicolumn{1}{c}{mask} &
\multicolumn{1}{c}{} &
\multicolumn{1}{c}{flag} &
\multicolumn{2}{c}{[G]} &
\multicolumn{1}{c}{} \\
\noalign{\smallskip}\hline \noalign{\smallskip}
$\gamma$\,Equ&2024-06-18& 60479.41 & 316 & \ion{Fe}{i}, \ion{La}{ii}, \ion{Pr}{iii}, \ion{Nd}{ii}& $< 10^{-10}$  & DD & $-$1027 & 44 & \\
29\,CMa & 2013-03-03 & 56354.31 & 507 & \ion{He}{i/ii}, \ion{N}{iii}, \ion{O}{ii/iii}, \ion{Si}{iii/iv}& $1\times10^{-6}$    & DD & $-$26 & 19 & ESPaDOnS \\
        & 2024-01-02 & 60311.04 & 328 & \ion{He}{i}, \ion{N}{iii}& $6\times10^{-4}$    & MD & $-$50 & 25 \\
        & 2024-01-04 & 60313.04 & 307 & \ion{He}{i/ii},  \ion{C}{iv}, \ion{O}{iii} & $7\times10^{-4}$    & MD & 27 & 7 \\
        & 2024-06-17 & 60478.95 & 230 & \ion{He}{i/ii} & $< 1\times10^{-10}$ & DD & $-$159 & 113 \\
        & 2024-06-18 & 60479.96 & 162 & \ion{He}{i/ii}, \ion{Si}{iv}     & $2\times10^{-5}$    & MD & 73 & 27 \\
        & 2024-06-19 & 60480.96 & 230 & \ion{He}{i/ii}, \ion{N}{iii}, \ion{Si}{iv} & $< 1\times10^{-10}$ & DD & 298 & 59 \\
        & 2025-04-15 & 60781.07 & 437 & \ion{He}{i/ii}, \ion{N}{iii}, \ion{O}{iii}, \ion{Si}{iv} & $< 1\times10^{-10}$ & DD & 56 & 16 \\
V402\,Pup  & 2025-04-15 & 60780.07 & 148 & \ion{He}{i/ii}                 &    $9\times10^{-5}$ & MD & \multicolumn{2}{l}{} & \\
           & 2025-04-16 & 60781.11 & 177 & \ion{He}{i/ii}, \ion{O}{iii}   & $2\times10^{-6}$     & DD & $-$353 & 67 & \\
TU\,Mus    & 2024-06-18 & 60479.05 & 90  & \ion{He}{i/ii}         & $10^{-5}$   & DD & 482 & 198 &  \\
           & 2024-06-18 & 60480.01 & 128 & \ion{He}{i/ii}          & $7\times10^{-4}$     & MD & \multicolumn{2}{l}{} & \\
           & 2025-04-14 & 60779.22 & 205 & \ion{He}{i/ii} \ion{C}{iv}  & $2\times10^{-6}$     & DD & \multicolumn{2}{l}{} & \\
SV\,Cen    & 2024-06-19 & 60480.09 & 91  &\ion{He}{i}, \ion{Si}{iii}   & $10^{-3}$     & MD & \multicolumn{2}{l}{} &  \\
V606\,Cen  & 2024-06-18 & 60479.15 & 55  & \ion{He}{i},\ion{C}{iii}, \ion{Si}{iii} &  $4\times10^{-4}$     & MD & \multicolumn{2}{l}{} & \\
           & 2024-06-19 & 60481.01 & 113 & \ion{He}{i/ii}, \ion{Si}{iii} & $2\times10^{-6}$     & DD & $-$419 & 168 & \\
           & 2025-04-14 & 60777.26 & 150 & \ion{He}{i/ii},\ion{Al}{iii}, \ion{Si}{iii} &   ---                & ND & \multicolumn{2}{l}{} & \\
           & 2025-04-16 & 60779.29 & 137 & \ion{He}{i/ii},\ion{Al}{iii}, \ion{Si}{iii} & $< 10^{-10}$   & DD & 1327 & 128  & \\
V889\,Cen  & 2024-06-20 & 60481.08 & 31  & \ion{He}{i/ii}  & $10^{-9}$     & DD & \multicolumn{2}{l}{} & \\
           & 2024-06-20 & 60482.03 & 24  & \ion{He}{i/ii}, \ion{C}{iv} & $6\times10^{-4}$     & MD & \multicolumn{2}{l}{} & \\
           & 2025-04-14 & 60779.14 & 46  &\ion{He}{ii}, \ion{C}{iv} & $6\times10^{-7}$     & DD & \multicolumn{2}{l}{} & \\
           & 2025-04-15 & 60780.22 & 33  & \ion{He}{i/ii}, \ion{C}{iv} &    ---               & ND & \multicolumn{2}{l}{} & \\
           & 2025-04-16 & 60781.19 & 49  & \ion{He}{i/ii}, \ion{C}{iv} & $< 10^{-10}$  & DD & \multicolumn{2}{l}{} & \\
V918\,Sco  & 2024-06-18 & 60479.10 & 289 & \ion{He}{i/ii}, \ion{C}{iv}, \ion{N}{iii}, \ion{Si}{iii}& $6\times10^{-6}$     & DD & 45 & 31 & \\
           & 2024-06-20 & 60481.34 & 245 & \ion{He}{i/ii}, \ion{N}{iii}, \ion{O}{iii}, \ion{Si}{iv}& $9\times10^{-8}$  & DD & \multicolumn{2}{l}{} &  \\
           & 2025-04-12 & 60777.38 & 612 &\ion{He}{i/ii}, \ion{C}{iv}, \ion{N}{iii}, \ion{O}{iii}& $3\times10^{-4}$     & MD & $-$11 & 6 & \\
           & 2025-04-15 & 60780.27 & 589 & \ion{He}{i/ii}, \ion{C}{iv}, \ion{O}{iii}, \ion{Si}{iii}& $< 10^{-10}$  & DD & \multicolumn{2}{l}{} & \\
HR\,6187   & 2024-06-17 & 60478.99 & 227 & \ion{He}{i/ii}, \ion{C}{iv},  \ion{N}{iii}& $6\times10^{-6}$     & DD & $-$369 & 58 & \\
$\mu^{01}$\,Sco & 2024-06-18 & 60479.35 & 322 & \ion{He}{i}, \ion{Si}{iii} & $10^{-5}$ & DD & 242 & 55 & \\
           & 2024-06-19 & 60480.14 & 323 & \ion{He}{i}, \ion{Si}{iii}    & $ 2\times10^{-4}$  & MD & $-$187 & 29 & \\
           & 2024-06-20 & 60481.32 & 278 &   \ion{He}{i}, \ion{Si}{iii}    &  $4\times10^{-6}$ & DD &\multicolumn{2}{l}{} & \\
           & 2024-06-20 & 60481.99 & 275 &  \ion{He}{i}        & $2\times10^{-9}$     & DD & \multicolumn{2}{l}{} & \\
           & 2025-04-13 & 60778.41 & 612 & \ion{He}{i}, \ion{Si}{ii/iii}  & $10^{-7}$     & DD & \multicolumn{2}{l}{} & \\
           & 2025-04-13 & 60778.41 & 583 & \ion{He}{i}, \ion{Si}{iii}      & ---                  & ND & \multicolumn{2}{l}{} & \\
           & 2025-04-13 & 60778.42 & 645 & \ion{He}{i}, \ion{Si}{ii/iii}      & $10^{-4}$   & MD & \multicolumn{2}{l}{} & \\
           & 2025-04-13 & 60778.42 & 641 & \ion{He}{i}, \ion{Si}{ii/iii}      & ---                  & ND & \multicolumn{2}{l}{} & \\
           & 2025-04-13 & 60778.43 & 585 & \ion{He}{i}, \ion{Si}{ii/iii}      & $< 10^{-10}$  & DD & \multicolumn{2}{l}{} & \\
           & 2025-04-14 & 60779.40 & 653 & \ion{He}{i}, \ion{Si}{iii}          & $ 2\times10^{-5}$ & MD & \multicolumn{2}{l}{} & \\
           & 2025-04-14 & 60779.41 & 637 & \ion{He}{i}, \ion{Si}{iii}          & $2\times10^{-4}$  & MD & \multicolumn{2}{l}{} & \\
  \hline
\end{tabular}
\tablefoot{
The first column lists the target name, followed by the date of the observation, the MJD 
values at the middle of the exposure, the signal to noise ratio ($S/N$) measured in the 
Stokes~$I$ spectra in the spectral region around 5000\,\AA, the line mask 
used, the FAP values, the detection flag -- where DD means definite 
detection, MD marginal detection, and ND no detection, -- the measured LSD 
mean longitudinal magnetic field strength, and a remark on the Stokes~$V$ 
feature where the field was diagnosed (if applicable). $\left< B_{\rm z} \right>$-values may not be included in all cases
due to the complexity of the composite spectra at some observing epochs.
}
\end{table*}

\addtocounter{table}{-1}

\begin{table*}
\caption{
Continued.
}
\centering
\begin{tabular}{lccrccc r@{$\pm$}l c}
\hline\hline \noalign{\smallskip}
\multicolumn{1}{c}{Target} &
\multicolumn{1}{c}{Date} &
\multicolumn{1}{c}{MJD} &
\multicolumn{1}{c}{$S/N$} &
\multicolumn{1}{c}{Line} &
\multicolumn{1}{c}{FAP}&
\multicolumn{1}{c}{Det.} &
\multicolumn{2}{c}{$\left< B_{\rm z} \right>$} &
\multicolumn{1}{c}{Remark} \\
\multicolumn{1}{c}{} &
\multicolumn{1}{c}{} &
\multicolumn{1}{c}{} &
\multicolumn{1}{c}{} &
\multicolumn{1}{c}{mask} &
\multicolumn{1}{c}{} &
\multicolumn{1}{c}{flag} &
\multicolumn{2}{c}{[G]} &
\multicolumn{1}{c}{} \\
\noalign{\smallskip}\hline \noalign{\smallskip}
V1294\,Sco & 2024-06-20 & 60481.30 & 106 & \ion{He}{i/ii}, \ion{O}{iii}& $< 1\times10^{-10}$  & DD & \multicolumn{2}{l}{} & \\
           & 2025-04-15 & 60780.39 & 282 & \ion{He}{i/ii}, \ion{C}{iv}, \ion{Si}{iii}& $< 10^{-10}$  & DD & 1559 & 131 & \\
           & 2025-04-16 & 60781.25 & 286 &\ion{He}{i/ii}, \ion{O}{iii} & $2\times10^{-7}$     & DD & \multicolumn{2}{l}{} & \\
V701\,Sco  & 2024-06-19 & 60480.18 & 144 & \ion{He}{i}, \ion{Si}{iii}   & $10^{-7}$     & DD & 678 & 150 & \\
           & 2025-04-12 & 60777.33 & 240 & \ion{He}{i}, \ion{Si}{iii} & $4\times10^{-4}$     & MD & $-$876 & 105 & \\
           & 2025-04-14 & 60779.36 & 204 & \ion{He}{i}, \ion{Si}{iii}& $< 10^{-10}$  & DD & \multicolumn{2}{l}{} & \\
           & 2025-04-15 & 60780.33 & 228 & \ion{He}{i}, \ion{Si}{iii}  & $< 10^{-10}$  & DD & \multicolumn{2}{l}{} & \\
MY\,Ser    & 2024-06-18 & 60479.31 & 200 & \ion{He}{i/ii}, \ion{C}{iii}, \ion{N}{iii}, \ion{O}{iii}, \ion{Si}{iv}& $10^{-7}$     & DD & $-$46 & 31 & \\
           & 2024-06-19 & 60480.30 & 232 &\ion{He}{i}, \ion{C}{iii/iv}, \ion{O}{iii} & $3\times10^{-8}$     & DD & 4008 & 402 & \\
           & 2024-06-20 & 60481.16 & 225 & \ion{He}{i/ii}, \ion{C}{iii}, \ion{O}{iii}& $< 10^{-10}$  & DD & \multicolumn{2}{l}{} & \\
           & 2025-04-16 & 60781.50 & 363 & \ion{He}{i/ii}, \ion{C}{iii}  & $2\times10^{-8}$     & DD &  $-$4875 & 394 & right \\
            &           &          &     & \ion{He}{i/ii},\ion{N}{iii}, \ion{O}{iii}  & $4\times10^{-5}$     & MD & 8619 & 662 & left \\
V356\,Sgr  & 2024-06-18& 60479.38 & 190 & \ion{He}{i}, \ion{Si}{ii/iii}, \ion{Fe}{ii}& $1.5\times10^{-7}$   & DD & 856  & 240 & \\
           & 2024-06-20	& 60481.39 & 223 & \ion{Mg}{ii}, \ion{Fe}{ii}& $7\times10^{-6}$     & DD & $-$718 & 245& \\
V337\,Aql  & 2024-06-18 & 60479.23 & 140 &\ion{Si}{iii}   & $2\times10^{-6}$     & DD & 430 & 235 & \\
           & 2024-06-20 & 60481.24 & 125 & \ion{He}{i/ii}, \ion{Si}{iii}  & $2\times10^{-5}$  & MD & $-$373 & 38 & \\
           & 2025-04-16 & 60781.37 & 266 & \ion{He}{i/ii} & $3\times10^{-4}$     & MD & 146 & 275 & right \\
          &             &          &     & \ion{He}{i/ii} & $9\times10^{-6}$     & DD & 25 & 68 & left \\
  \hline
\end{tabular}
\end{table*}

In the following, we give a brief description of the targets in our sample and
discuss the results of the magnetic field measurements for each target individually.
These results are also summarised in Table~\ref{tab:obsall}.

\begin{figure}
    \centering
    \includegraphics[width=0.242\textwidth]{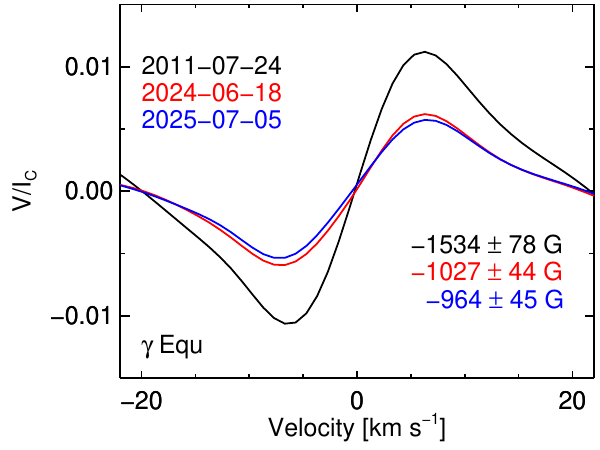}
    \includegraphics[width=0.242\textwidth]{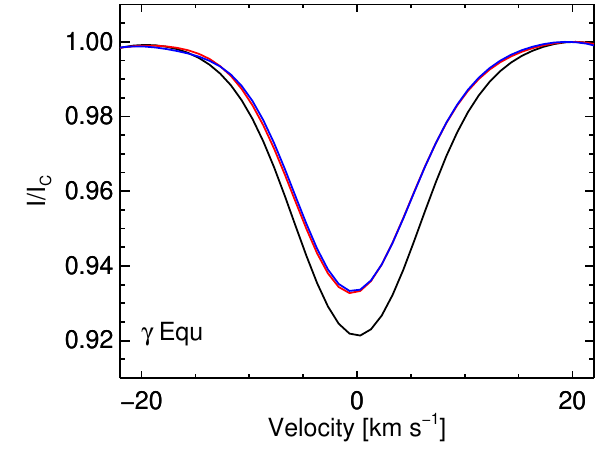}
    \caption{
LSD Stokes profiles of the standard star $\gamma$\,Equ.
{\it Left:} LSD Stokes~$V$ profiles for the HARPS\-pol observations
of $\gamma$\,Equ from 2011 to 2025.
{\it Right:} Intensities of the LSD Stokes~$I$ profiles of $\gamma$\,Equ at different observational epochs.
}
\label{fig:gamma}
\end{figure}

\paragraph*{$\gamma$\,Equ (=HD\,201601):}

The typical strongly magnetic Ap star $\gamma$\,Equ was used 
as a standard star during our observations in 2024.
Because of its high brightness and the extremely long rotation
period $P_{\rm rot} > 97$\,yr \citep{Bychkov2016}, this star is frequently used as a magnetic standard to test the functionality
of spectropolarimetric devices at different 
telescopes. We used in our LSD analysis HARPS\-pol observations of $\gamma$\,Equ
acquired on July~24 2011, on June~18 2024, and on July~5 2025,
applying a mask containing lines belonging to iron and rare earth elements. In
Fig.~\ref{fig:gamma} we demonstrate that the strength of the longitudinal magnetic field is gradually decreasing
from $-$1534\,G in 2011 to $-$964\,G in 2025.
The intensities of the LSD Stokes~$I$ profiles presented in this figure on the right side are also decreasing over the same
time interval.

\paragraph*{29\,CMa (=HD\,57060):}

\begin{figure*}
    \centering
\includegraphics[width=0.23\textwidth]{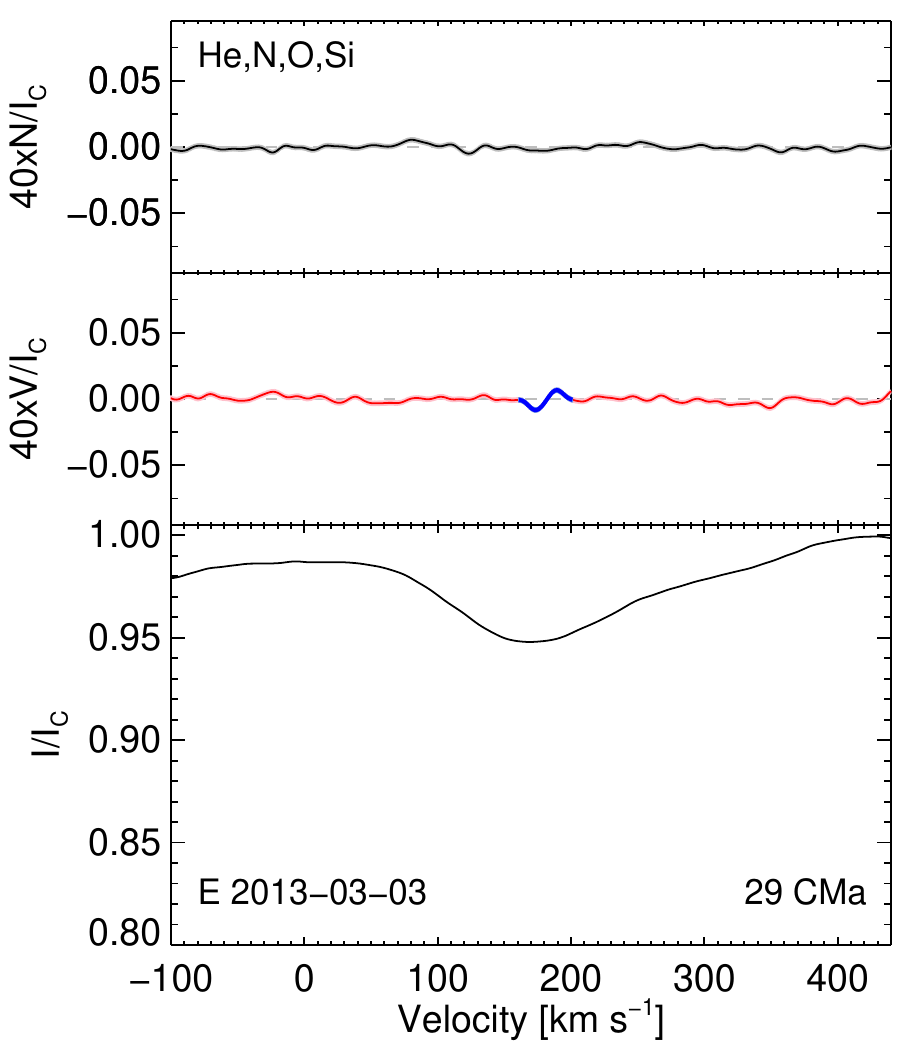}
 \includegraphics[width=0.23\textwidth]{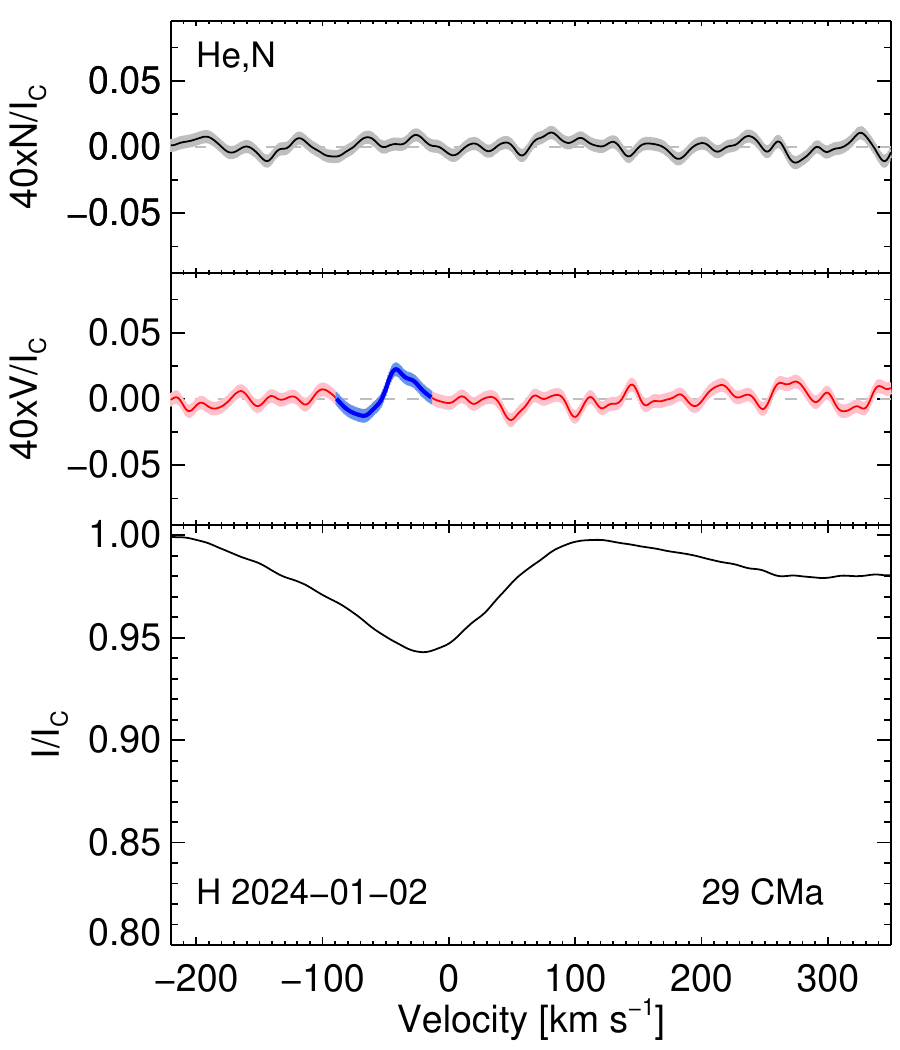}
\includegraphics[width=0.23\textwidth]{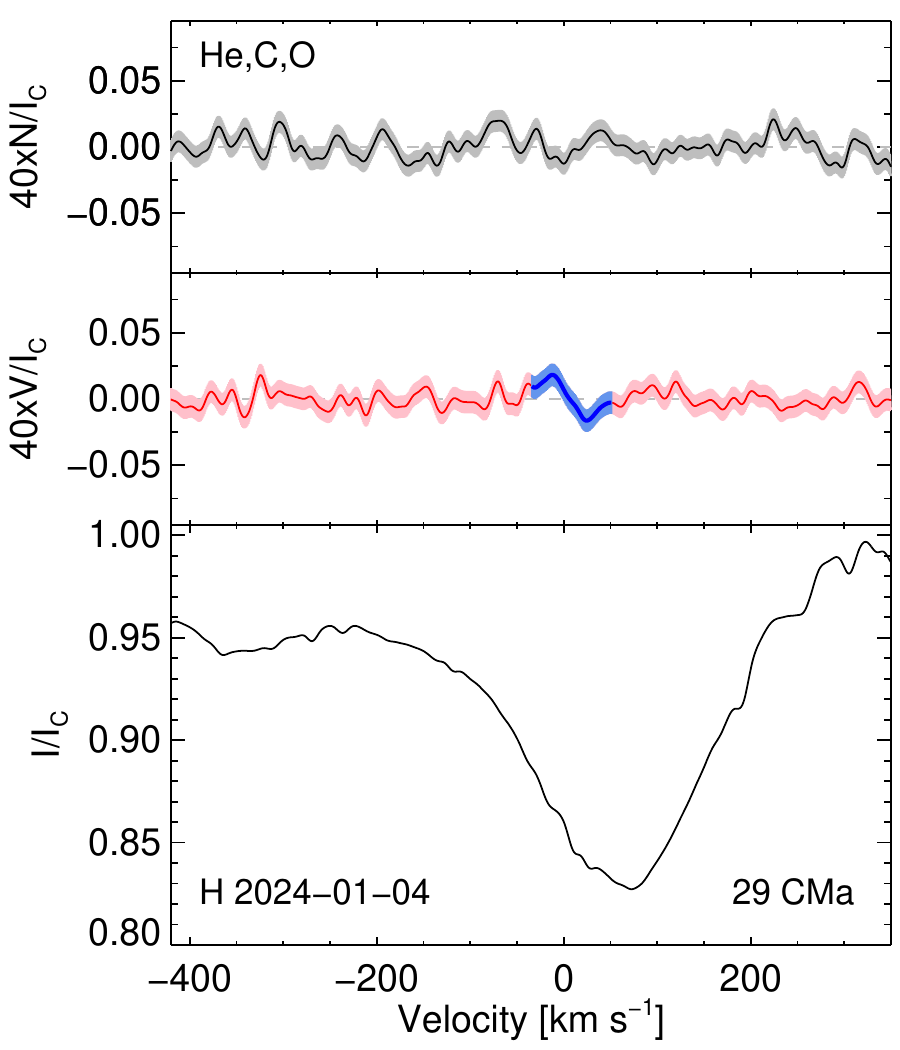}
\includegraphics[width=0.23\textwidth]{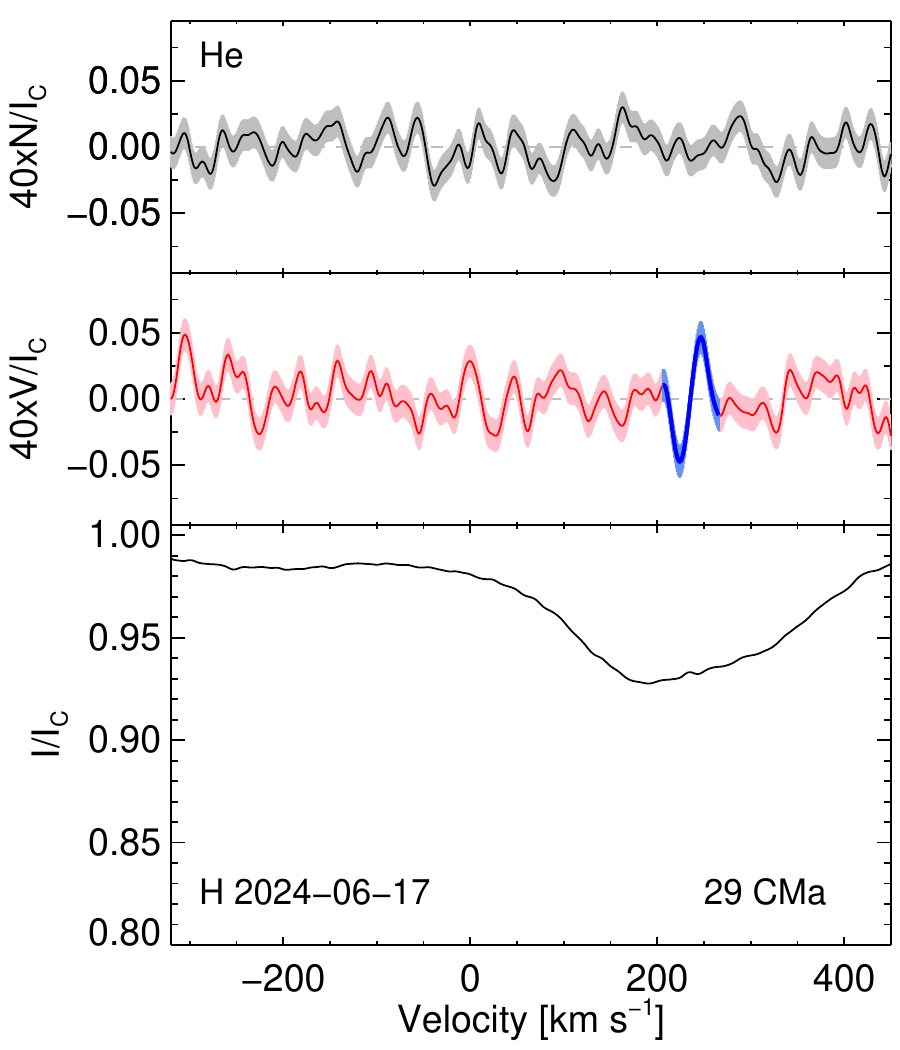}
\includegraphics[width=0.23\textwidth]{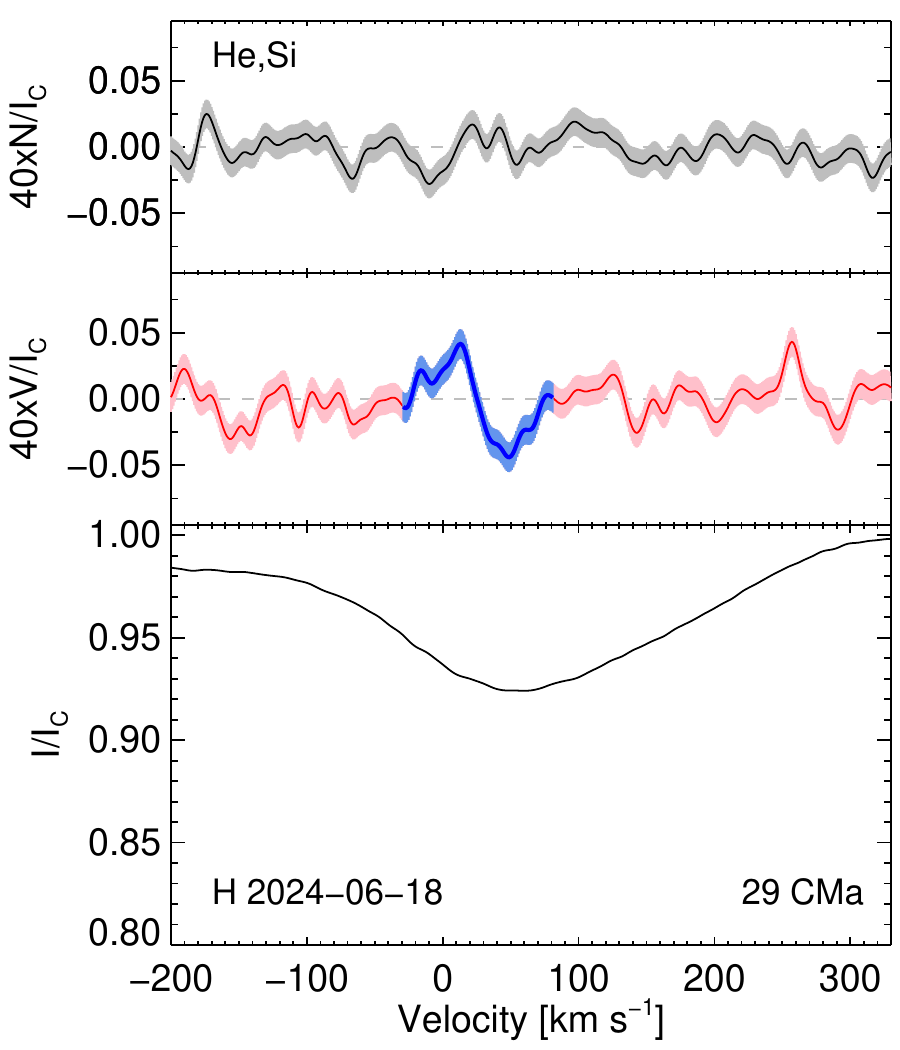}
\includegraphics[width=0.23\textwidth]{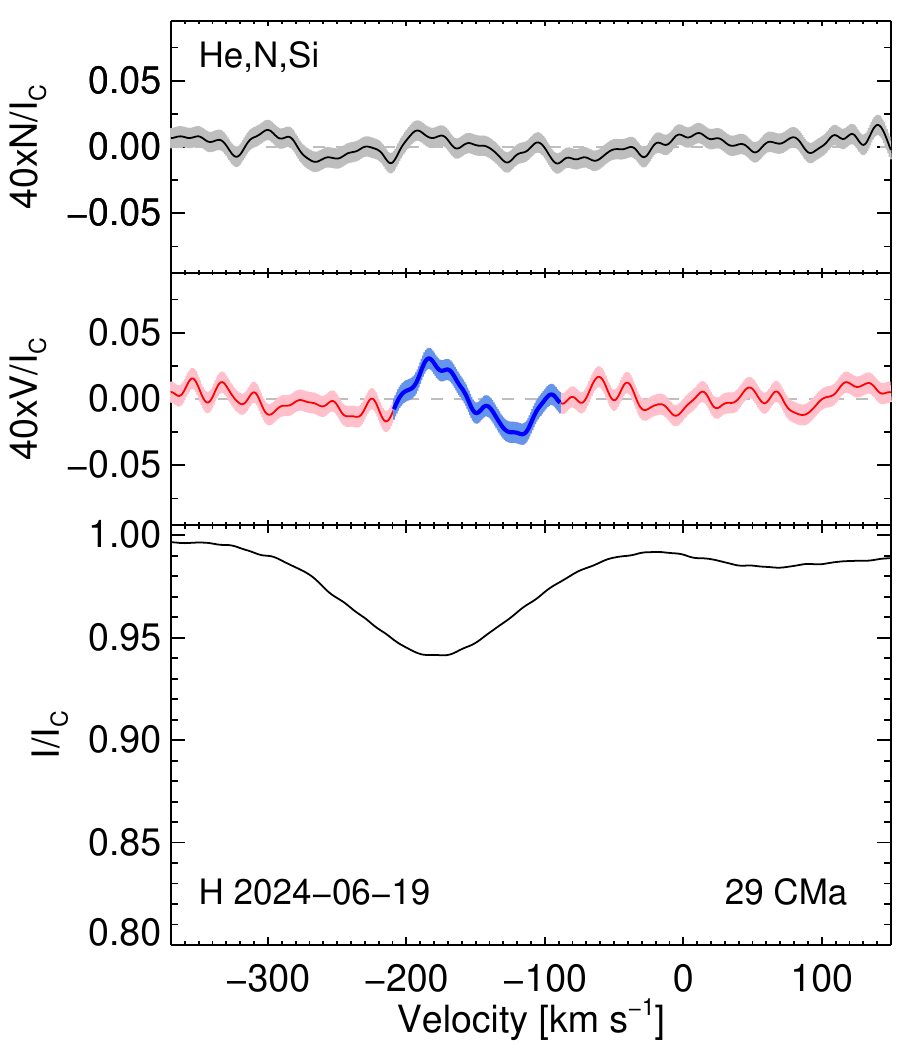}
\includegraphics[width=0.23\textwidth]{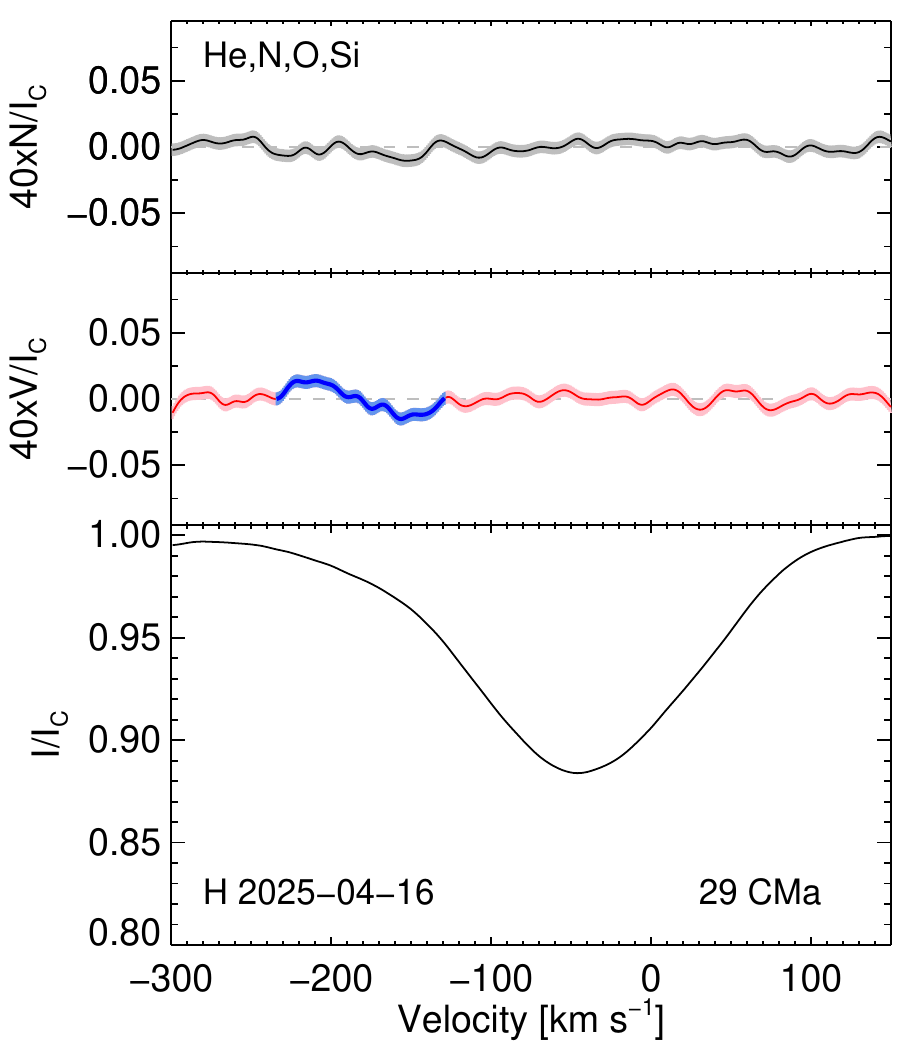}
    \caption{
Results of our LSD analysis of 29\,CMa.
Stokes~$I$, Stokes~$V$, and diagnostic null $N$ spectra
obtained for different line masks using one archival ESPaDOnS observation and six HARPS\-pol observations are shown.
The identified Zeeman signatures are presented in blue colour.
}
    \label{fig:29CMa}
\end{figure*}

\begin{figure*}
    \centering
    \includegraphics[width=0.23\textwidth]{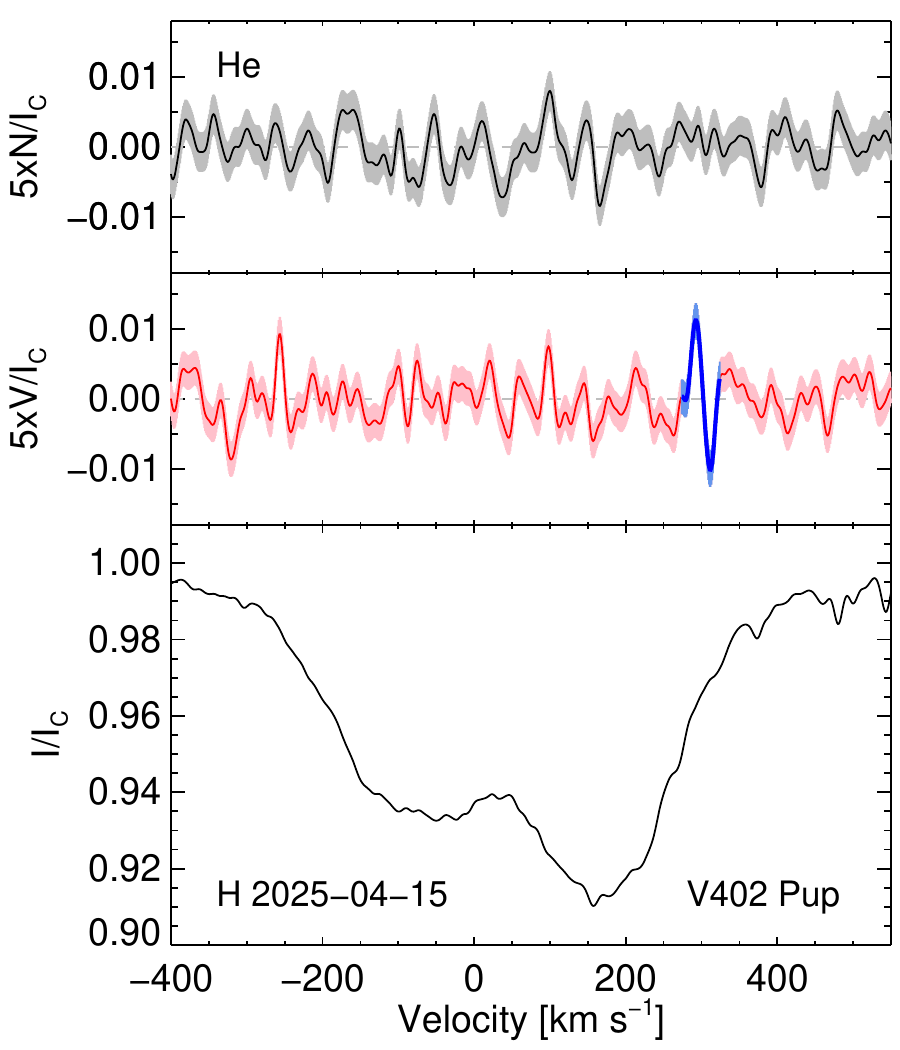}
    \includegraphics[width=0.23\textwidth]{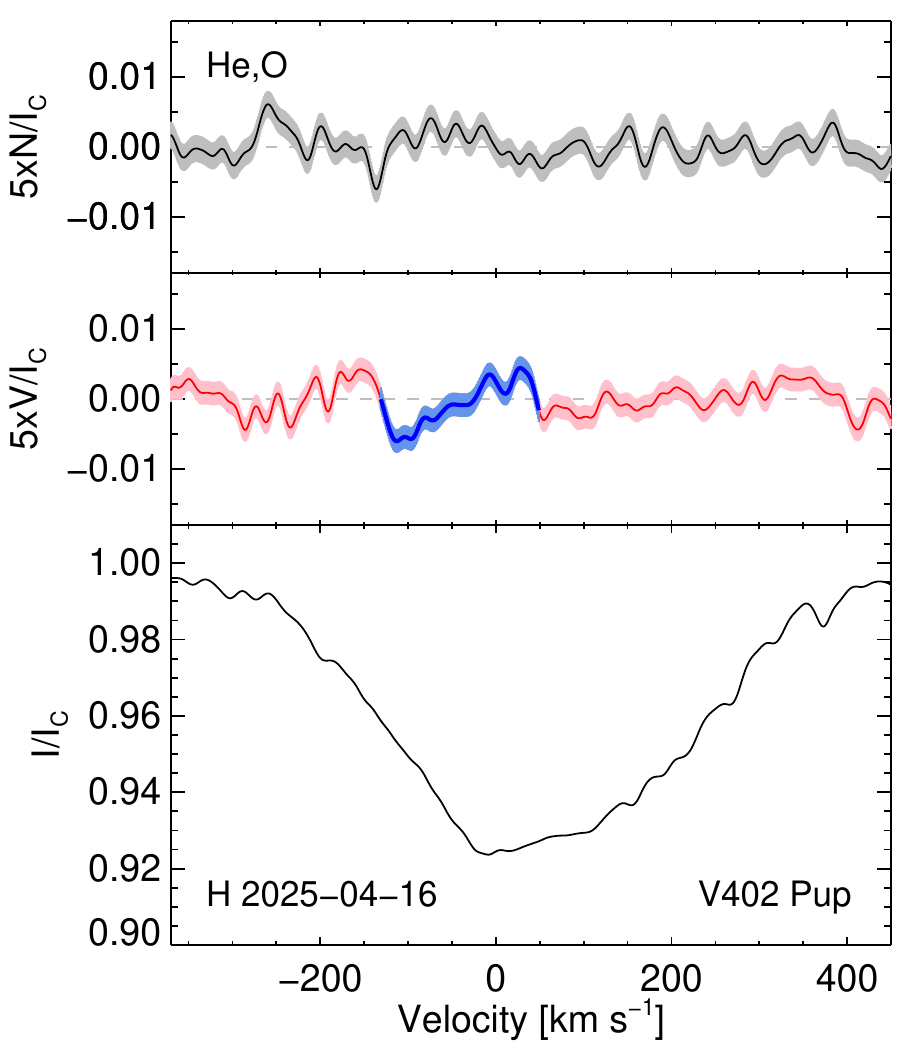}
    \caption{
As Figure~\ref{fig:29CMa}, but for V402\,Pup.
}
    \label{fig:HD64315}
\end{figure*}

\begin{figure*}
    \centering
    \includegraphics[width=0.23\textwidth]{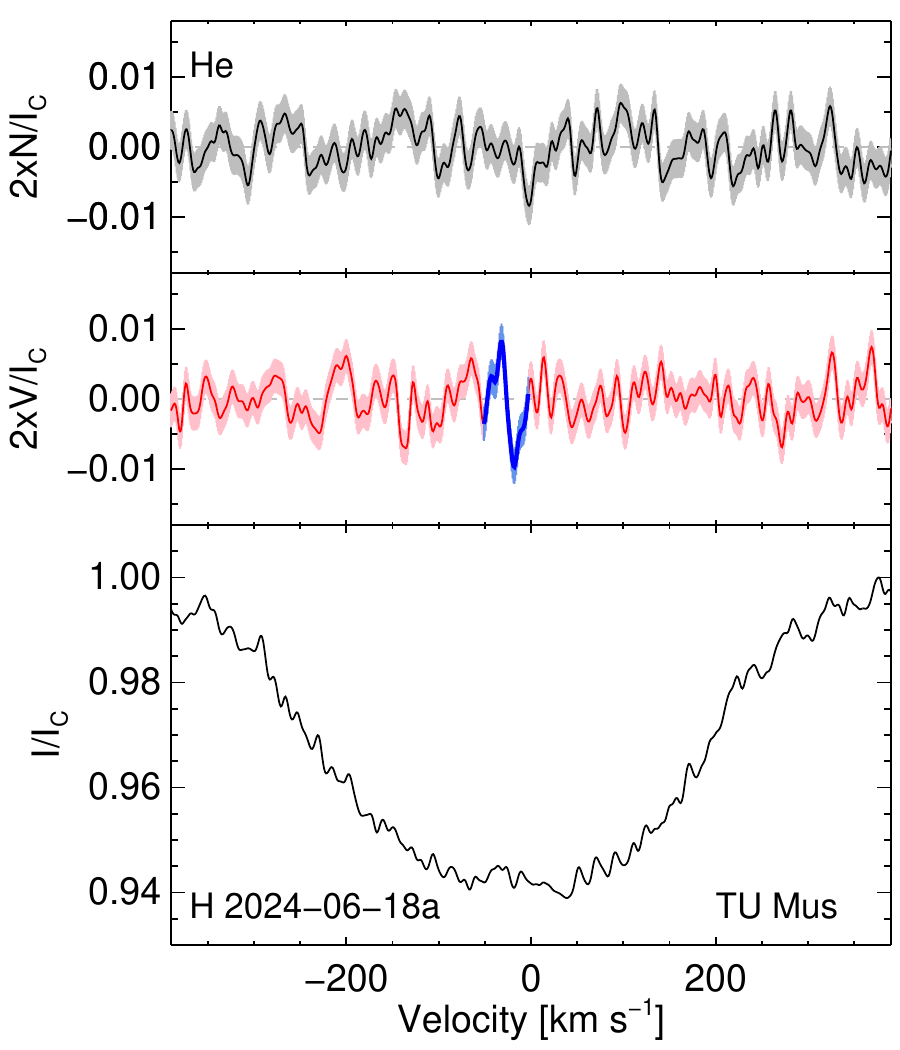}
      \includegraphics[width=0.23\textwidth]{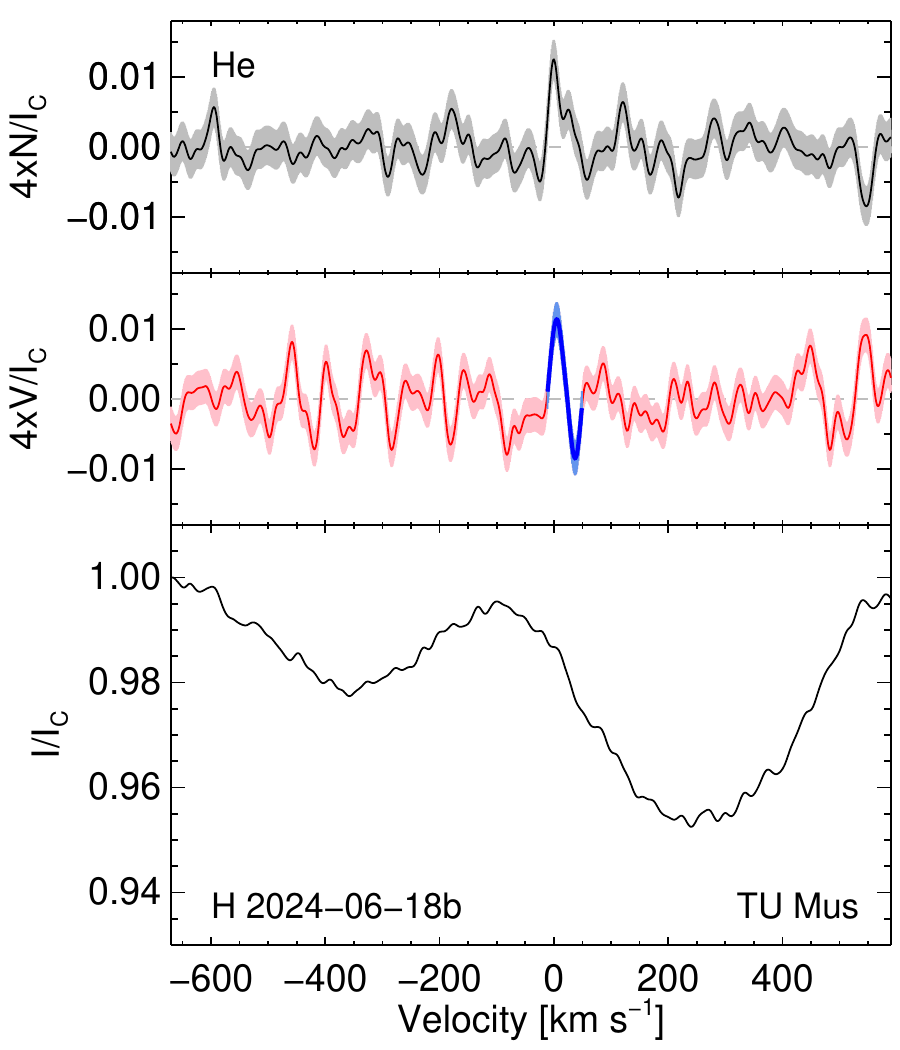}
      \includegraphics[width=0.23\textwidth]{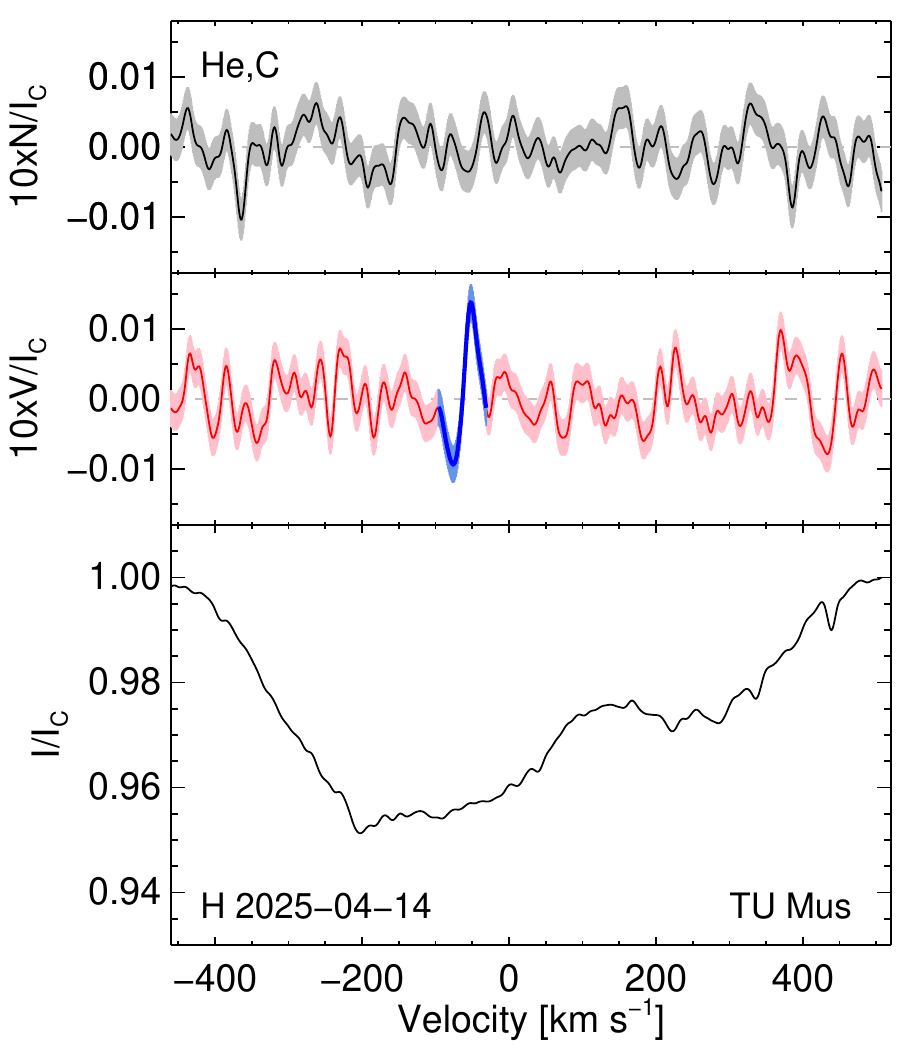}
      \caption{As Figure~\ref{fig:29CMa}, but for TU\,Mus.}
    \label{fig:TUMus}
\end{figure*}

Using archival IUE spectra, \citet{Bagnuolo1994} reported for this eclipsing binary a spectral classification O7.5-8\,Iabf
for the primary and O9.7\,Ib for the secondary and respective masses 19\,$M_{\odot}$ and 16\,$M_{\odot}$,
i.e.\ a mass ratio of $q = M_2/M_1 = 0.84$ \citep{Sen2022}.
The system is expected to be in a contact configuration and the
light curve analysis of photometric observations obtained during the Hipparcos mission 
yielded an orbital period of 4.39336\,d \citep{Antokhina2011}. However, since spectroscopic material was not available at that time,
it was not possible to determine the mass ratio from the light curve solution alone. 
Using BRITE and SMEI photometry in conjunction with concurrent spectroscopy for the analysis of this system,
\citet{Pablo2018} confirmed previously reported asymmetries in the light curve (e.g.\ \citealt{Leung1978}),
which are difficult to model effectively. The authors also found the presence of an unexplained frequency at
twice that of the orbit.
So far, no convincing radial velocity curve of the secondary has been presented in the literature
and attempts to fit both the light curve and the radial velocity curve simultaneously have resulted in unreliable combined fits,
bringing its orbital configuration into question.
\citet{Burssens2020} reported 29\,CMa as an SB1 from IACOB/OWN spectroscopy, in addition to wind variability
in emission. Further, they detected in the one-sector TESS data large amplitude eclipses with a period of 4.40(6)\,d.

Five HARPS\-pol observations were acquired in January and June 2024 and one observation in April 2025. An additional ESPaDOnS observation
of this system from 2013 was retrieved from the CFHT (the Canada-France-Hawaii Telescope) scientific archive.
Clear Zeeman signatures in the LSD Stokes~$V$ profile calculated using various masks including \ion{He}{i/ii},  \ion{C}{iv},
\ion{N}{iii}, \ion{O}{ii/iii}, and \ion{Si}{iii/iv} lines are well visible in all observations and correspond to the
more  massive component.
As is demonstrated in Fig.~\ref{fig:29CMa} and in Table~\ref{tab:obsall},
definite detections are achieved in the ESPaDOnS observation, and the HARPS\-pol observations
on June~17 and 19 2024 and April~15 2025.
The longitudinal magnetic field is variable and shows a change of polarity, with the strongest positive field
$\left< B_{\rm z} \right>=298\pm59$\,G (${\rm FAP}<10^{-10}$) measured on June~19 2024 and the strongest negative field
$\left< B_{\rm z} \right>=-159\pm113$\,G  (${\rm FAP}<10^{-10}$) measured on June~17 2024.
The presence of a magnetic field in 29\,CMa was also confirmed in the recent low-resolution observation carried out on November~10 2024
using the ESO multi-mode instrument FORS2.
The longitudinal magnetic field $\left< B_{\rm z} \right>=-183\pm59$\,G was detected at a
significance of 3.1$\sigma$ ({\sl in preparation}).

\paragraph*{V402\,Pup (=HD\,64315):}

\citet{Lorenzo2017} reported that V402\,Pup is a multiple system consisting of two binary systems, one of which, V402\,Pup BaBb,
is an eclipsing binary. The two binary systems are separated by about 0.09\,\arcsec{}.
The components of the eclipsing binary
V402\,Pup BaBb with an orbital period of 1.019\,d have spectral types O9.5\,V$+$O9.5\,V and almost identical masses
of 14.6\,$M_{\odot}$. 
Both components in the eclipsing system are overfilling their respective Roche lobes,
sharing a common surface.
In a recent overview of massive contact binary observations, \citet{Abdul-Masih2025} reported a fillout factor of $f=1.31$.
The non-eclipsing binary V402\,Pup AaAb is a detached system with an orbital period of 2.710\,d and is composed of
two stars with spectral types around O6\,V. \citet{Lorenzo2017} suggest a minimum mass of
10.8\,$M_{\odot}$ for the primary and 10.2\,$M_{\odot}$ for the secondary, and a mass ratio $q = 0.94$.

The presence of four components in the system V402\,Pup 
complicates the search for the presence of a magnetic field. \citet{Lorenzo2017} reported that the lines corresponding
to V402\,Pup BaBb are mostly hidden within the complex and broad profiles generated by the more luminous components of the
system V402\,Pup AaAb, but each of the four components contributes to every spectrum in the \ion{He}{i},
\ion{He}{ii}, and Balmer lines.
The system was observed with HARPS\-pol twice in April 2025.
As is presented in Fig.~\ref{fig:HD64315} and Table~\ref{tab:obsall}, using a mask with the \ion{He}{i/ii} lines
for the first observation on April~15 2025, we detect in the LSD Stokes~$V$ spectrum a narrow Zeeman signature with ${\rm FAP}=9\times10^{-5}$
corresponding to a marginal detection. For the second observation on April~16 2025 showing all components overlapped,
we detect a rather broad Zeeman signature with ${\rm FAP}=2\times10^{-6}$ using a mask containing \ion{He}{i/ii} and \ion{O}{iii} lines.
For this observation we measure a definite longitudinal magnetic field  $\left< B_{\rm z} \right>=353\pm67$\,G.
Due to the complexity of this multiple system, it is not clear which component(s) in which system possess a magnetic field.

\paragraph*{TU\,Mus (=HD\,100213):}

\begin{figure*}
    \centering
    \includegraphics[width=0.23\textwidth]{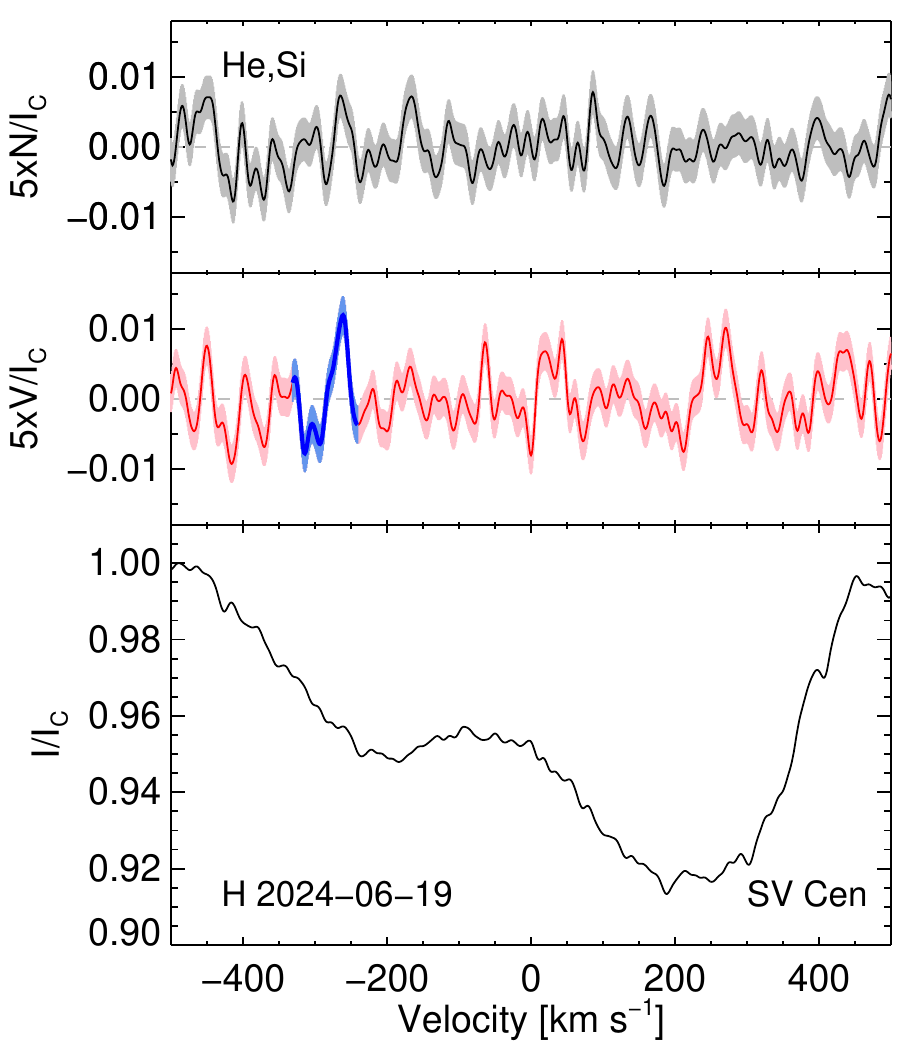}
    \caption{As Figure~\ref{fig:29CMa}, but for SV\,Cen.}
    \label{fig:SVCen}
\end{figure*}

\begin{figure*}
    \centering
    \includegraphics[width=0.23\textwidth]{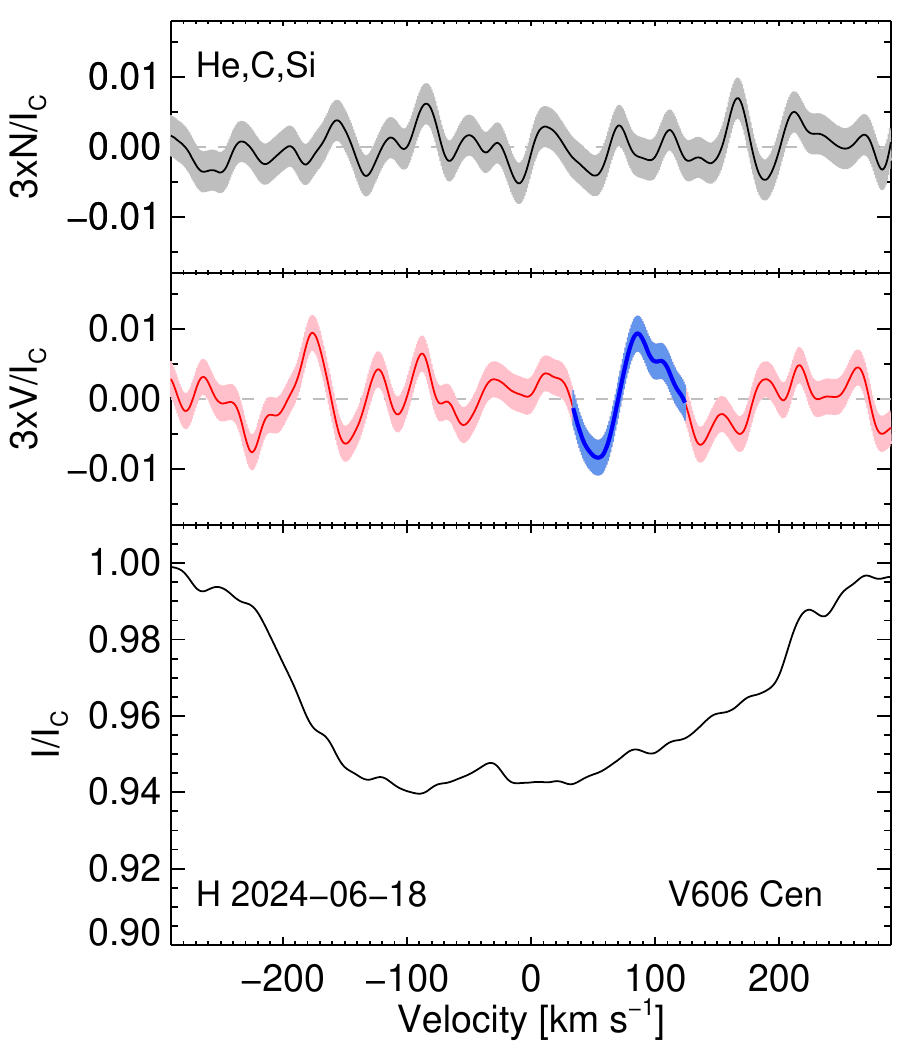}
  \includegraphics[width=0.23\textwidth]{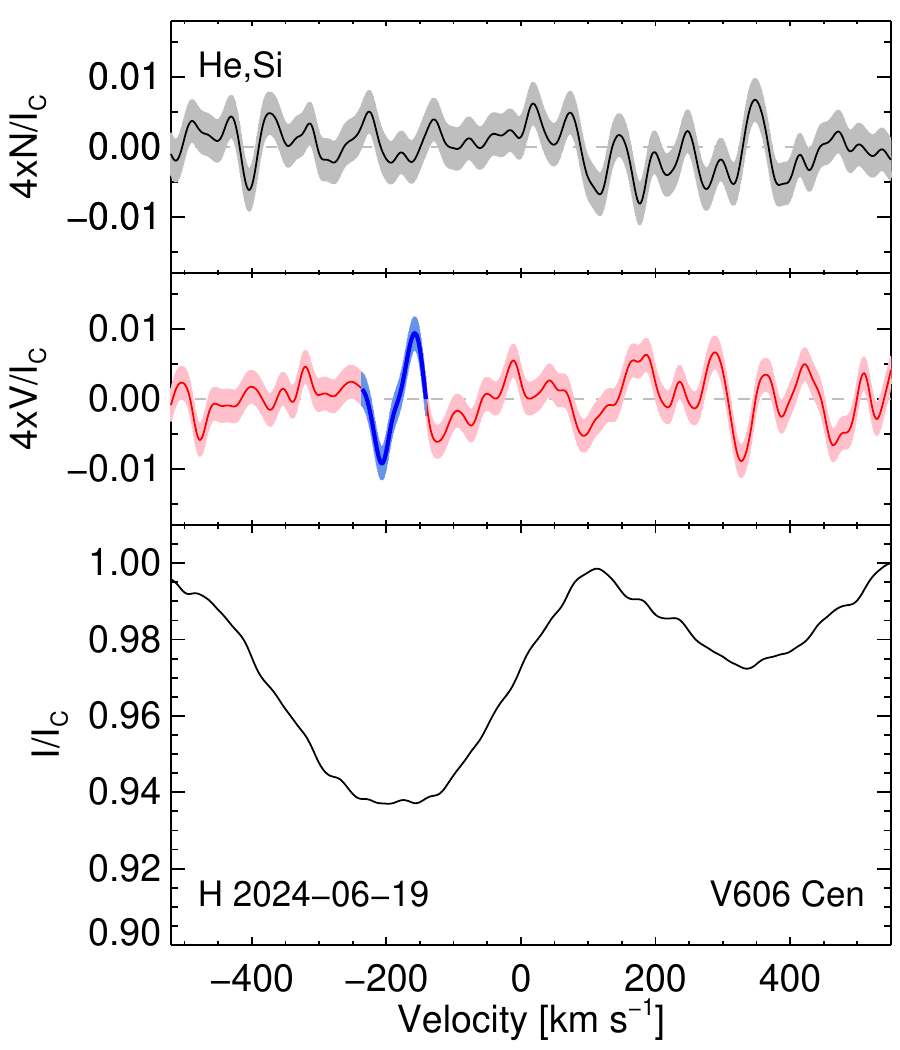}
  \includegraphics[width=0.23\textwidth]{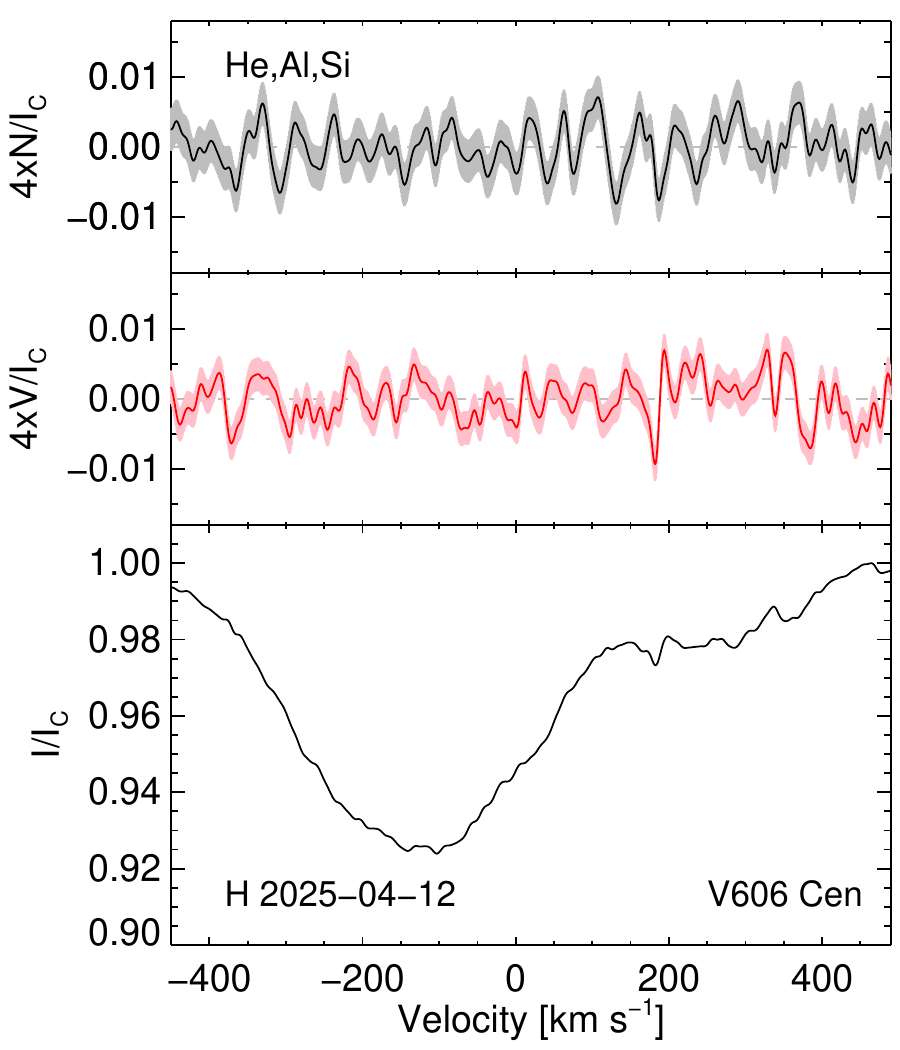}
  \includegraphics[width=0.23\textwidth]{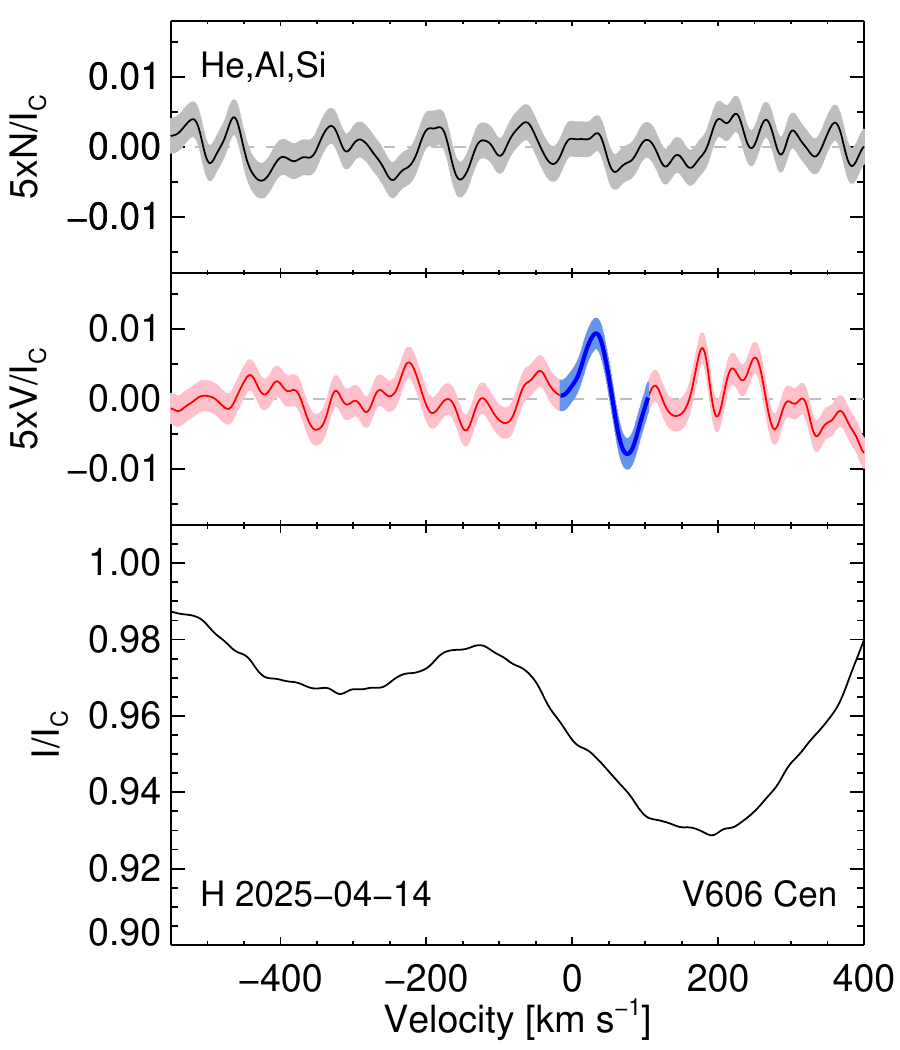}
    \caption{As Figure~\ref{fig:29CMa}, but for V606\,Cen.}
    \label{fig:V606Cen}
\end{figure*}

According to \citet{Penny2008}, the spectral types of the components in the eclipsing binary with an orbital period
of 1.387\,d are O7\,V$+$O8\,V with  
corresponding masses of 16.7\,$M_\odot$ and 10.4\,$M_\odot$, respectively. Their analysis indicated that both stars are currently 
experiencing RLOF. In a recent overview of 
massive contact binary observations, \citet{Abdul-Masih2025} reported a fillout factor of $f=1.12$. With a mass ratio of
0.623, TU\,Mus is one of the most unequal mass contact systems \citep{Abdul-Masih2022}.
The periodic changes in the orbital period of the system detected by \citet{Qian2007}
suggest that it contains a third much fainter component with the lowest mass in the system of 1.55\,$M_\odot$.
This study also revealed the presence of a distant fourth visual companion
at a separation of 16\,\arcsec{}
southeast of TU\,Mus. The existence of additional components in TU\,Mus probably plays an
important role for their formation and evolution by removing angular momentum from the central system \citep{Qian2007}.

TU\,Mus was observed three times, twice in June 2024 and once in April 2025.
As is presented in Fig.~\ref{fig:TUMus} and Table~\ref{tab:obsall}, using a line list containing \ion{He}{i/ii} lines, we detect a
definite magnetic field
$\left< B_{\rm z} \right>=482\pm198$\,G with ${\rm FAP}=10^{-5}$ in the first observation on June~18 2024. In this orbital phase
the binary components appear overlapped. The components are well separated during the second observation one day later,
but the Zeeman signature detected using \ion{He}{i/ii} lines is marginal with $7\times10^{-4}$.
The position of the  definitely detected Zeeman signature on April~14 2025 with  ${\rm FAP}=2\times10^{-6}$ using a mask with 
\ion{He}{i/ii} and \ion{C}{iv} lines probably indicates that the more massive component in the eclipsing binary possesses a magnetic field.

\paragraph*{SV\,Cen (=HD\,102552):}

This system consists of two components with spectral types B1.5\,V and B6.5\,III \citep{Drechsel1994},
with masses of 8.56\,$M_{\odot}$ and 6.05\,$M_{\odot}$ ($q = 0.71$) \citep{Rensbergen2011}, respectively.
Both components in a tight 1.66\,d orbit apparently undergo a very rapid orbital evolution.
The configuration of this system has been widely debated in the literature. 
Some studies claim that it is a contact system (e.g.\ \citealt{Rucinski1992,Drechsel1994}),
while others claim that it is semi-detached (e.g.\ \citealt{Degreve1994,Deschamps2013,Davis2014}).
While contact systems evolve on a nuclear timescale, 
the recent analysis of SV\,Cen by \citet{Vrancken2024} suggests
that SV\,Cen is evolving on a thermal timescale.
This indicates that the configuration is more likely semi-detached instead of contact.
According to the work of these authors,
most of the studied contact systems have a period stability of at least $\sim$1\,Myr up to 20\,Myr,
while for SV\,Cen it is about 0.05\,Myr.

One spectropolarimetric observation for this relatively faint target was carried out on June~19 2024.
As is presented in Fig.~\ref{fig:SVCen} and Table~\ref{tab:obsall}, the Zeeman signature is only marginally detected
(${\rm FAP}=10^{-3}$) using a mask with \ion{He}{i} and \ion{Si}{iii} lines.
The position of this signature relative to the Stokes~$I$ spectrum showing separated components
suggests an association with the less massive component.

\paragraph*{V606\,Cen (=HD\,115937):}

\begin{figure*}
    \centering
    \includegraphics[width=0.195\textwidth]{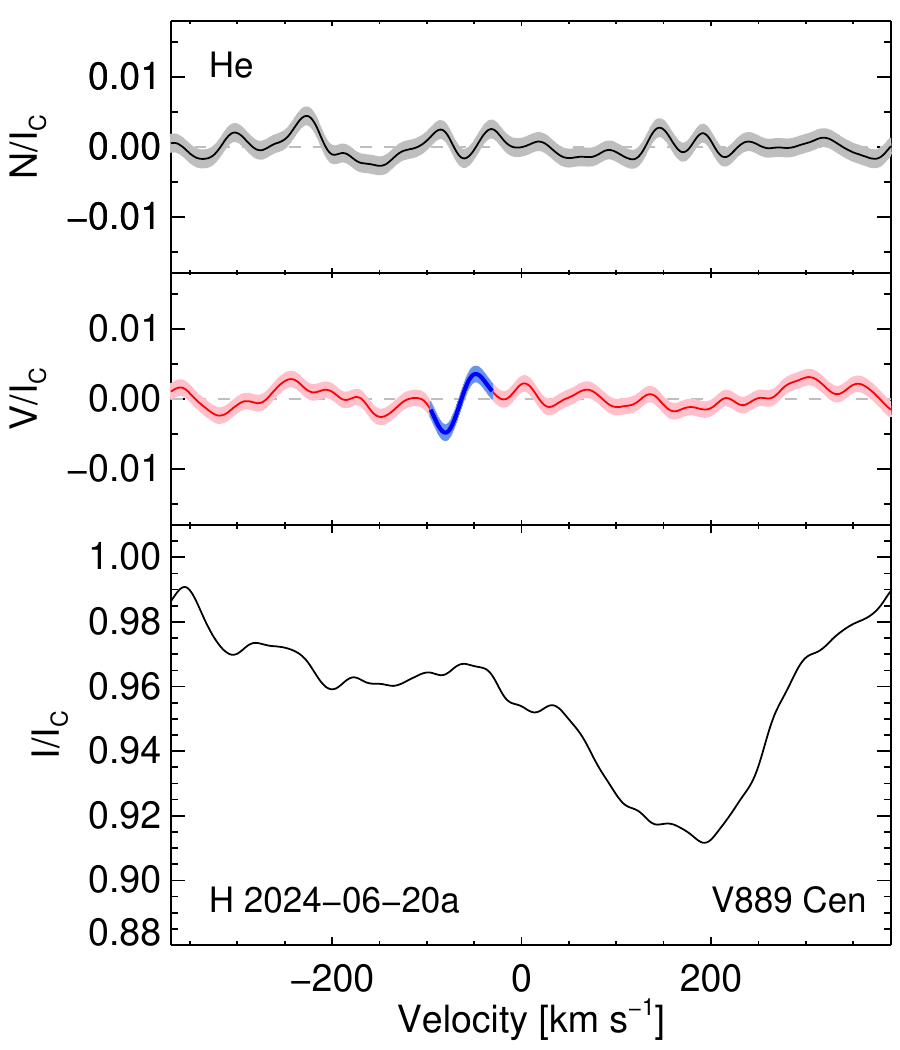}
    \includegraphics[width=0.195\textwidth]{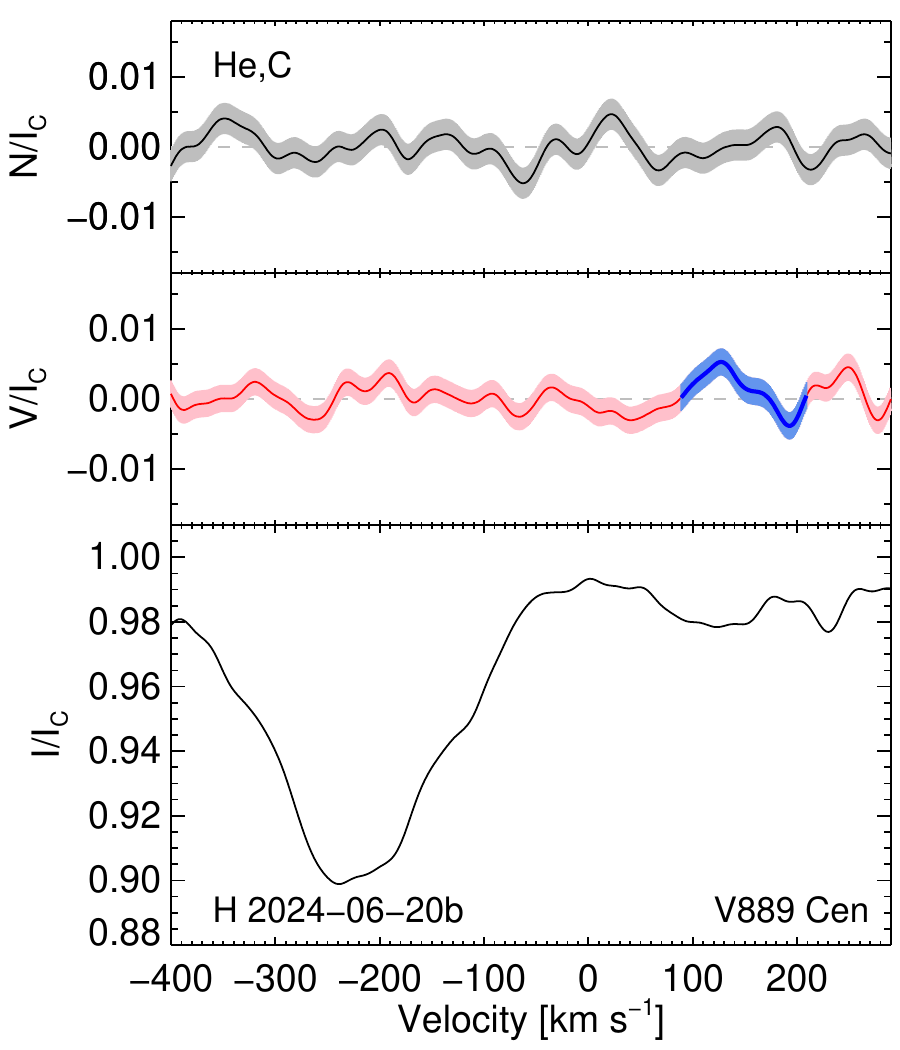}
    \includegraphics[width=0.195\textwidth]{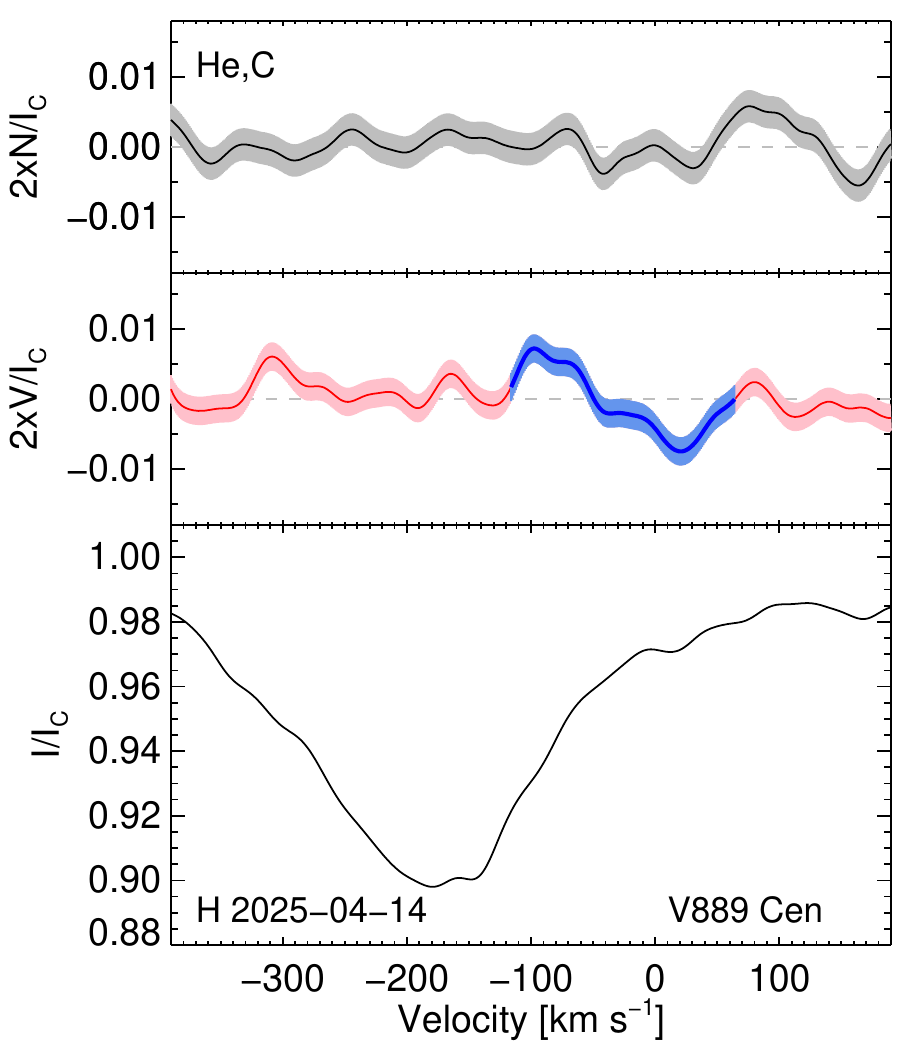}
    \includegraphics[width=0.195\textwidth]{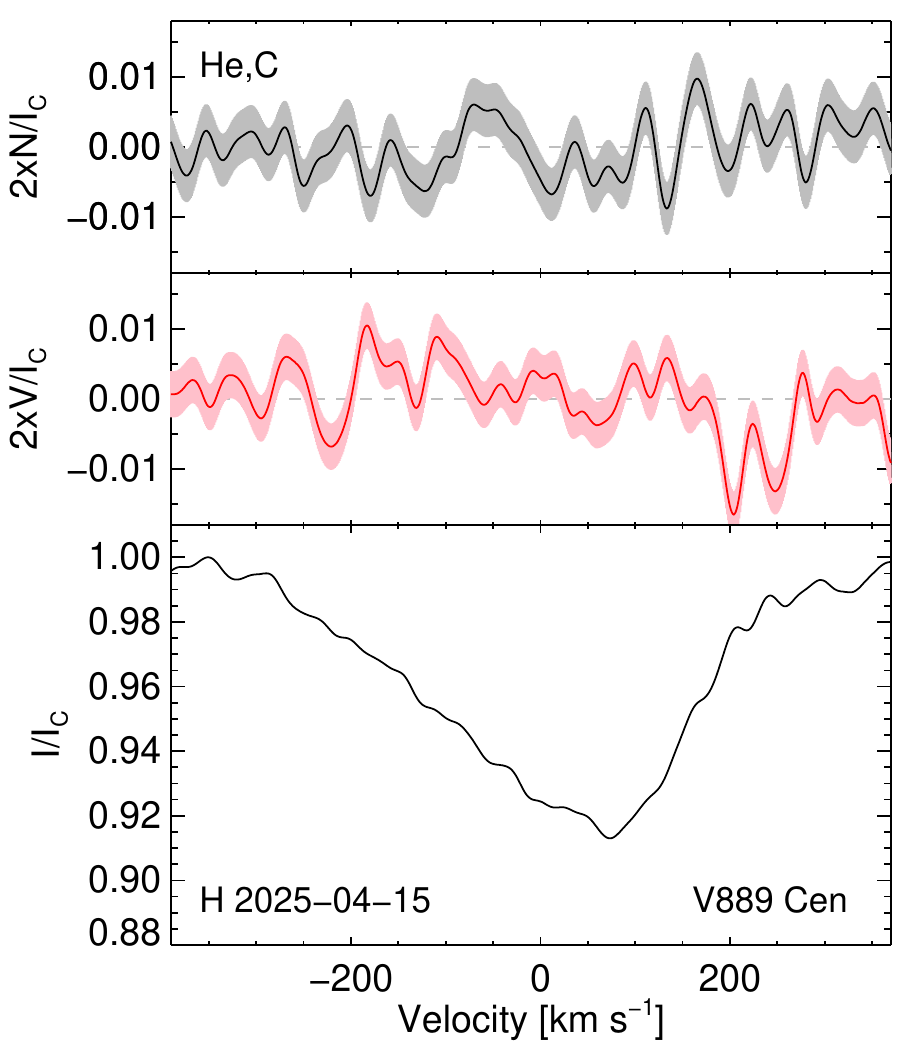}
    \includegraphics[width=0.195\textwidth]{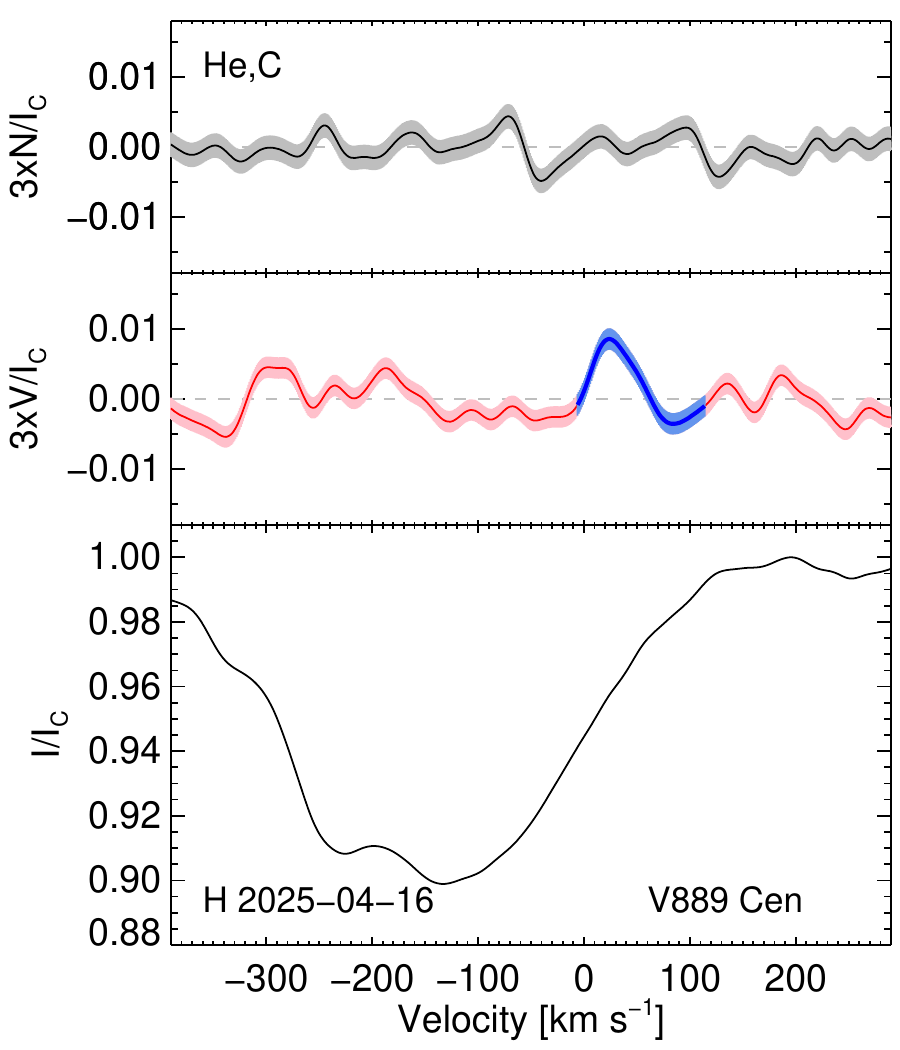}
        \caption{As Figure~\ref{fig:29CMa}, but for V889\,Cen.}
    \label{fig:V889Cen}
\end{figure*}

\begin{figure*}
    \centering
    \includegraphics[width=0.23\textwidth]{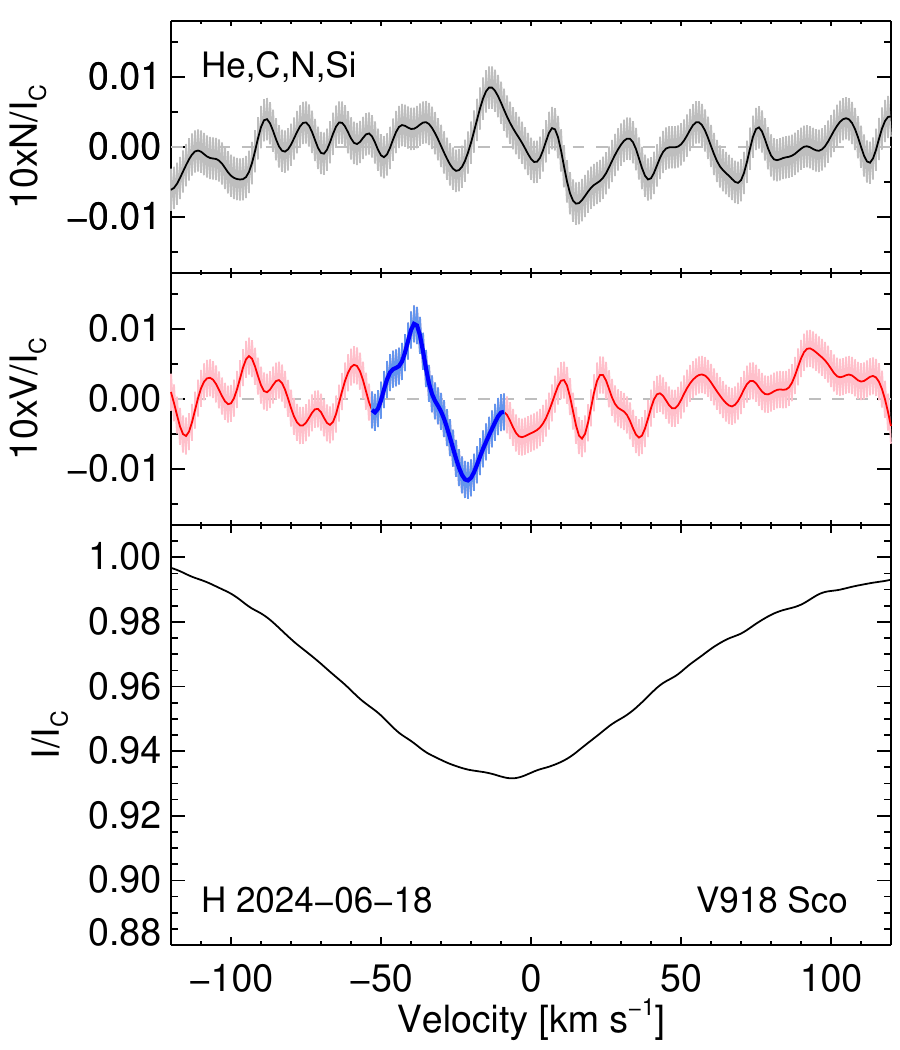}
     \includegraphics[width=0.23\textwidth]{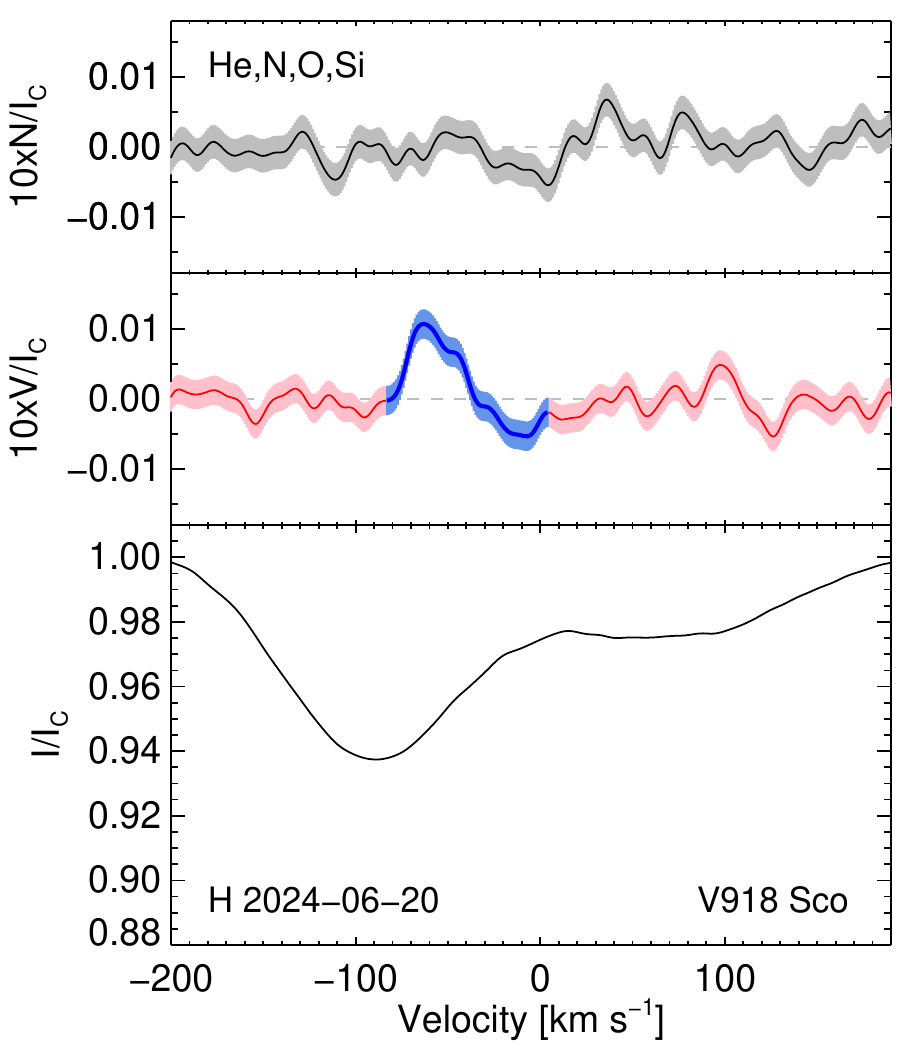}
     \includegraphics[width=0.23\textwidth]{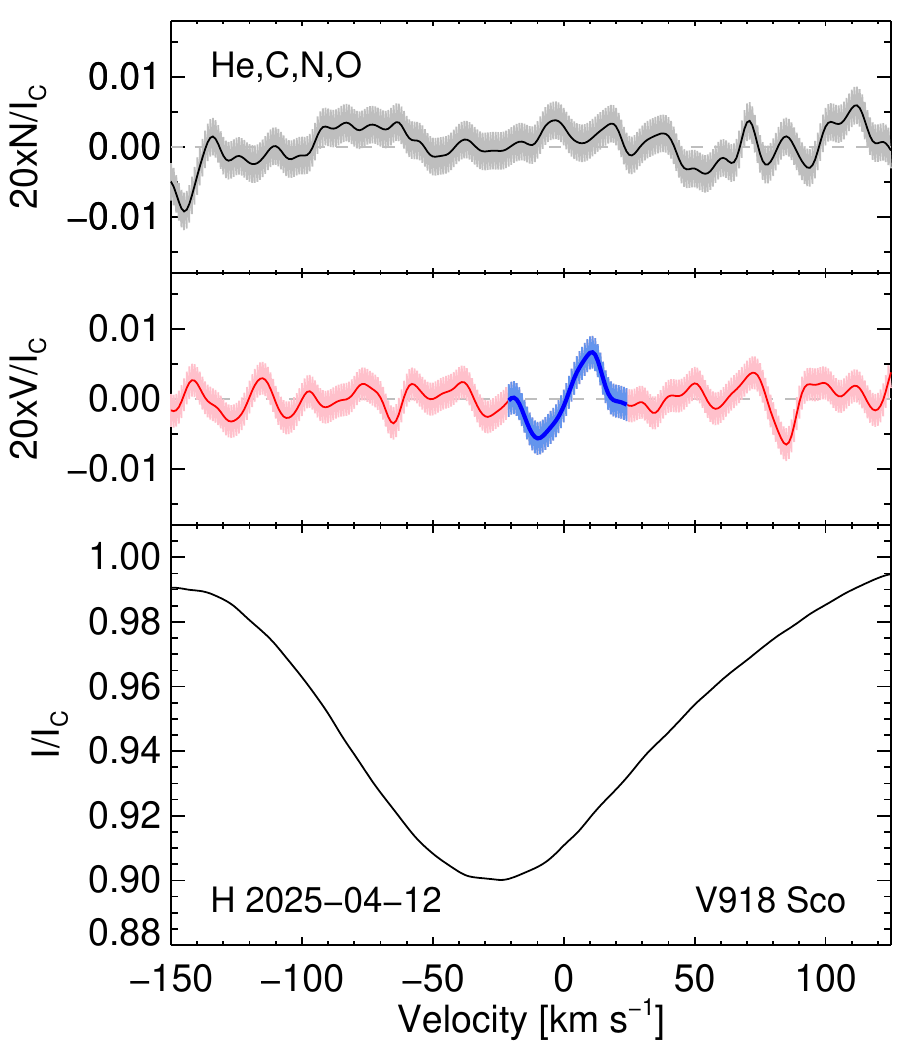}
     \includegraphics[width=0.23\textwidth]{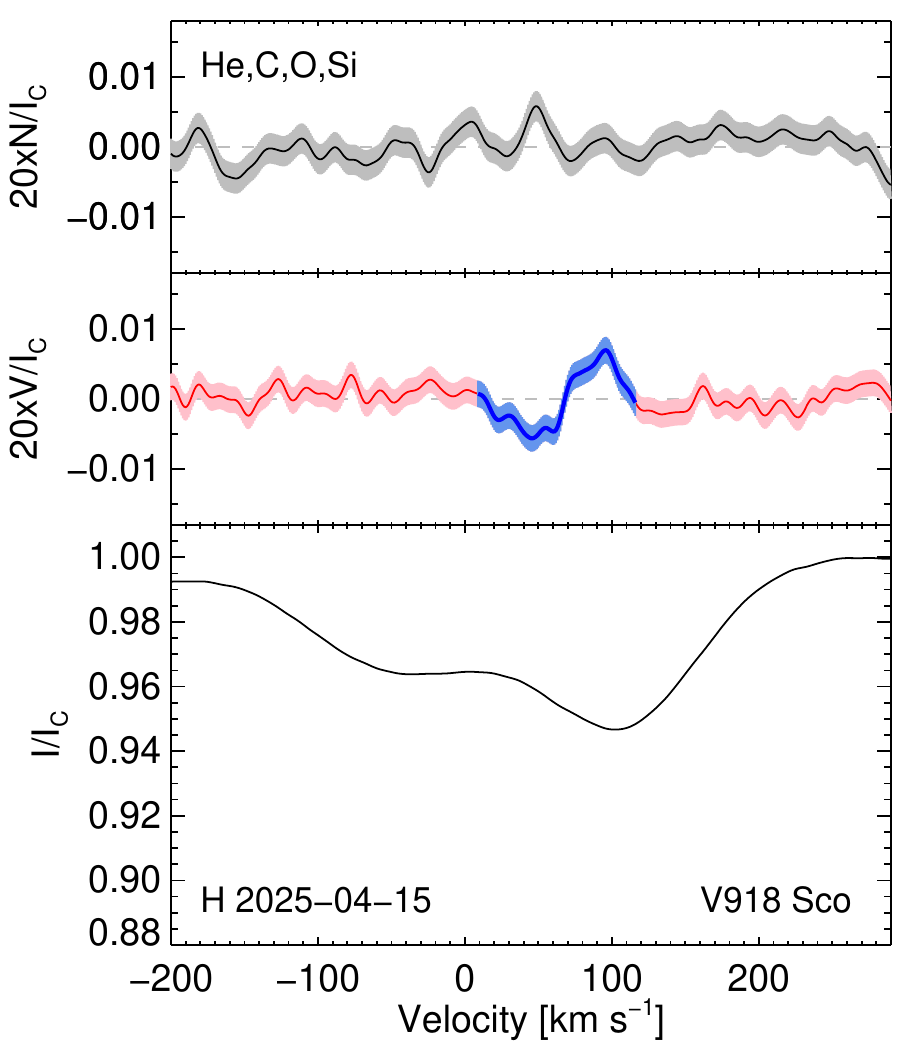}
    \caption{As Figure~\ref{fig:29CMa}, but for V918\,Sco.}
    \label{fig:HD149404}
\end{figure*}

The components of the rather faint eclipsing contact binary V606\,Cen
with an orbital period of 1.495\,d have spectral types B0-B0.5\,V
for the primary with a mass of 14.7\,$M_{\odot}$
and B2-B3\,V for the secondary component with a mass of 8.0\,$M_{\odot}$ \citep{Lorenz1999}.
By analysing the continuous light curve obtained by TESS,
\citet{Li2022} found that it is a marginal contact binary
with a very low fillout factor of about 2\%.
Both the marginal contact configuration and the continuous period decrease suggest that
V606 Cen is a newly formed contact binary via case~A mass transfer. Their results also indicated that
V606\,Cen is a member in a hierarchical triple system:
the cyclic change in the O--C diagram was explained by the light-travel time effect via the presence of a third body,
with the tertiary having the lowest mass in the system of 4.51\,$M_{\odot}$.
The tertiary is orbiting the central eclipsing
binary with $q = 0.5484$ in an excentric orbit with $e = 0.33$. 

This system was observed twice each in June 2024 and April 2025.
As is presented in Fig.~\ref{fig:V606Cen} and Table~\ref{tab:obsall}, while only a marginal detection of a Zeeman signature
(${\rm FAP}=4\times10^{-4}$) was achieved
for the first observation on June~18 2024, showing overlapped binary components, we achieved a definite detection 
$\left< B_{\rm z} \right>=-419\pm168$\,G with ${\rm FAP}=2\times10^{-6}$ on 
June~19 2024 using a mask with \ion{He}{i/ii}, and \ion{Si}{iii} lines. A second definite detection,
$\left< B_{\rm z} \right>=1327\pm128$\,G with ${\rm FAP}<10^{-10}$
was obtained on April~14 2025 using a mask containing \ion{He}{i/ii}, \ion{Al}{iii},
and \ion{Si}{iii} lines. 
The positions of the Zeeman signatures relative to the composite
Stokes~$I$ spectrum indicate that the magnetic field is detected in the more massive component.
No detection was achieved on April~14 2025 with a mask containing \ion{He}{i/ii}, \ion{Al}{iii},
and \ion{Si}{iii} lines.

\paragraph*{V889\,Cen (=LSS\,3074):}

With $m_{\rm V}=11.7$, this system is the faintest in our sample.
Thus the $S/N$ values for the obtained spectra are very low, in the range
from 24 to 49. The orbital parameters of this short-period spectroscopic binary with masses of 14.8\,$M_{\odot}$ for the primary
and 17.2\,$M_{\odot}$ for the secondary ($q = 0.86$) and spectral types O4f$^{+}$ and O6-7:(f):, respectively,
were reported by \citet{Raucq2017}.
The authors assumed a circular orbit and an orbital period of 2.1852 days.
An analysis of the light curve indicated that V889\,Cen is a candidate for a contact binary with $f=1.05$
and has lost a significant portion of its mass to its surroundings.
According to \citet{Abdul-Masih2022}, the photometric analysis strongly favours a contact configuration
over a semi-detached configuration.
\citet{Raucq2017} detected a strong overabundance in
nitrogen and a strong carbon and oxygen depletion in the atmospheres of both primary and secondary, and
a strong helium enrichment  of the primary star.

V889\,Cen was observed two times in June 2024 and three times in April 2025.
As is presented in Fig.~\ref{fig:V889Cen} and Table~\ref{tab:obsall},  using a line mask containing \ion{He}{i/ii} lines
we achieved a definite detection of a Zeeman signature with ${\rm FAP}<10^{-9}$ on June~20 2024 and a marginal detection
using a mask containing \ion{He}{i/ii} and \ion{C}{iv} lines with ${\rm FAP}=6\times10^{-4}$ one day later.
In the observations carried out in 2025, using a mask with \ion{He}{i/ii} and \ion{C}{iv} lines, we achieved definite
detections of a Zeeman signature with ${\rm FAP}=6\times10^{-7}$ and ${\rm FAP}<10^{-10}$ on April~14 and 16 2025, respectively.
No Zeeman signature was detected on April~15 2025.
The positions of the detected Zeeman signatures relative to the composite
Stokes~$I$ spectrum indicate that the magnetic field is present in the less massive component. However, due to the fact that
in our low-$S/N$ spectra the secondary component is not well separated, we are not able to measure
the longitudinal magnetic field corresponding to the detected Zeeman signatures.

\paragraph*{V918\,Sco (=HD\,149404):}

\begin{figure*}
    \centering
    \includegraphics[width=0.23\textwidth]{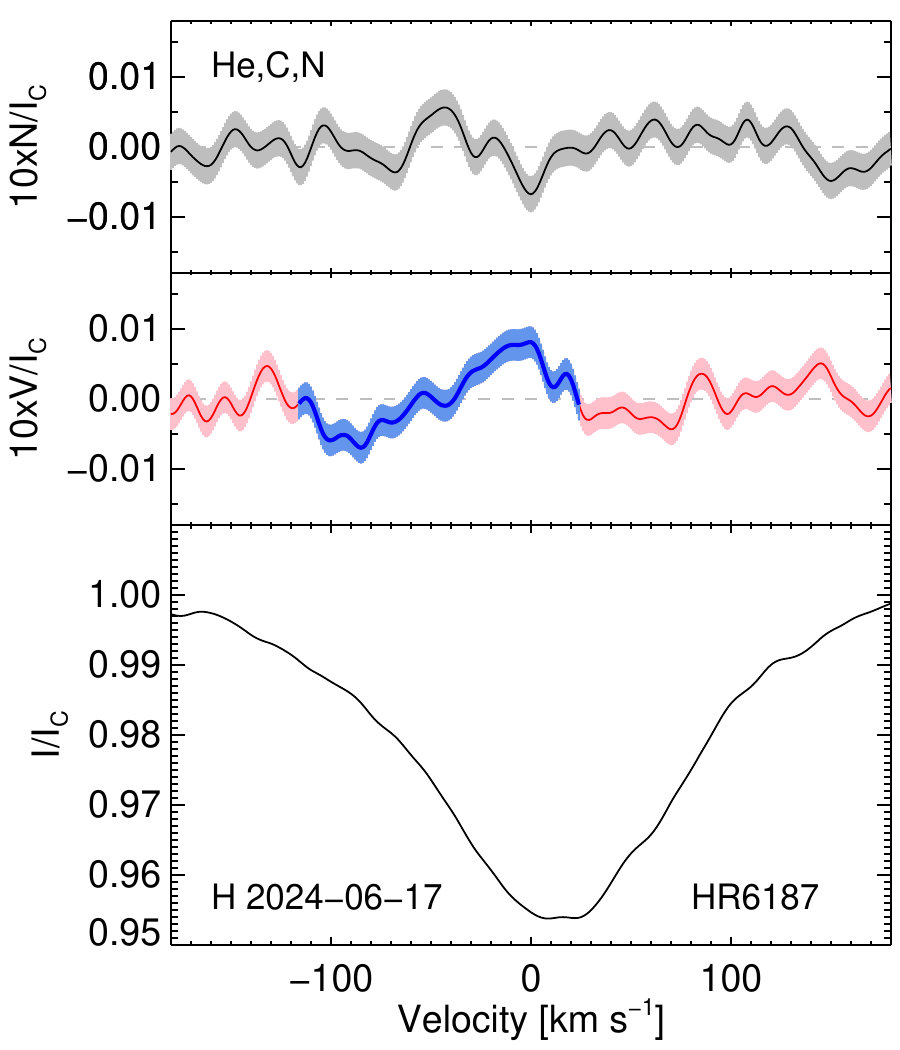}
    \caption{As Figure~\ref{fig:29CMa}, but for HR\,6187.}
    \label{fig:HD150136}
\end{figure*}

According to \citet{Rauw2001}, the ON9.7\,I secondary in this system is on a circular orbit with a period of 9.81\,d
around the O7.5\,I(f) primary and  seems to be the
most evolved component. \citet{Raucq2016} reported 50.5\,$M_{\odot}$ for the primary and 31.9\,$M_{\odot}$ for the secondary.
and that the current evolutionary status of V918\,Sco could best be
explained if the system has undergone a RLOF episode in the past.
The authors detected a large overabundance in nitrogen and a carbon and oxygen depletion in the secondary star and also found
a slight nitrogen enhancement in the primary’s spectrum. Furthermore, they inferred an asynchronous
rotation of the two stars of the system.
It is also possible that both components emit powerful,
radiatively driven stellar winds.
\citet{Thaller1998} suggested that the double-peaked structure of
the H$\alpha$ emission observed in this system is due to colliding wind interaction. Also \citet{Rauw2001} favoured a model where the
H$\alpha$ emissions arise in the arms of a colliding wind shock region.

The presence of a magnetic field in V918\,Sco was recently studied  by
\citet{Hubrig2023}, who used for the LSD analysis one HARPS\-pol observation obtained in 2016 and one ESPaDOnS
observation acquired in 2014. For both observations the authors achieved marginal detections with ${\rm FAP}=5\times10^{-4}$
and ${\rm FAP}=7\times10^{-4}$, respectively.
As is shown in Fig.~\ref{fig:HD149404} and Table~\ref{tab:obsall}, this system was observed four more times.
We achieved definite detections in both observations obtained in
June 2024 and in the second observation acquired in April 2025. The longitudinal magnetic field is rather weak with
$\left< B_{\rm z} \right>=45\pm31$\,G and ${\rm FAP}=6\times10^{-6}$ measured on June~20 2024 using a mask containing
\ion{He}{i/ii}, \ion{C}{iv}, \ion{N}{iii}, and \ion{Si}{iii} lines.
The first observation in April 2025 showed a marginal detection $\left< B_{\rm z} \right>=-11\pm6$\,G with ${\rm FAP}=4\times10^{-4}$ using
a mask with \ion{He}{i/ii}, \ion{C}{iv}, \ion{N}{iii}, and \ion{O}{iii} lines.

\paragraph*{HR\,6187  (=HD\,150136):}

\begin{figure*}
    \centering
    \includegraphics[width=0.23\textwidth]{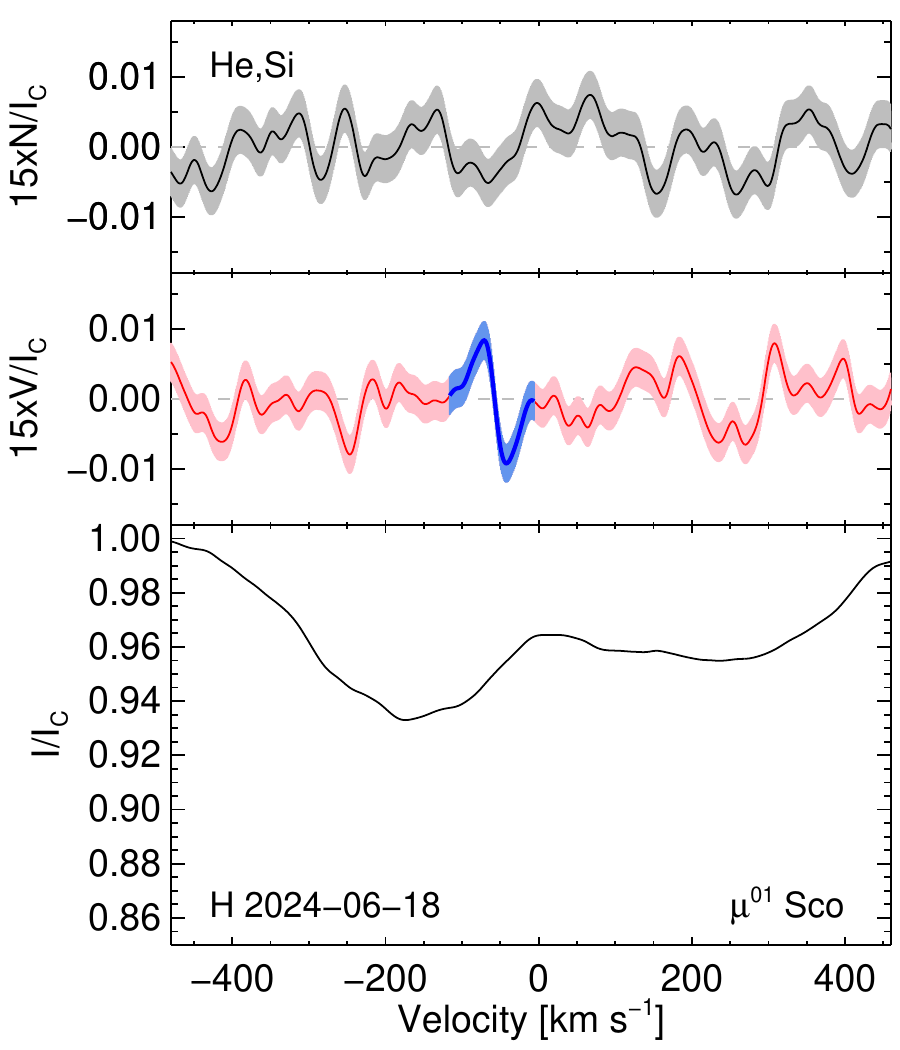}
    \includegraphics[width=0.23\textwidth]{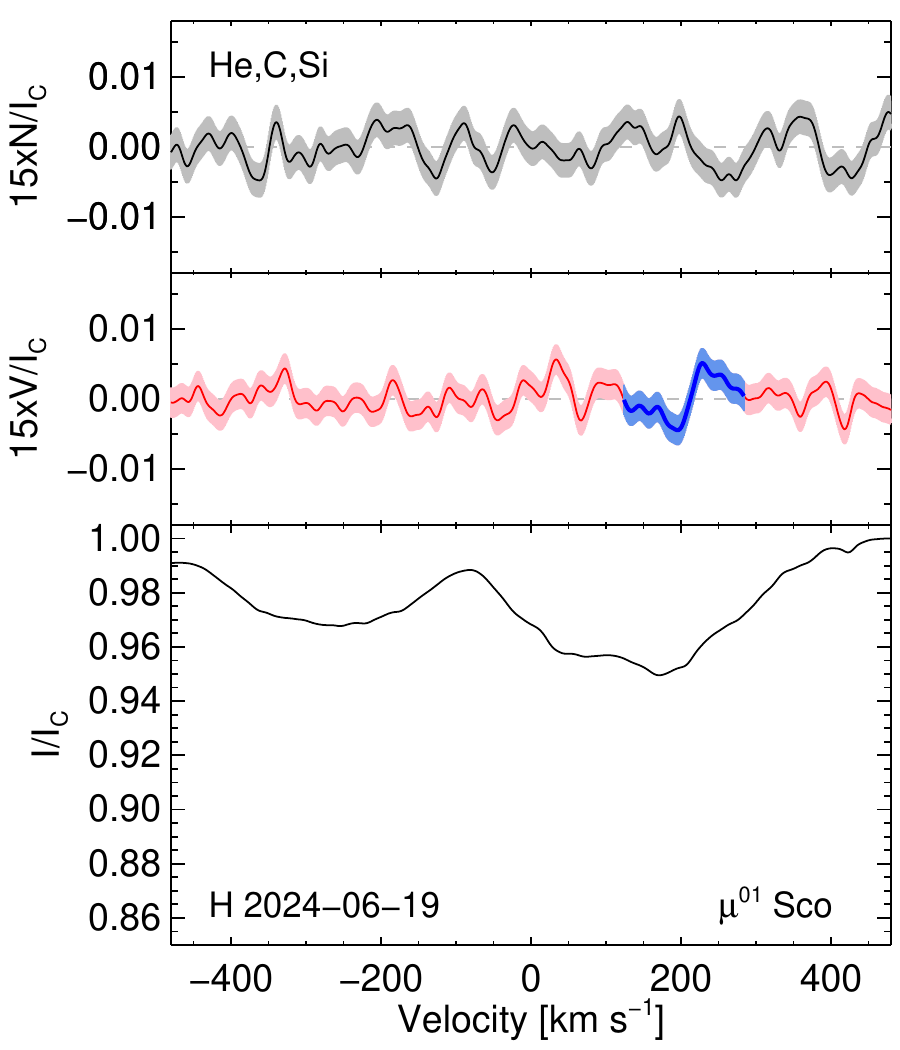}
     \includegraphics[width=0.23\textwidth]{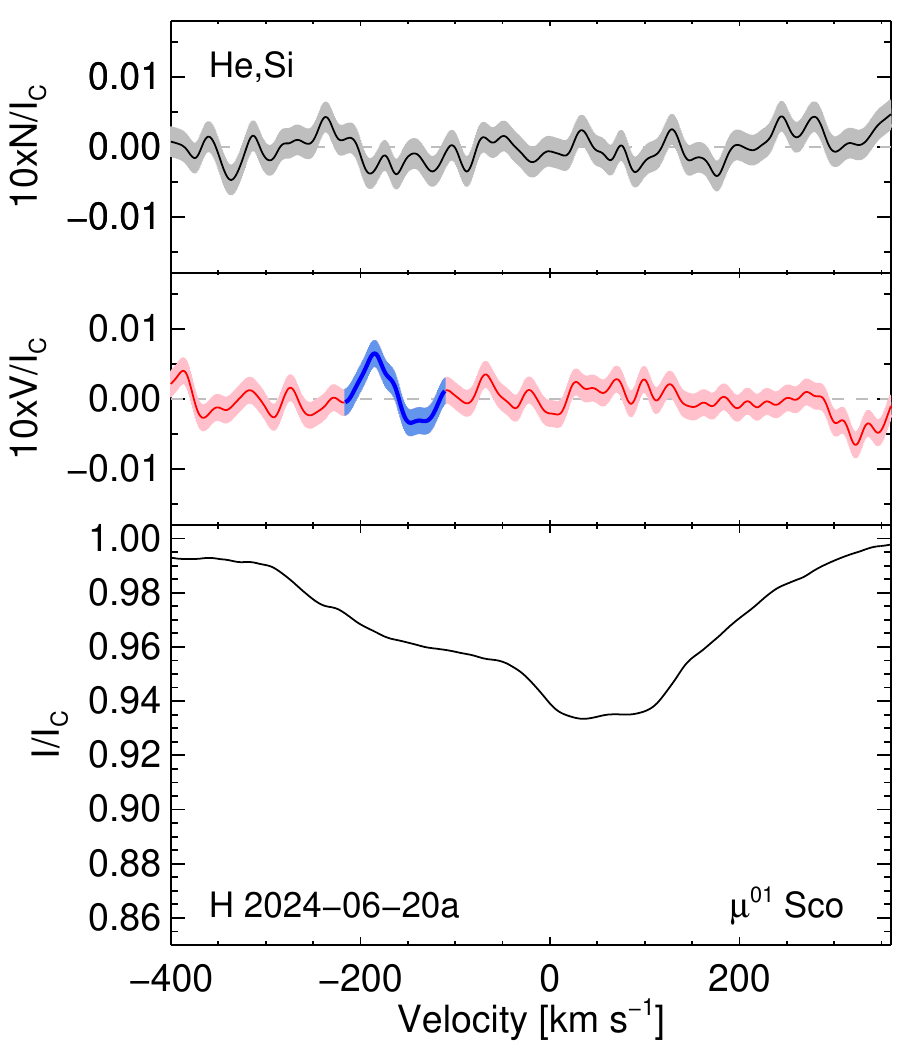}
    \includegraphics[width=0.23\textwidth]{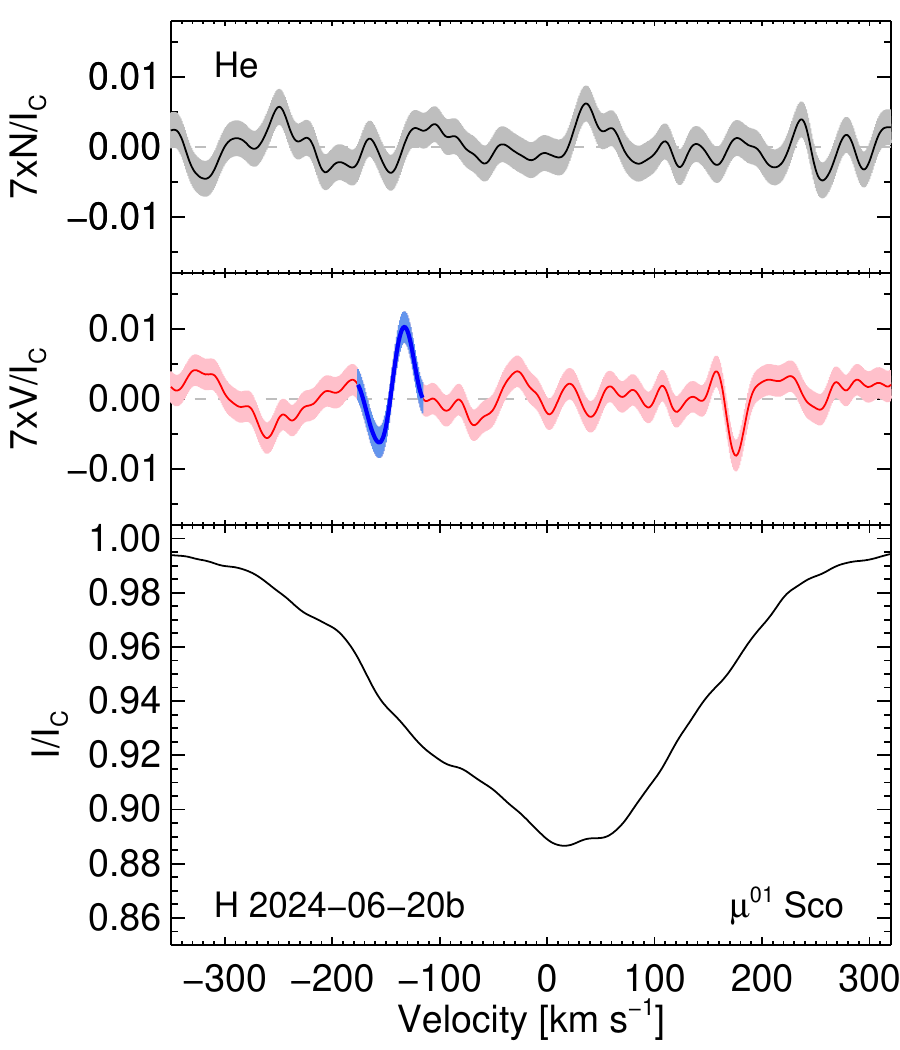}
    \includegraphics[width=0.23\textwidth]{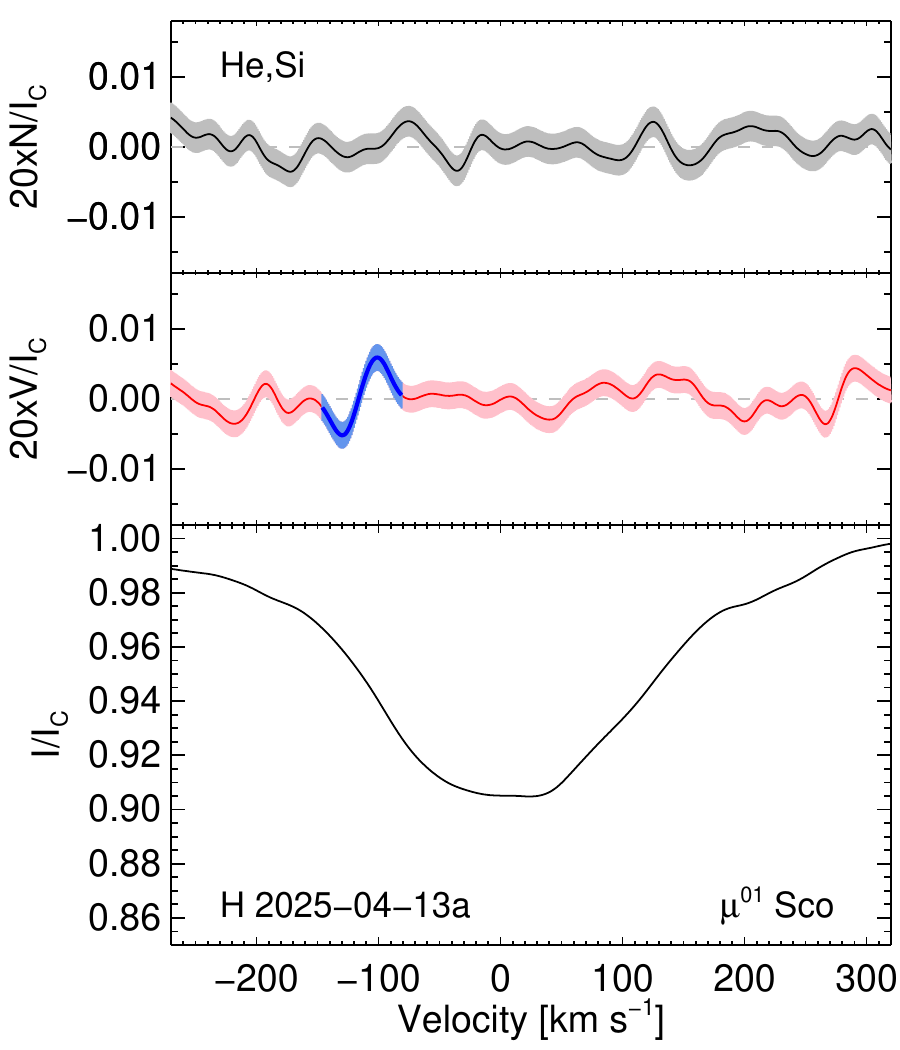}
    \includegraphics[width=0.23\textwidth]{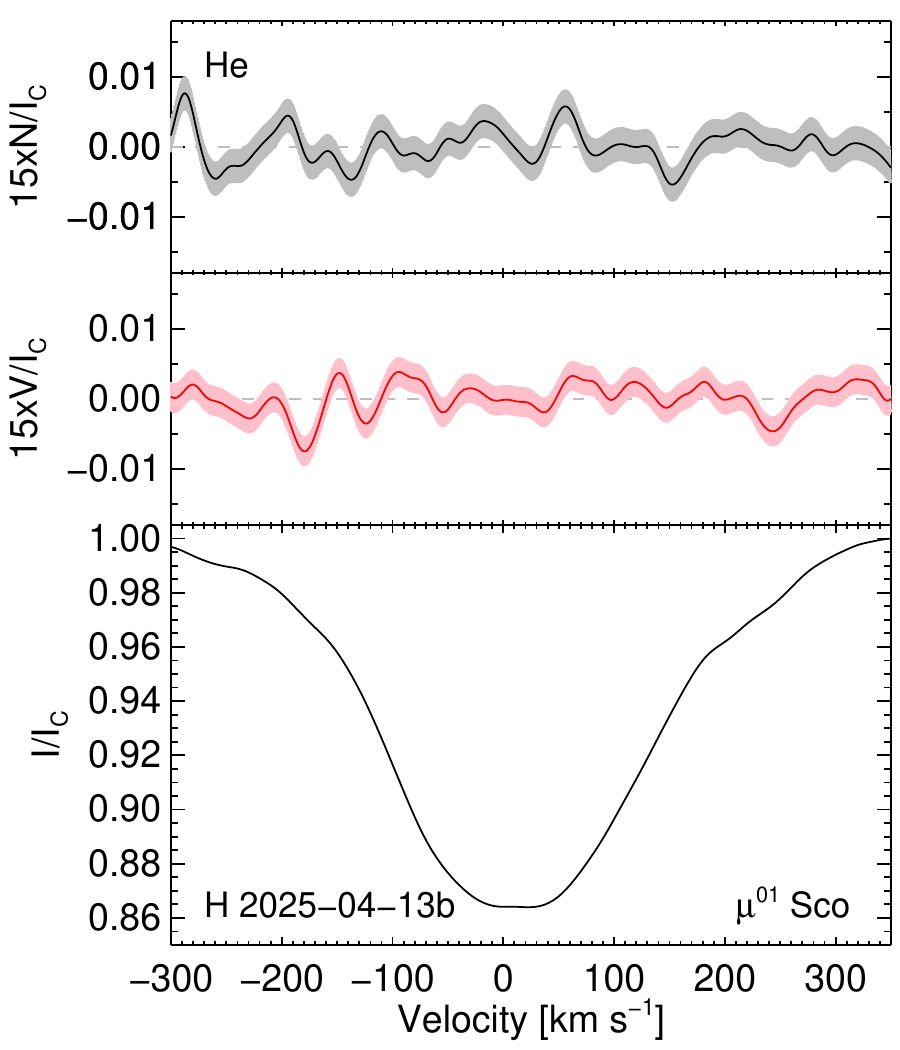}
     \includegraphics[width=0.23\textwidth]{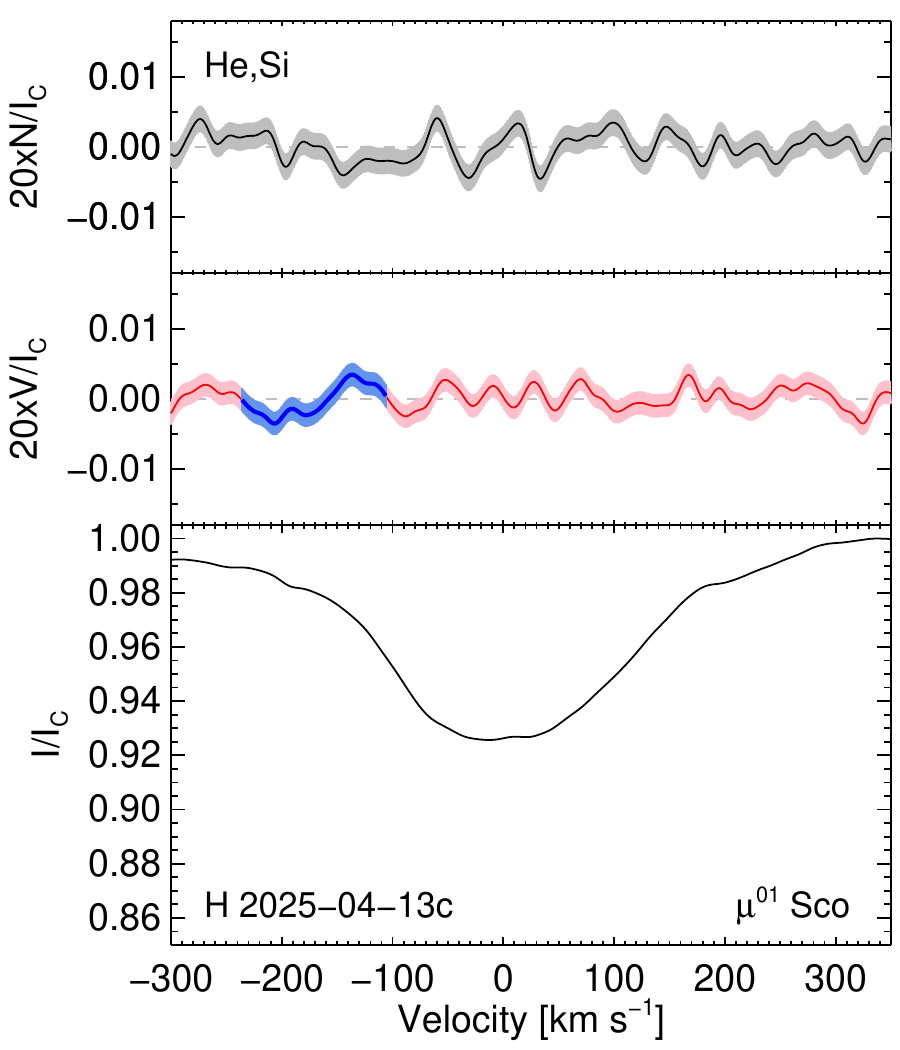}
     \includegraphics[width=0.23\textwidth]{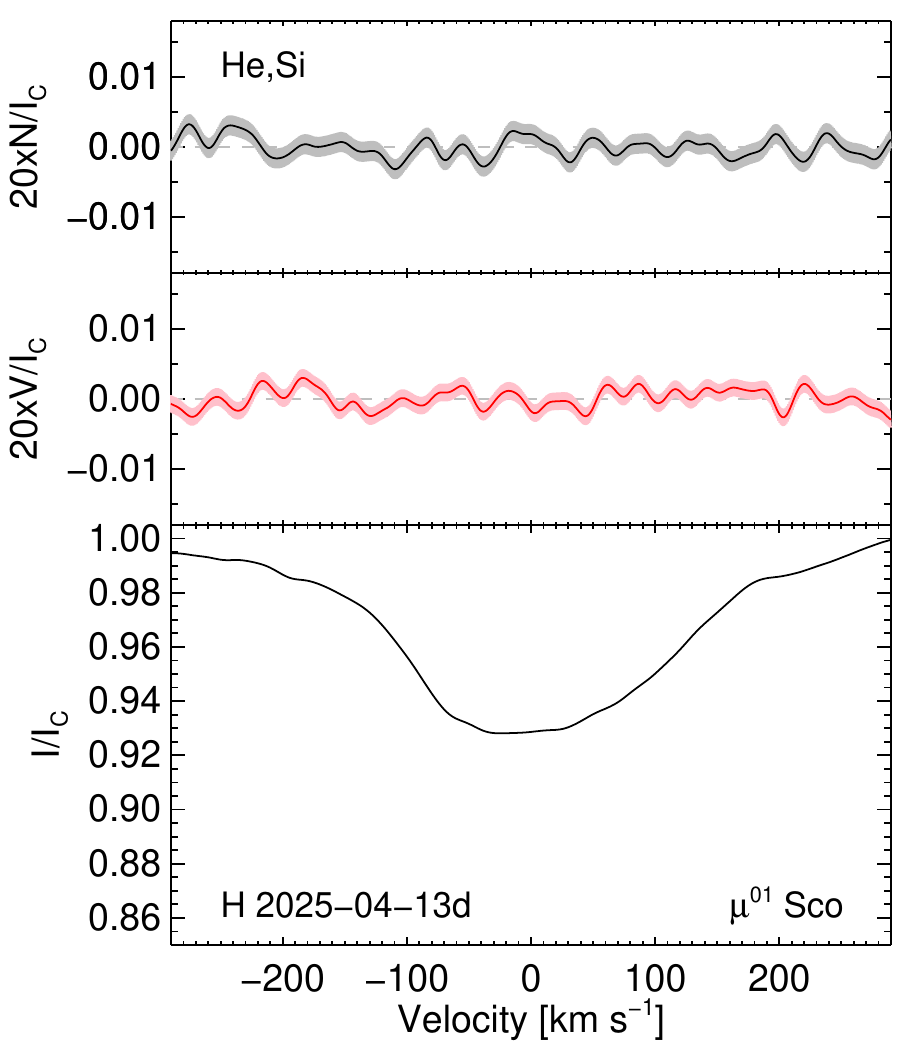}
    \includegraphics[width=0.23\textwidth]{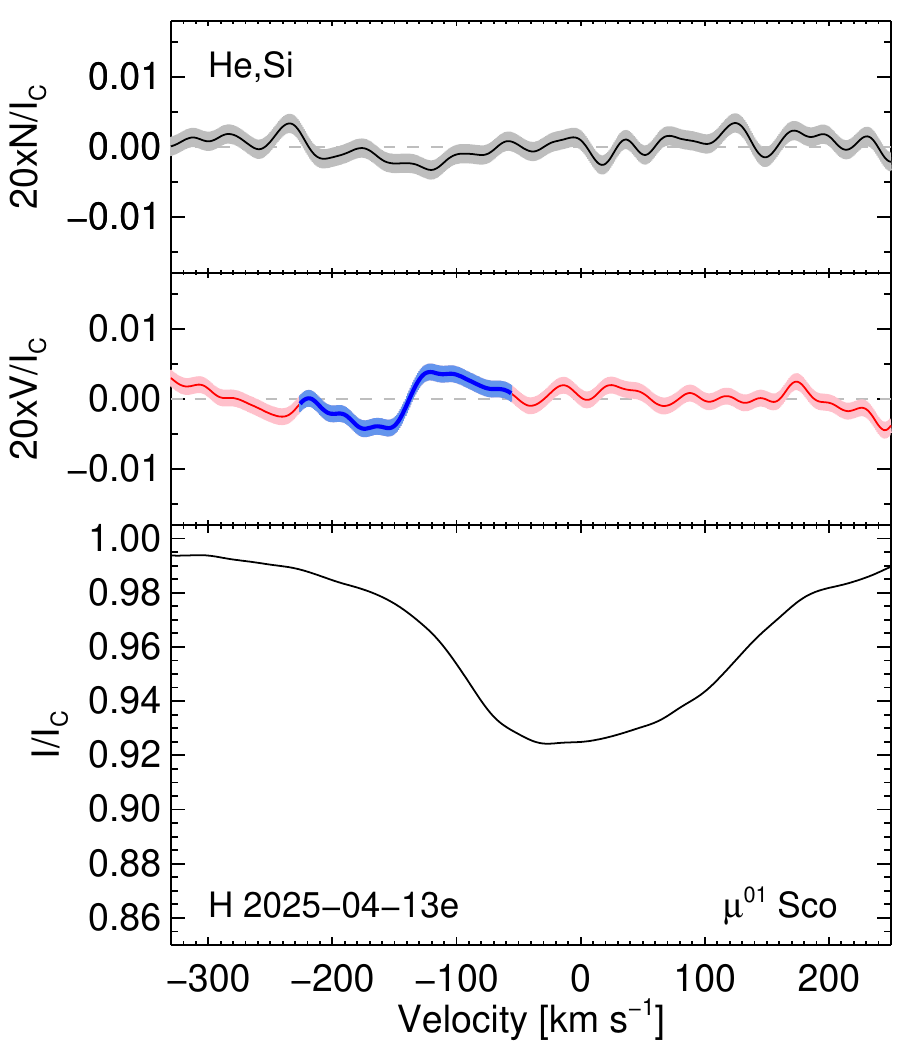}
    \includegraphics[width=0.23\textwidth]{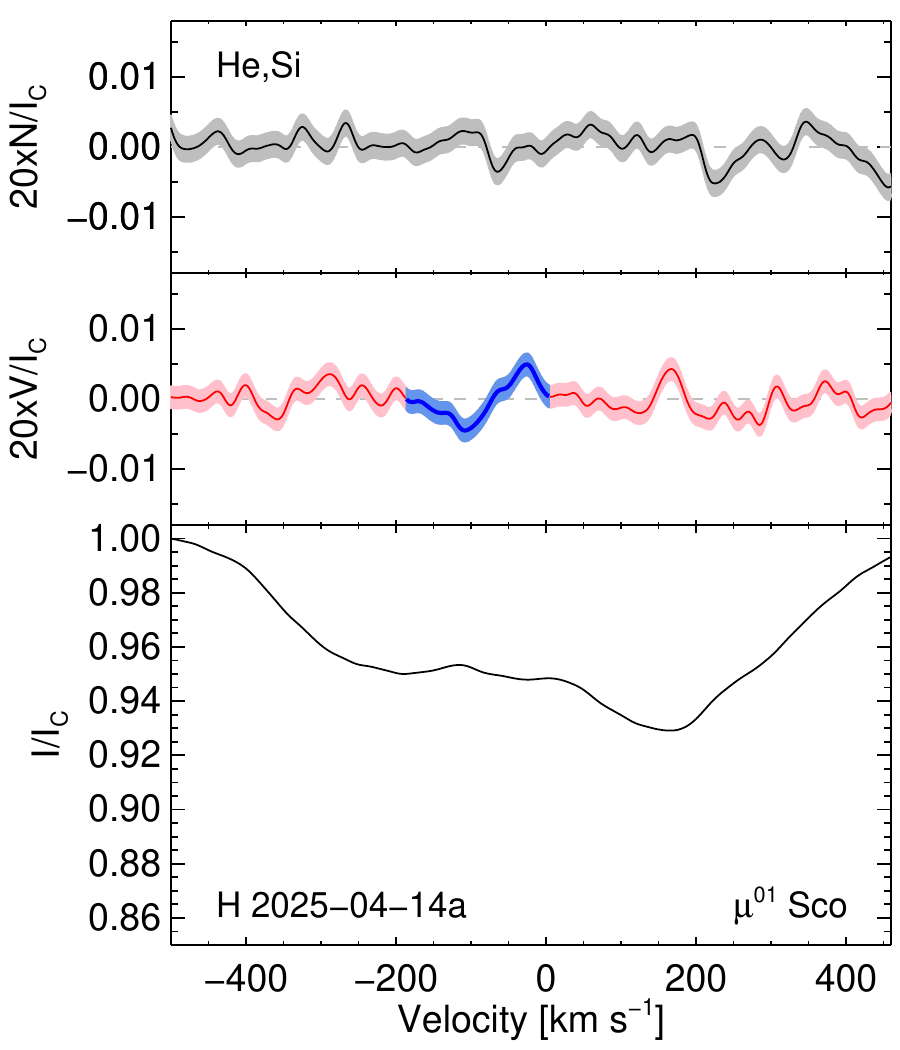}
    \includegraphics[width=0.23\textwidth]{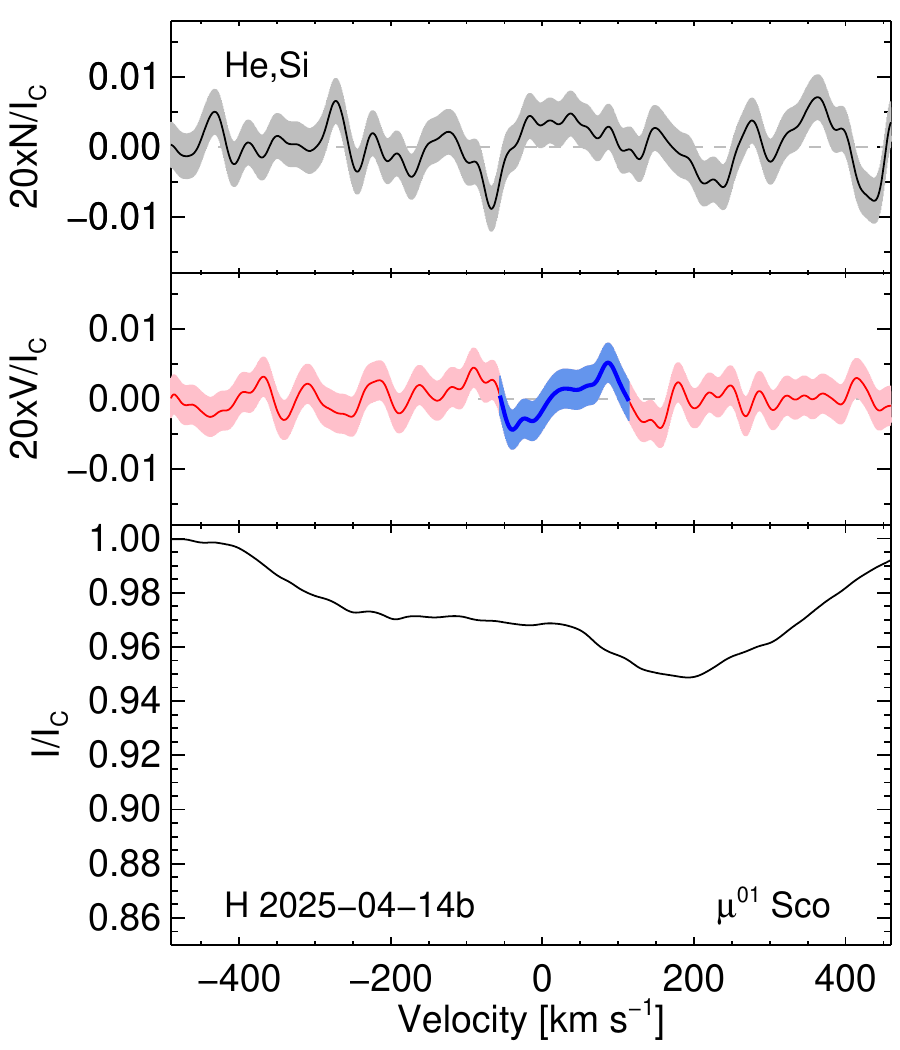}
    \caption{As Figure~\ref{fig:29CMa}, but for $\mu^{01}$\,Sco.}
    \label{fig:mu01Sco}
\end{figure*}

This triple hierarchical system is formed by an inner binary with
an O3 V((f*))--O3.5 V((f+)) primary and an O5.5--6 V((f)) secondary orbiting around each
other with a period of 2.67455\,d, and an O6.5--7 V((f)) physically bound tertiary
component located on a much longer orbit of about 8.2\,yr \citep{Mahy2018}.
By combining spectroscopy, interferometry, and photometry for this triple system,
\citet{Mahy2018} showed that the masses estimated through their analysis (42.8\,$M_\odot$,
29.5\,$M_\odot$, and 15.5\,$M_\odot$) are much smaller than those expected
from evolutionary models (55.4\,$M_\odot$, 30.6\,$M_\odot$, and
22.4\,$M_\odot$). This indicates that a merge or mass transfer took place in this system.

Only one HARPS\,pol observation has been acquired for this triple system.
As is demonstrated in Fig.~\ref{fig:HD150136} and Table~\ref{tab:obsall}, we achieved a definite detection
$\left< B_{\rm z} \right>=369\pm58$\,G with ${\rm FAP}=6\times10^{-6}$ using a mask with \ion{He}{i/ii}, \ion{C}{iv}, and \ion{N}{iii}.
Since in the LSD Stokes~$I$ spectrum all
components appear overlapped, it is not obvious which component hosts the magnetic field.

\paragraph*{$\mu^{01}$\,Sco (=HR\,6247):}

The components of this bright eclipsing system ($m_{\rm V}=2.98$) with an orbital period of 1.446\,d have spectral types B1.5\,V and B8-B3 and
masses 8.49\,$M_{\odot}$ and 5.33\,$M_{\odot}$ , respectively \citep{Antwerpen2010}. Using photometric and spectroscopic
data, \citet{Budding2015} suggested for the masses of the components 8.3\,$M_{\odot}$ and 4.6\,$M_{\odot}$ ($q = 0.55$).
The authors confirmed the results of previous studies, which reported that the
secondary star is larger than the primary, and certainly larger than a
normal main-sequence relationship would imply for the given primary radius and the secondary’s much lower mass.
Accordingly, the system must be in a state of interactive evolution.
Earlier studies suggested appreciable mass-loss from the system in the relatively recent past (e.g.\ \citealt{Stickland1996}).
This conclusion was based on the assumption of Algol-like conditions.
However, the work of \citet{Budding2015} showed that the relatively high masses of the components, the high primary
surface temperature, and the proximity of the components do not indicate an Algol-like nature of this binary.
Using Gaia data, \citet{Gratton2023} reported that the system has a distant companion
with a mass of 0.48\,$M_{\odot}$ at a separation of 44\,\arcsec{}.

Our observations were carried out on four different epochs in June 2024 and on seven different epochs in April 2025.
As is presented in Fig.~\ref{fig:mu01Sco} and Table~\ref{tab:obsall}, out of the eleven observations, definite detections
were achieved in five observations, marginal detections in four observations and non-detections in two observations. Given the
spectral types of the components, we used line masks containing \ion{He}{i} and \ion{Si}{ii/iii} lines.
For one observation on June~18 2024 with a definite detection, we were able to estimate for the more massive
component a mean longitudinal magnetic field
$\left< B_{\rm z} \right>=242\pm55$\,G with ${\rm FAP}=10^{-5}$. 
A longitudinal magnetic field  $\left< B_{\rm z} \right>=-187\pm29$\,G for the same component
was measured for one observation in June~19 2024 with a marginal
detection (${\rm FAP}=2\times10^{-4}$).
A mask containing \ion{He}{i} and \ion{Si}{iii} lines was used for both measurements.

\paragraph*{V1294\,Sco (=HD\,152218A):}

\begin{figure*}
    \centering
    \includegraphics[width=0.23\textwidth]{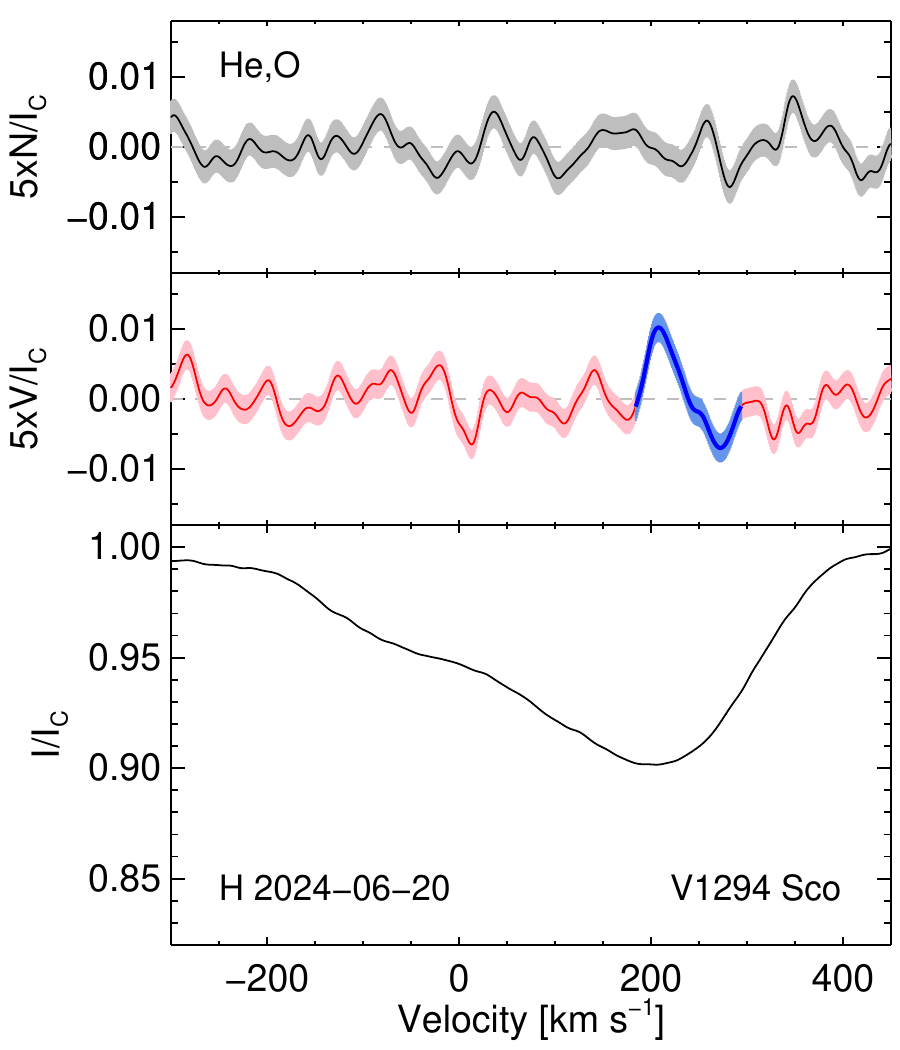}
    \includegraphics[width=0.23\textwidth]{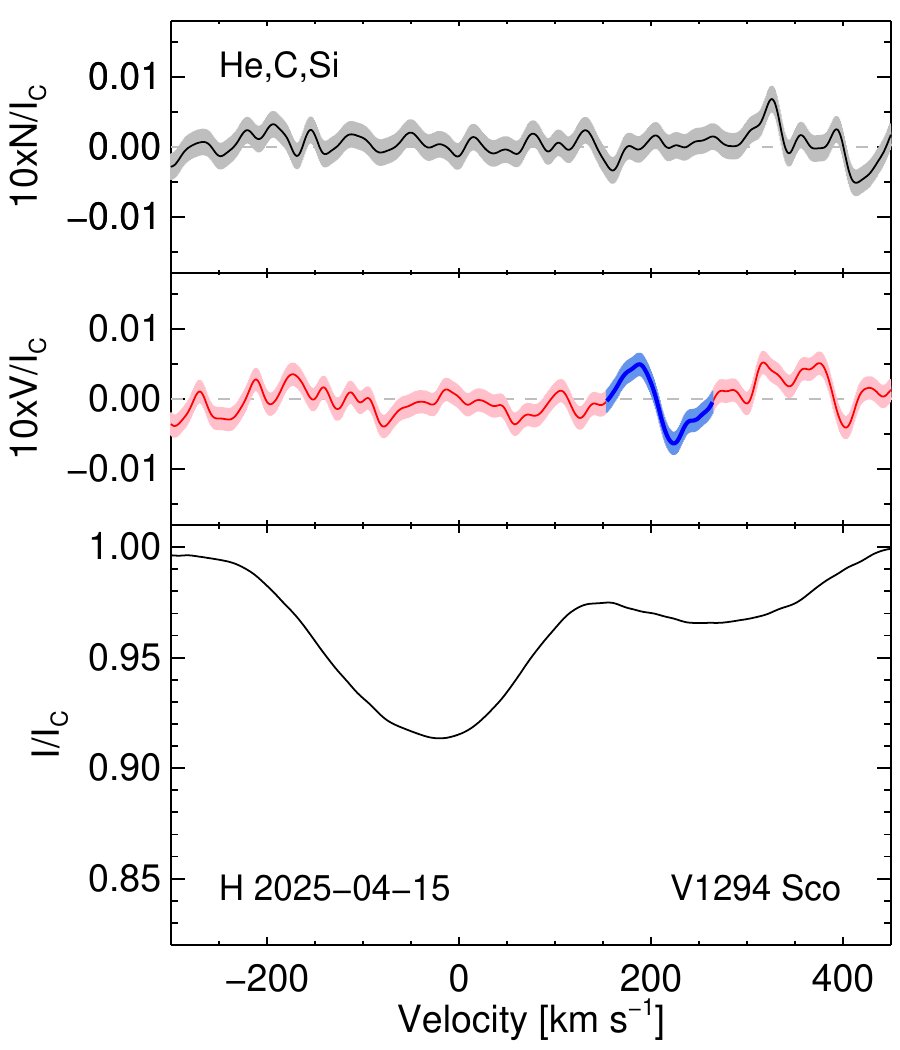}
    \includegraphics[width=0.23\textwidth]{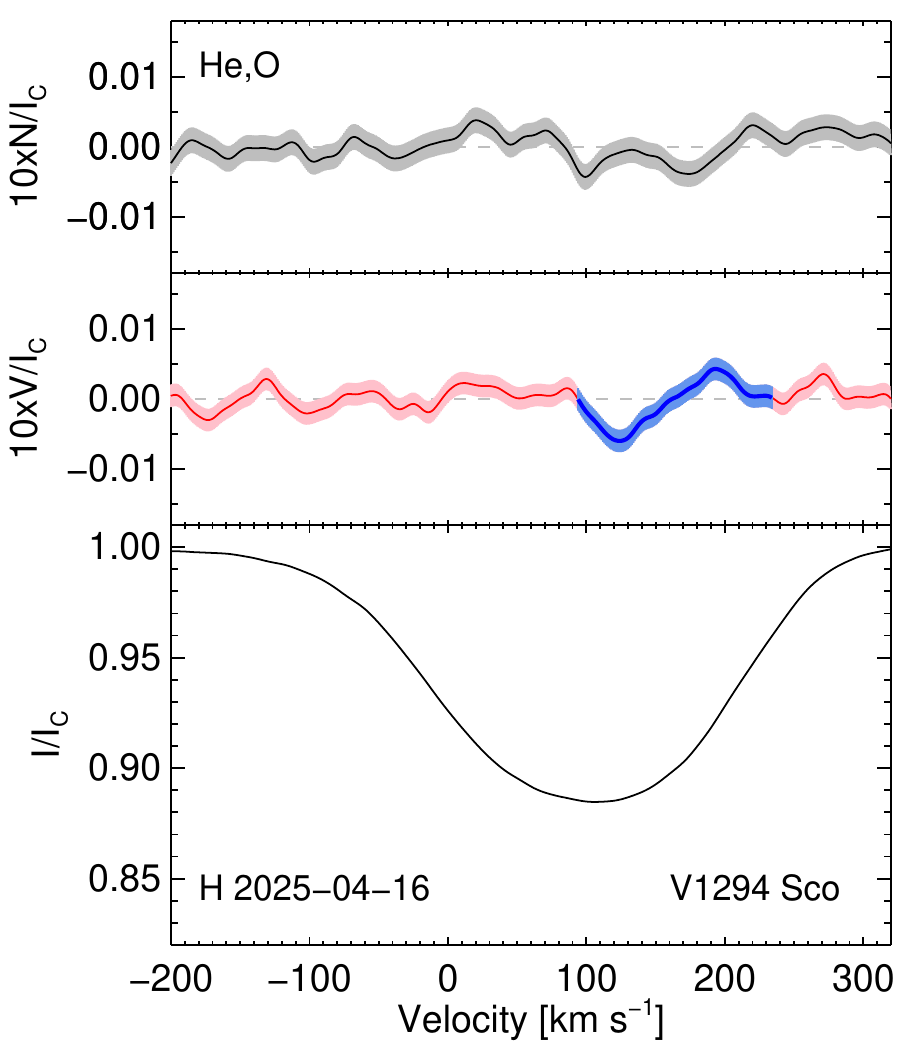}
    \caption{As Figure~\ref{fig:29CMa}, but for V1294\,Sco.}
    \label{fig:HD152218A}
\end{figure*}

According to \citet{Rosu2022}, V1294\,Sco is an eclipsing eccentric binary system
with an orbital period of 5.6\,d and components with masses 
of 20.6\,$M_{\odot}$ and 15.5\,$M_{\odot}$  ($q = 0.75$) and corresponding spectral types O9\,IV for the primary and
O9.7\,V for the secondary \citep{Sana2008}.
The rotation periods for the primary and the secondary  are 2.69 and 2.56\,d, respectively. 
The authors reported the lack of a secondary eclipse in
photometric observations, which induces degeneracies between the primary and secondary
Roche lobe filling factors and the inclination of the system.
Due to the quite high eccentricity of the system of about 0.28 and the absence of an enhanced
nitrogen abundance, \citet{Rauw2016} suggested that the components in V1294\,Sco did not interact in the past.

This system was already observed with HARPS\-pol in the past twice and reported as magnetic by \citet{Hubrig2023}.
While the first observation in 2013 with both components blended with each other yielded a non-detection, the second 
HARPS\-pol observation from 2016 revealed a clear Zeeman feature with ${\rm FAP}<10^{-10}$
corresponding to a longitudinal magnetic field $\left< B_{\rm z} \right>=-307\pm94$\,G in the secondary less massive component.
As is presented in Fig.~\ref{fig:HD152218A} and Table~\ref{tab:obsall}, we confirm the magnetic nature of this system.
We achieved in all newly obtained three HARPS\-pol observations definite detections
and measure on April~15 2025 a rather strong mean longitudinal magnetic field in the secondary component,
$\left< B_{\rm z} \right>=1559\pm131$\,G with ${\rm FAP} < 10^{-10}$,
using a line mask with \ion{He}{i/ii}, \ion{C}{iv}, and \ion{Si}{iii} lines.
The binary components in the two other observations
appeared partly overlapped, preventing us from measuring the magnetic field.

\paragraph*{V701\,Sco (=HD\,317844):}

\begin{figure*}
    \centering
    \includegraphics[width=0.23\textwidth]{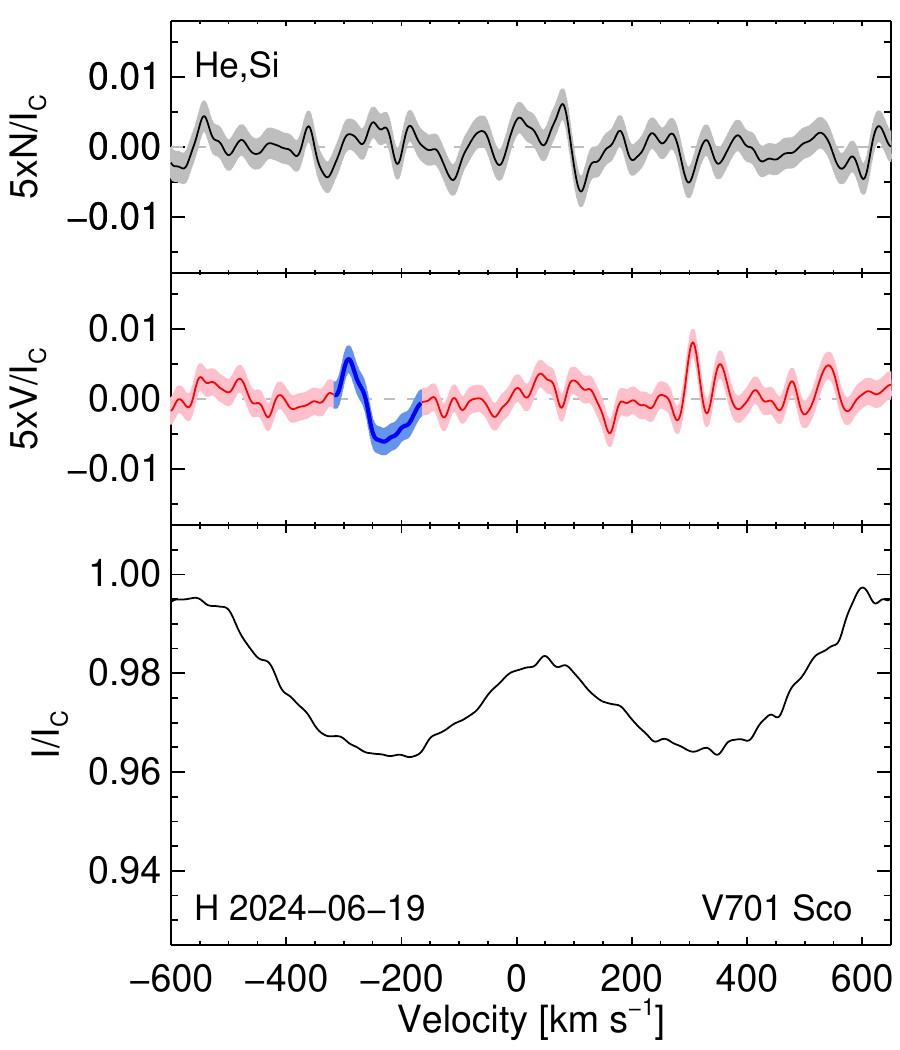}
    \includegraphics[width=0.23\textwidth]{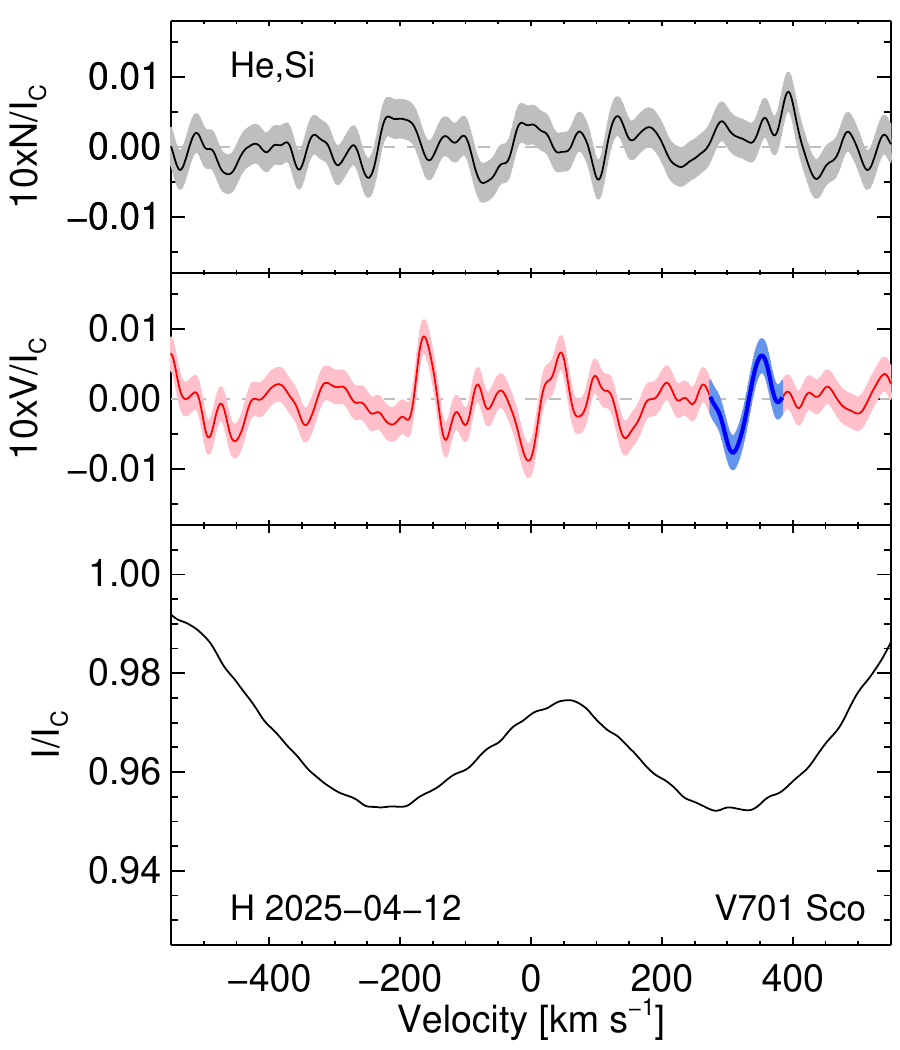}
        \includegraphics[width=0.23\textwidth]{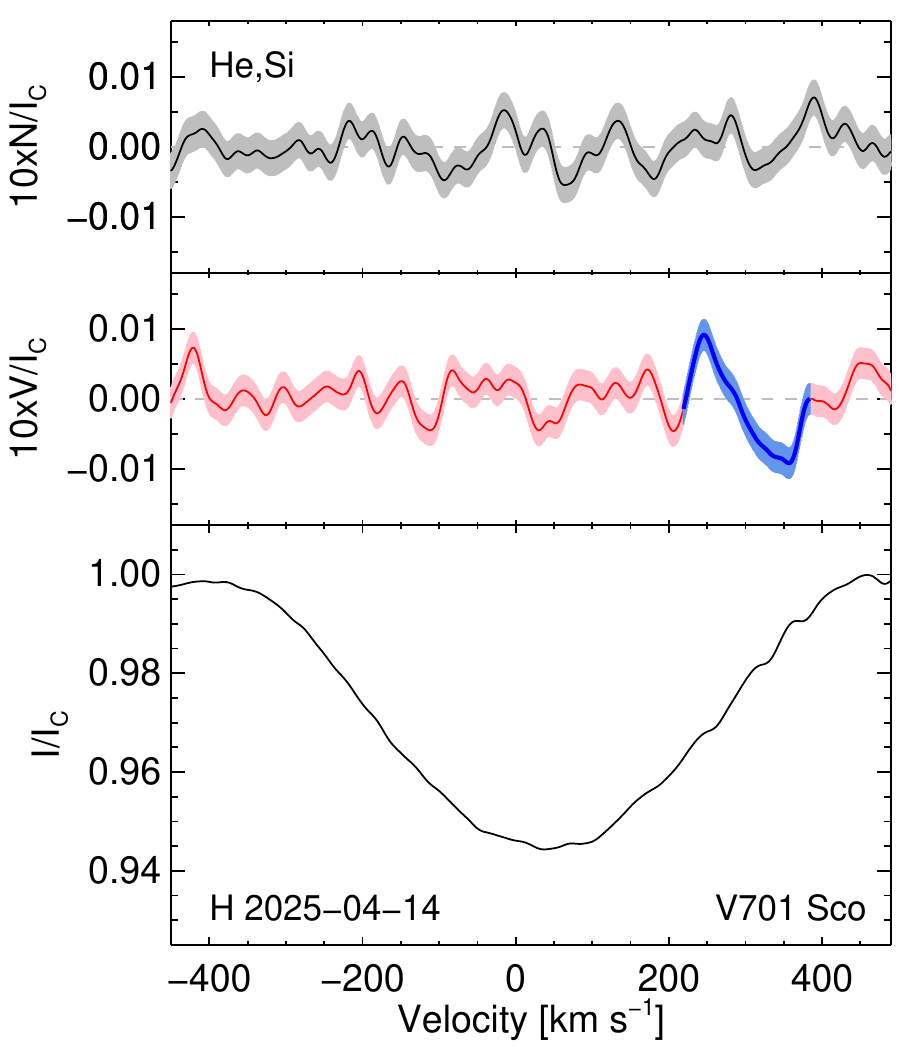}
      \includegraphics[width=0.23\textwidth]{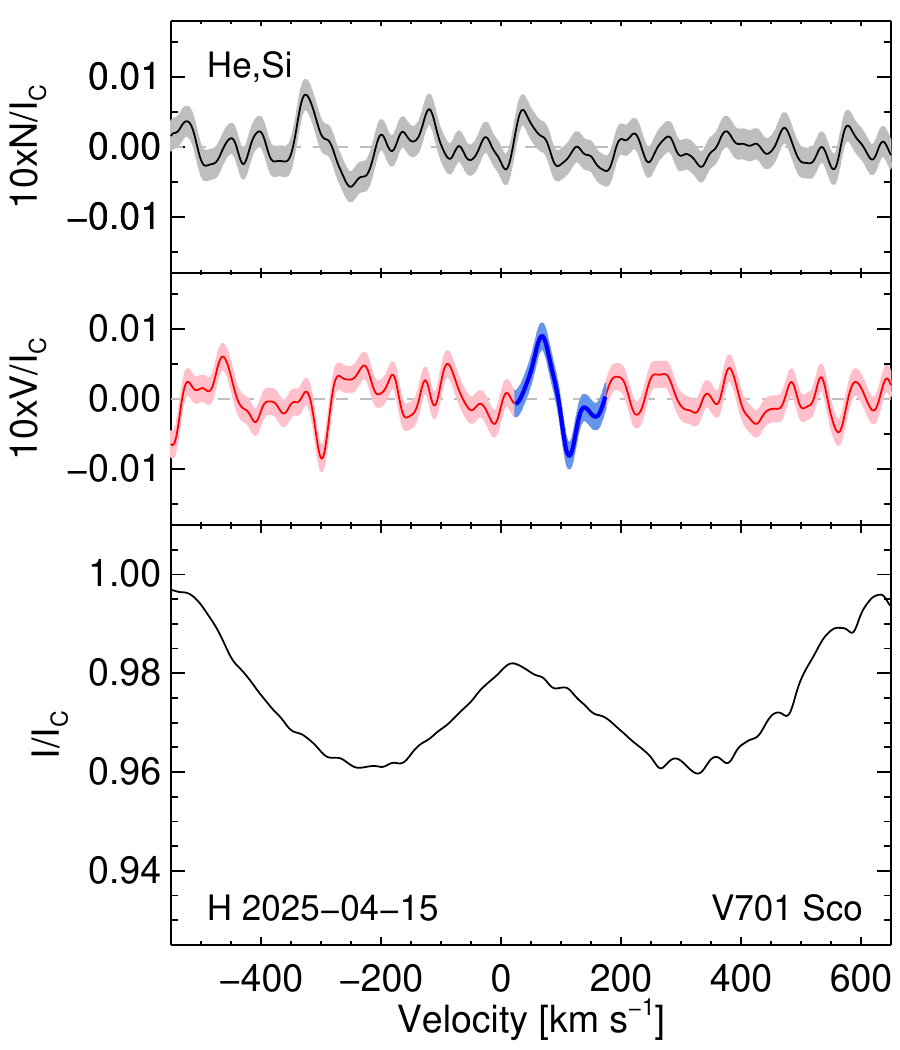}
      \caption{As Figure~\ref{fig:29CMa}, but for V701\,Sco.}
    \label{fig:V701Sco}
\end{figure*}

\begin{figure*}
    \centering
    \includegraphics[width=0.195\textwidth]{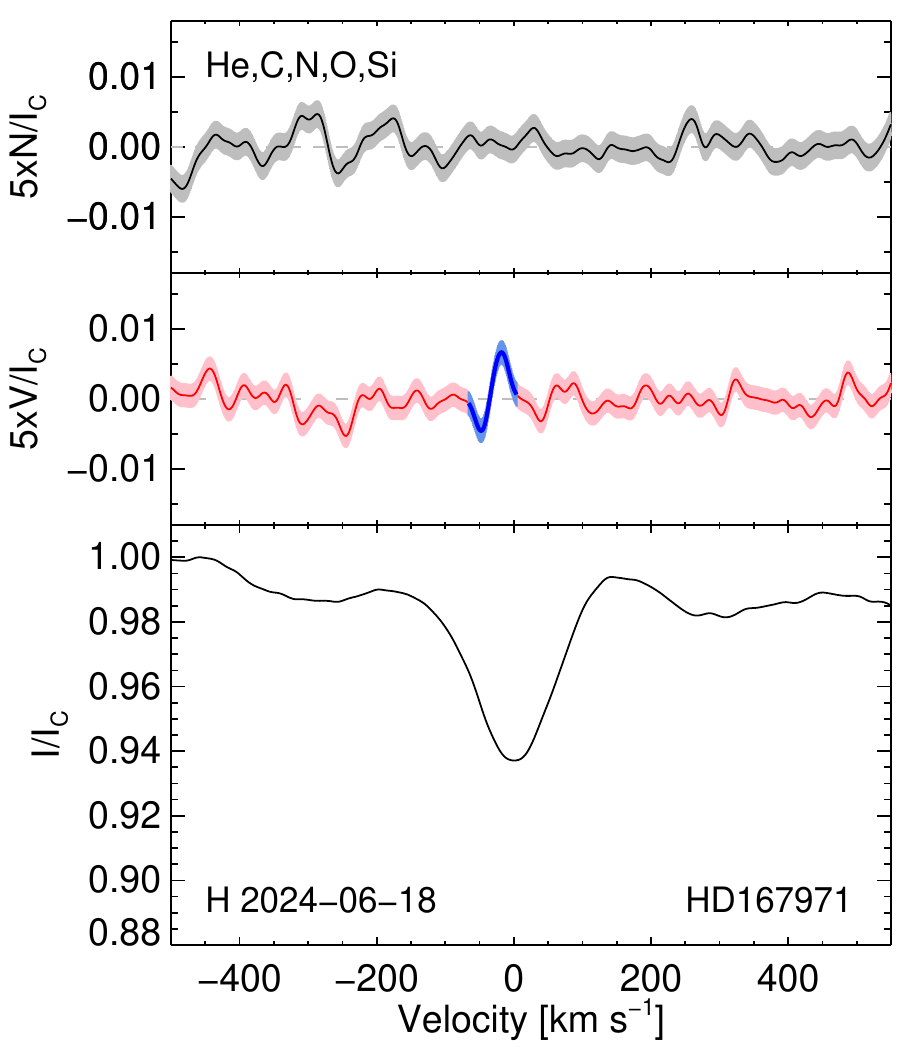}
    \includegraphics[width=0.195\textwidth]{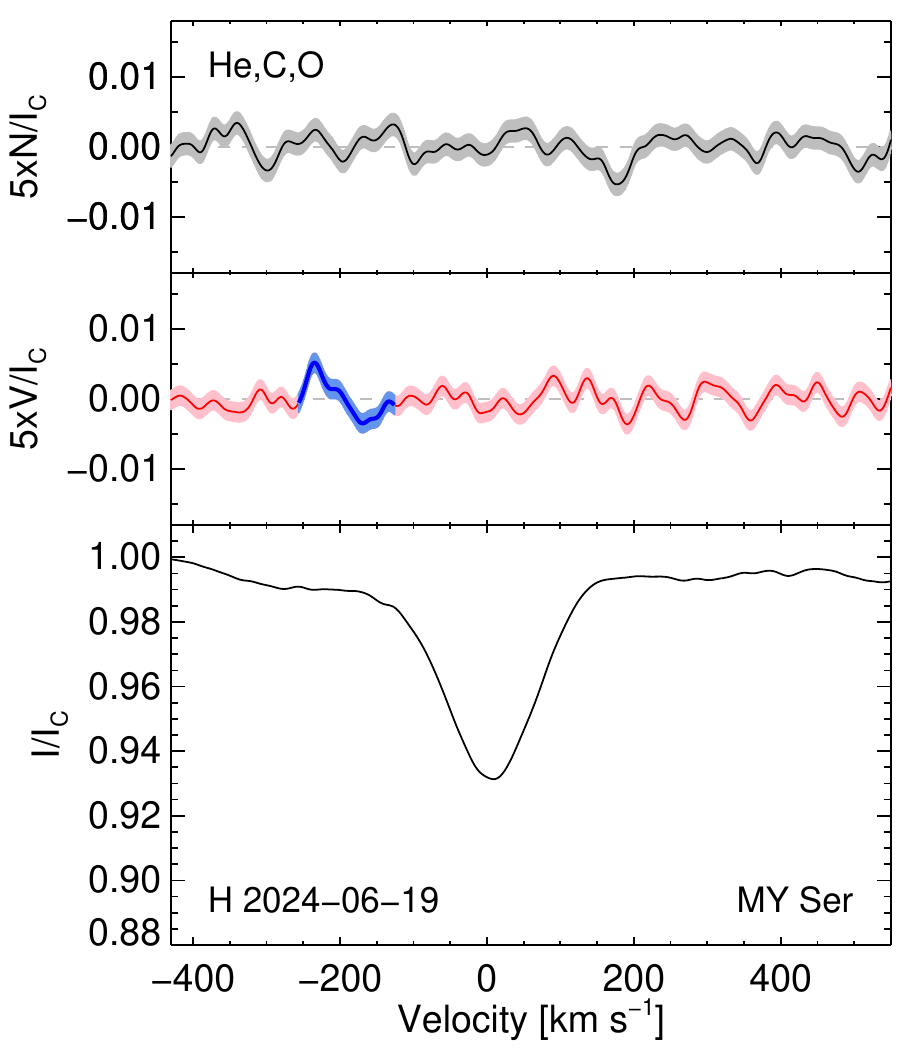}
    \includegraphics[width=0.195\textwidth]{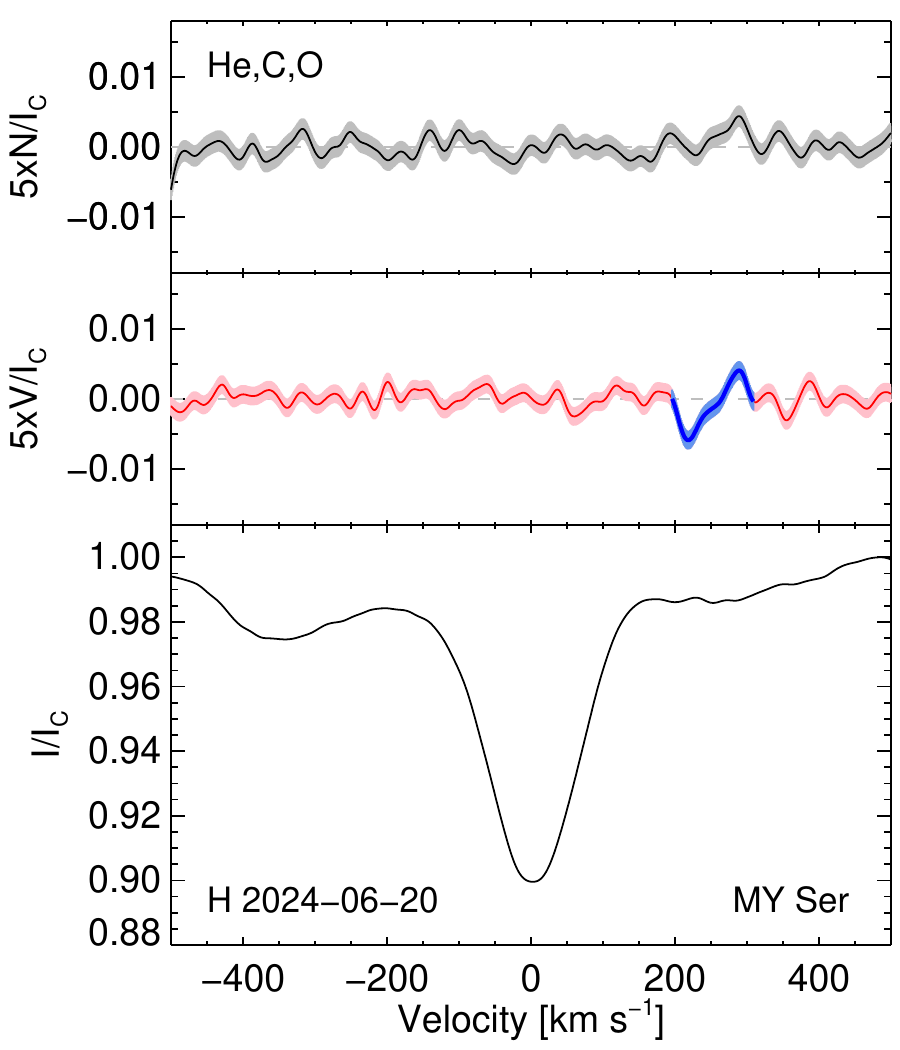}
 \includegraphics[width=0.195\textwidth]{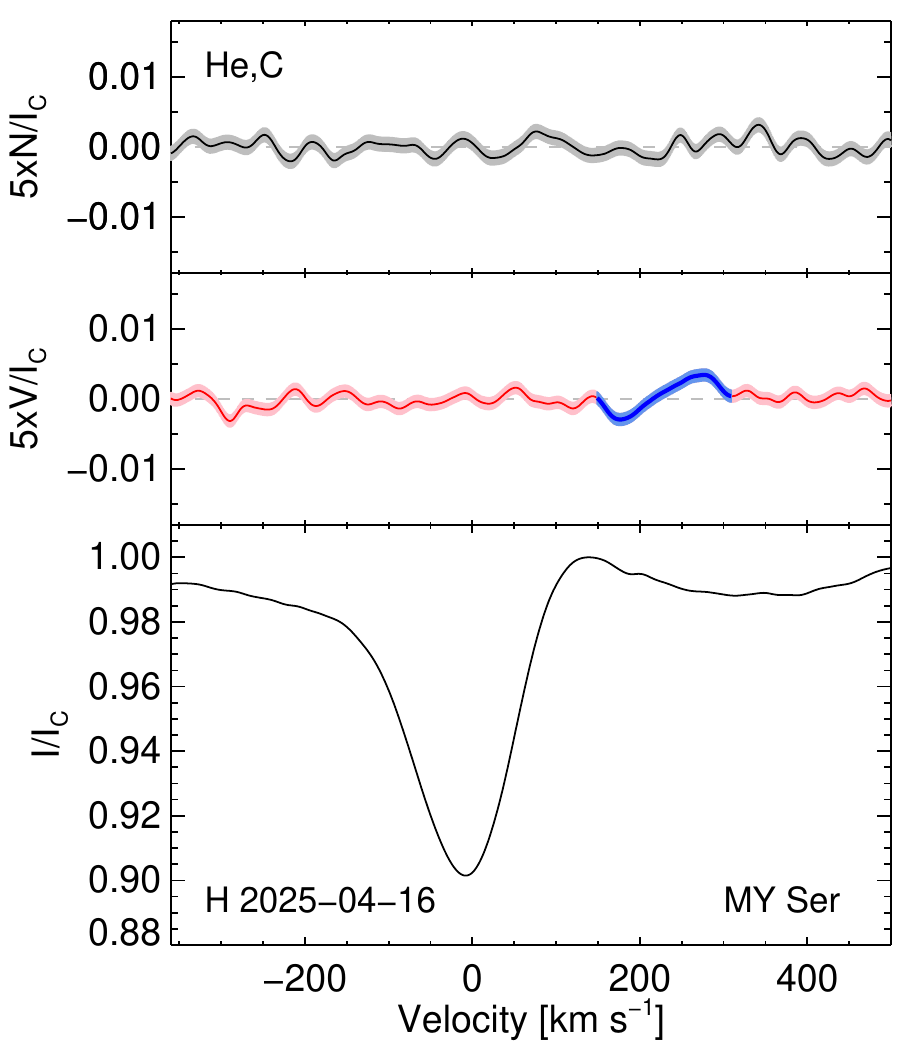}
    \includegraphics[width=0.195\textwidth]{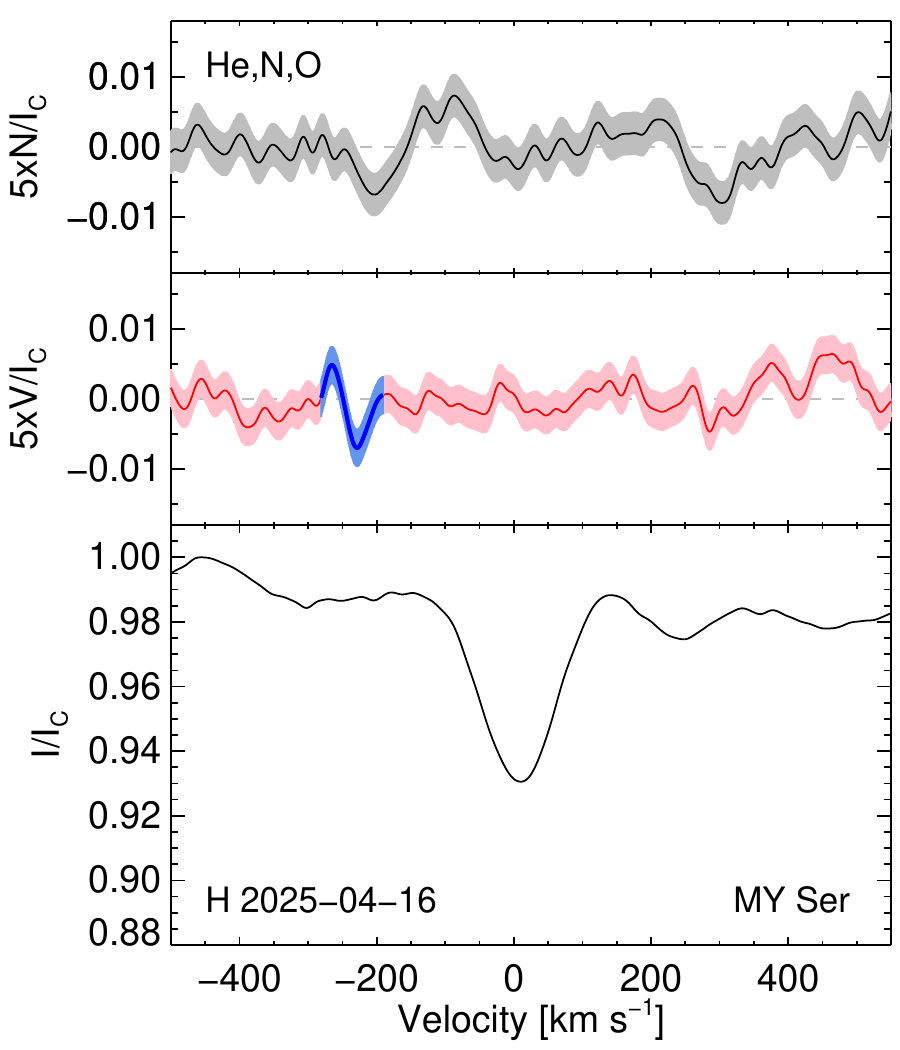}
    \caption{As Figure~\ref{fig:29CMa}, but for MY\,Ser.}
    \label{fig:HD167971}
\end{figure*}

This contact eclipsing binary with an orbital period of 0.761875\,d is composed of two B1-1.5V type stars
\citep{Qian2006} with stellar masses of 9.78\,$M_{\odot}$ and 9.74\,$M_{\odot}$, respectively, and is still
on the stage of a slow case~A mass transfer \citep{Yang2019}.
\citet{Eggen1961} suggested that the two almost identical components of V701\,Sco are zero-age-main-sequence stars in the
young open galactic cluster NGC\,6383.
According to \citet{Qian2006}, this contact binary
has been formed by the fission of the third body by bringing angular momentum for the central
system and creating a contact configuration of identical components.
A fillout factor of $f=1.55$ and the possible
presence of an additional component was reported by \citet{Yang2019}.

This system was observed once in June 2024 and three times in April 2025.
As is presented in Fig.~\ref{fig:V701Sco} and Table~\ref{tab:obsall}, we achieved definite detections of the Zeeman signatures for three
observations. A longitudinal magnetic field $\left< B_{\rm z} \right>=678\pm150$\,G (${\rm FAP}=10^{-7}$) was measured 
on June~19 2024. We detected a change of
polarity for the marginal detection $\left< B_{\rm z} \right>=-876\pm105$\,G (${\rm FAP}=4\times10^{-4}$),
achieved on April~12 2025.
For all observations we used a mask containing \ion{He}{i} and \ion{Si}{iii} lines.
Taking into account the orbital period of 0.761875\,d, we conclude that both B-type components are magnetic.

\paragraph*{MY\,Ser (=HD\,167971):}

\begin{figure*}
    \centering
    \includegraphics[width=0.23\textwidth]{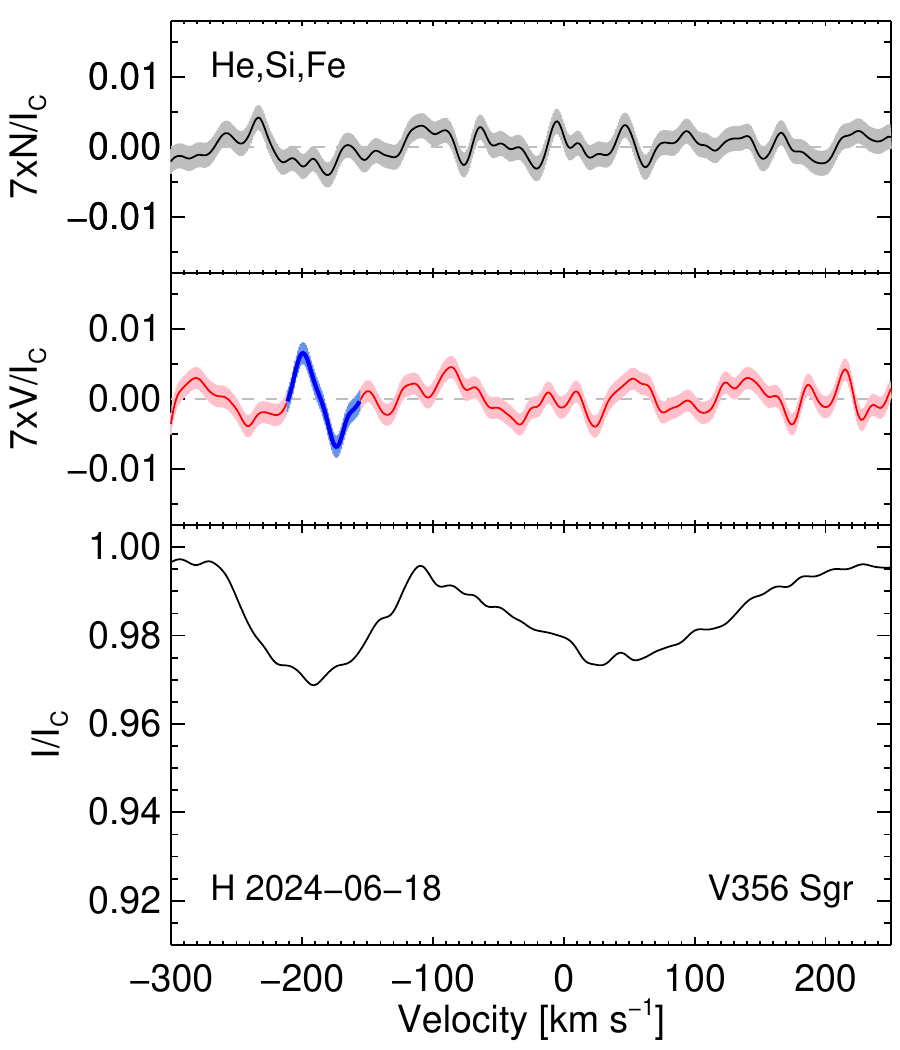}
   \includegraphics[width=0.23\textwidth]{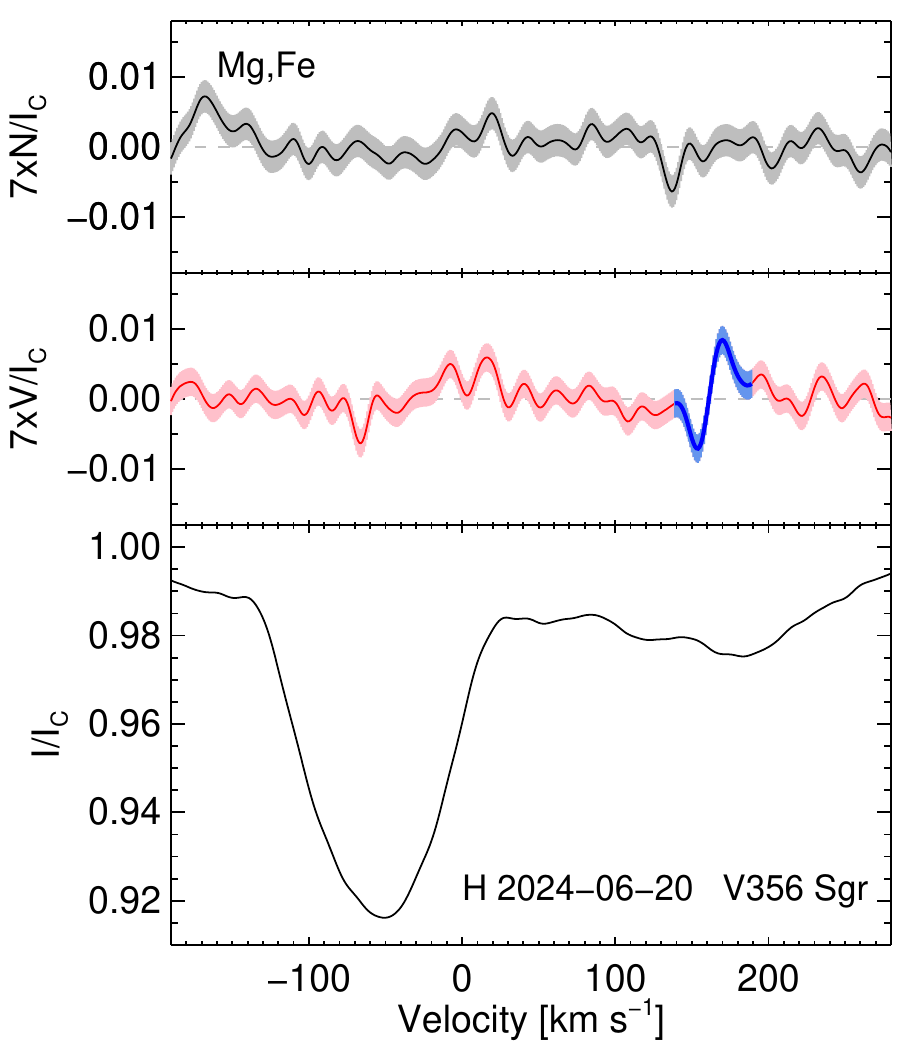}
    \caption{As Figure~\ref{fig:29CMa}, but for V356\,Sgr.}
    \label{fig:V356Sgr}
\end{figure*}

\begin{figure*}
    \centering
    \includegraphics[width=0.23\textwidth]{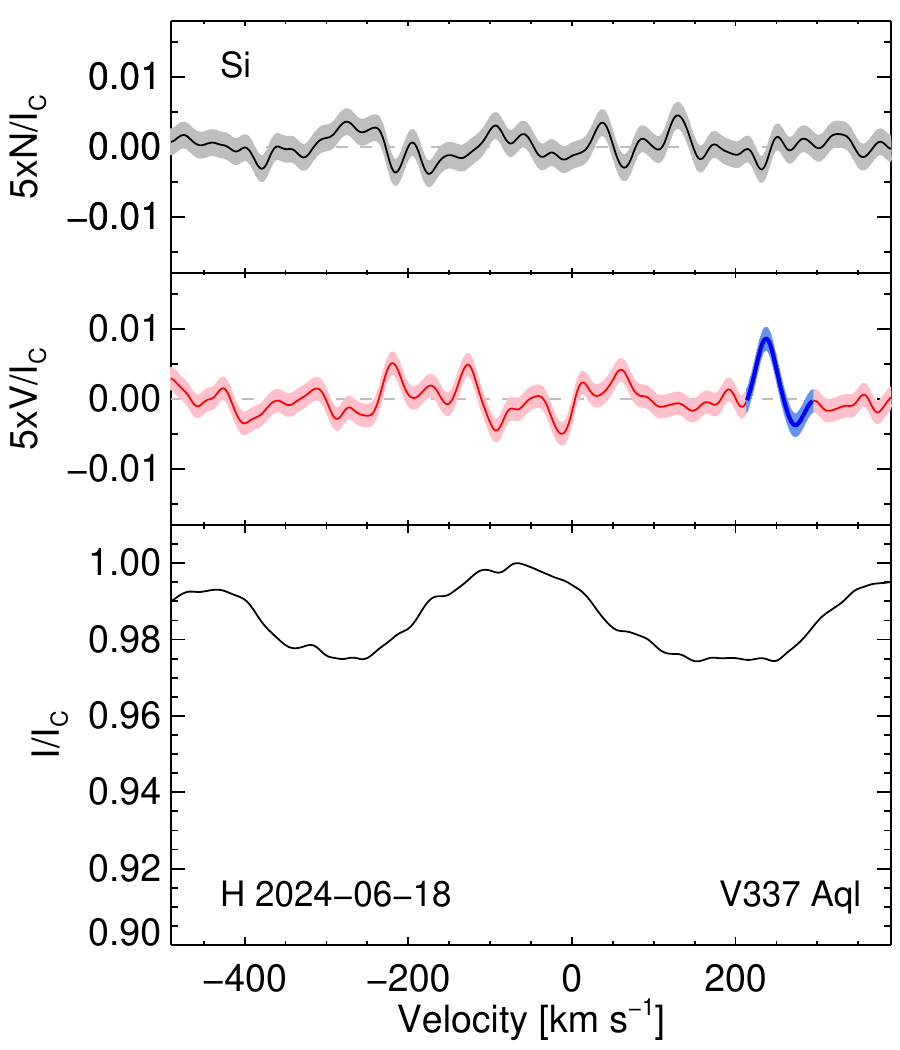}
    \includegraphics[width=0.23\textwidth]{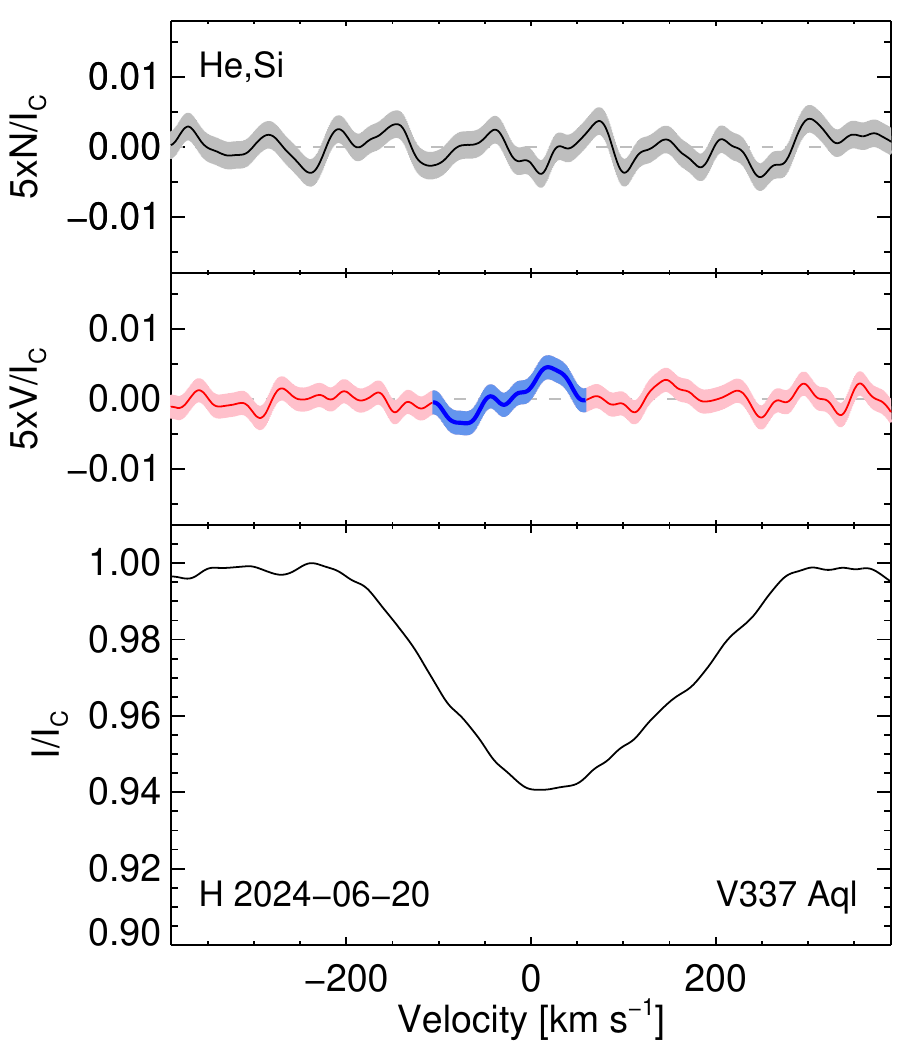}
     \includegraphics[width=0.23\textwidth]{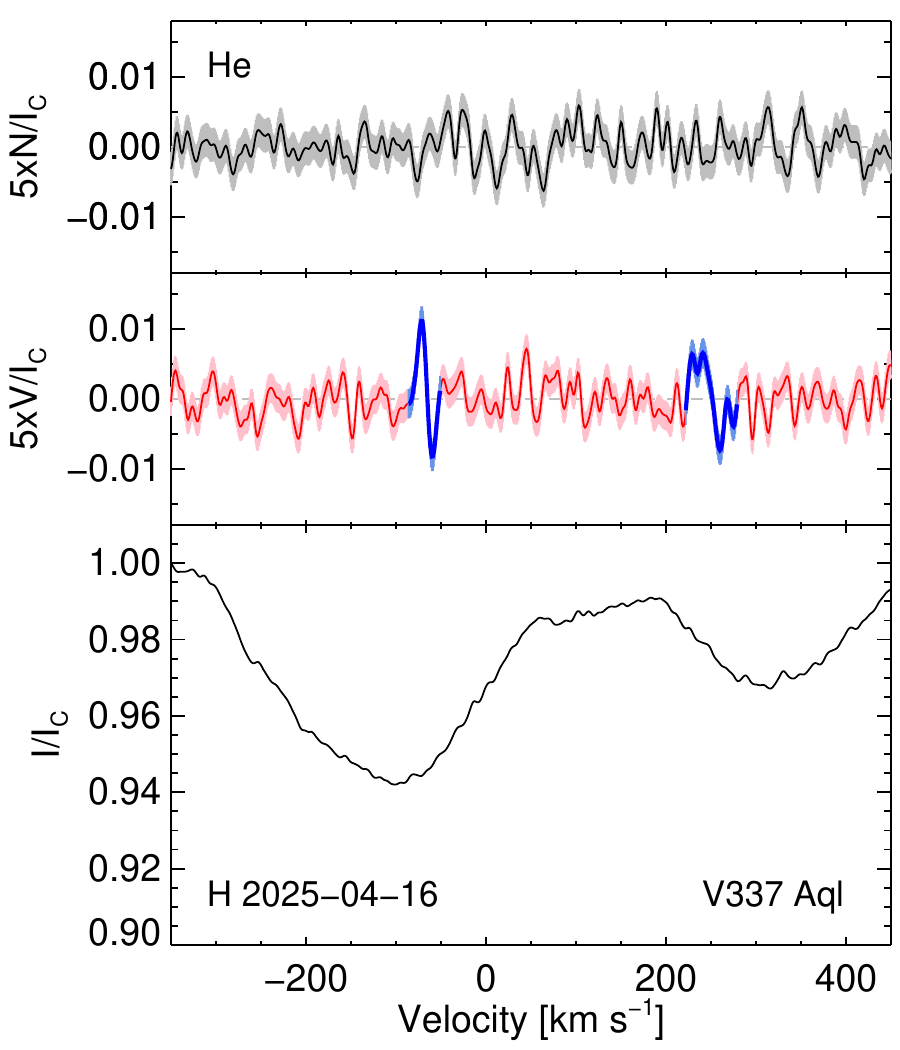}
    \caption{As Figure~\ref{fig:29CMa}, but for V337\,Aql.}
    \label{fig:V337Aql}
\end{figure*}

According to \citet{Ibanoglu2013},
this SB3 system is one of the rare massive O-type triple systems where the secondary and the tertiary components compose
an eclipsing binary. The spectral types of the stars in the close binary
are O7.5\,III (component Aa) and O9.5\,III (component Ab), and O9.5\,I (component B) for the tertiary.
The  corresponding masses are
32.23\,$M_{\odot}$ for the primary and 30.59\,$M_{\odot}$ for the secondary ($q = 0.95$) in the close binary, and
29\,$M_{\odot}$ for the tertiary.
The brighter component Aa/Ab of
this triple system is a contact eclipsing binary with an orbital period of 3.3216\,d, just filling the entire outer
contact surface with a fillout factor of 1.99. 
The outer pair AaAb/B has a period of about 21.2\,yr.
MY\,Ser was also classified as a particle-accelerating colliding-wind binary (PACWB) exhibiting synchrotron radio
emission \citep{Becker2013}. This system is also known to be one of the brightest synchrotron radio emitters.

The presence of a rather strong magnetic field was already reported by \citet{Hubrig2023},
who used two ESPaDOnS archival observations from 2013 and 2017.  A Zeeman signature with ${\rm FAP}<10^{-10}$
in the first observation corresponding to $\left< B_{\rm z} \right>= 1324\pm582$\,G was detected in the weakest red
component B. Zeeman signatures for two components were detected in the
second ESPaDOnS observation where a marginal
detection with ${\rm FAP}=9\times10^{-4}$ corresponding to $\left< B_{\rm z} \right>=-977\pm437$\,G was achieved for the blue
component Aa and a definite detection with ${\rm FAP}=6\times10^{-6}$ corresponding to $\left< B_{\rm z} \right>=-57\pm33$\,G
was achieved for the central component B, showing the deepest intensity profile in the Stokes~$I$ spectrum.  No detection was
achieved for the third component.

Four new HARPS\-pol observations were acquired on three consecutive nights in June 2024 and on one night in April 2025.
As is presented in Fig.~\ref{fig:HD167971}, similar to previous observations, all three components and corresponding Zeeman signatures are well visible in all
LSD Stokes~$I$ and $V$ spectra. Our result for the magnetic field measurement in the central component B in the first observation obtained on June~18 2024,
$\left< B_{\rm z} \right>=-46\pm31$\,G with ${\rm FAP}=1\times10^{-7}$ and using a line mask containing
\ion{He}{i/ii}, \ion{C}{iii}, \ion{N}{iii}, \ion{O}{iii}, and \ion{Si}{iv} lines,
is in agreement with the result by  \citet{Hubrig2023}, suggesting that
this component is weakly magnetic. The field measurement in the blueshifted  weak component Ab using  a mask containing
\ion{He}{i}, \ion{C}{iii/iv}, and \ion{O}{iii} lines for the second observation obtained on June~19 2024
revealed the presence of an extremely strong longitudinal field, $\left< B_{\rm z} \right>=4008\pm402$\,G with ${\rm FAP}=2\times10^{-8}$.
Taking into account the orbital period of MY\,Ser of 3.3216\,d and the
radial velocity variability presented in figure~2 in the study of \citet{Ibanoglu2013}, this strong field belongs to the more massive binary component Aa.
Applying a mask with \ion{He}{i/ii}, \ion{C}{iii}, and \ion{O}{iii} lines,
the detected redshifted  Zeeman signature in the third observation on June~20 2024
corresponds to the lower-mass component Ab in the eclipsing binary. We achieved for this observing epoch
a definite detection with ${\rm FAP}<10^{-10}$, but are not able to estimate the field strength due to the weakness of this component.
A definite magnetic field for the same component Ab, $\left< B_{\rm z} \right>=-4875\pm394$\,G with ${\rm FAP}=2\times10^{-8}$,
was measured in the observation acquired on April~16 2025 using a mask with \ion{He}{i/ii} and \ion{C}{iii} lines.
In this last observation,
the lower-mass component Ab is shifted from the position of the central component B to the red, while the more massive component Aa
is shifted to the blue.
Using a mask with \ion{He}{i/ii}, \ion{N}{iii}, and \ion{O}{iii} lines, we achieved for the more massive component Aa a
marginal detection (${\rm FAP}=4\times10^{-5}$) and measure a longitudinal field
$\left< B_{\rm z} \right>=8619\pm662$\,G.
In view of the previously reported  detection of strong synchrotron radio emission in this system, it is quite possible
that its generation is directly related to the presence 
of extremely strong magnetic fields in both binary components.

\paragraph*{V356\,Sgr (=HD\,173787):}

According to \citet{Lomax2017}, this system is considered as an eclipsing RLOF system 
and is morphologically similar to the $\beta$~Lyr system. Both
systems, V356\,Sgr and $\beta$~Lyr, are undergoing RLOF, which has caused discs to form around their gainers.
However, in contrast to the $\beta$~Lyr system, the accretion disc in the
V356\,Sgr system is optically thin. It was suggested that the
3\,$M_{\odot}$ A2 supergiant secondary star has filled its
Roche lobe and is currently transferring matter to the brighter, B3\,V
primary star with a mass of 11\,$M_{\odot}$ (e.g.\ \citealt{Rensbergen2011}).
The orbital period of 8.9\,d is reported in the work of \citet{Hamme1990}.

V356\,Sgr was observed with HARPS\-pol on two nights in June 2024.
As is shown in Fig.~\ref{fig:V356Sgr} and Table~\ref{tab:obsall}, we detected definite longitudinal magnetic fields
in both observations. Using a line mask with \ion{He}{i}, \ion{Si}{ii/iii}, and \ion{Fe}{ii} lines  we measured
$\left< B_{\rm z} \right>=856\pm240$\,G with ${\rm FAP}=1.5\times10^{-7}$ for the B3\,V component in the first observation
obtained on June~18 2024.
Using a mask with \ion{Mg}{ii} and \ion{Fe}{ii} lines, we obtained $\left< B_{\rm z} \right>=-718\pm245$\,G
with ${\rm FAP}=7\times10^{-6}$
for the A2 component in the second observation obtained on June~20 2024.

\paragraph*{V337\,Aql (=HD\,177284):}

The simultaneous light and radial velocity curves solution by \citet{Soydugan2014} indicated for this $\beta$~Lyr
type eclipsing binary system with an orbital period of 2.7339\,d a mass of 17.44\,$M_\odot$ for the primary and
7.83\,$M_\odot$ for the secondary. The authors suggested that this B0$+$B3 system is a near-contact semi-detached
binary, in which a primary star is inside its Roche lobe with a filling ratio of 92\% and the
secondary star fills its Roche lobe. 

We obtained three HARPS\-pol observations, two in June 2024 and one
in April 2024.
As is presented in Fig.~\ref{fig:V337Aql} and Table~\ref{tab:obsall}, using a mask  containing \ion{Si}{iii} lines, we detect a
definite magnetic field
$\left< B_{\rm z} \right>=430\pm235$\,G with ${\rm FAP}=2\times10^{-6}$ in the primary component. 
Both components appeared overlapped during the second observation on June~20 2024. We achieved for this observation a marginal
detection $\left< B_{\rm z} \right>=-373\pm38$\,G with ${\rm FAP}=2\times10^{-5}$
using a mask with \ion{He}{i/ii} and \ion{Si}{iii} lines. In the third observation acquired on April~16 2025,
we detect Zeeman signatures in both components,
but a definite field detection using a line mask with \ion{He}{i/ii} lines,
$\left< B_{\rm z} \right>=25\pm68$\,G with ${\rm FAP}=9\times10^{-6}$,
corresponds to the more massive component shifted to the blue.
For the lower-mass component shifted to the red, we achieve a marginal detection $\left< B_{\rm z} \right>=146\pm275$\,G
with ${\rm FAP}=2\times10^{-4}$.
More observations are necessary to verify whether both components in this system are magnetic.

\section{Discussion}\label{sec:disc}

Despite the progress achieved in previous surveys of the magnetism in massive stars,
the origin of their magnetic fields remains the least understood topic.
The presented study aims to detect magnetic fields and to characterise their strengths
in massive binary and multiple systems with components at different stages of interaction.
Our sample comprises eight close systems with different contact configurations, one post-RLOF system, one merger, three semi-detached
systems, and one detached system.
As the longitudinal magnetic field is strongly dependent on the viewing angle between the field orientation
and the observer and is modulated as the star rotates,
we intended to observe each system at least at two different observing epochs.
The number of available spectropolarimetric observations ranged from two observations (e.g.\ for TU\,Mus and V356\,Sgr) up to
eleven observations for $\mu^{01}$\,Sco.
The semi-detached system SV\,Cen
and the triple system HR\,6187, suggested to comprise a merger, were observed only once.
As is shown in Table~\ref{tab:obsall}, the analysis of 53 high-resolution HARPS\-pol spectra
acquired over the last two years reveals the definite presence of a magnetic field in all studied systems apart
from the rather faint system SV\,Cen, for which only a marginal detection was achieved.
The lowest longitudinal magnetic field strength of the order of only a few tens of Gauss was measured
for the post-RLOF SB2 system V918\,Sco with two supergiant components \citep{Rauw2001}.
The measured mean longitudinal magnetic field strengths for all other targets are of the order of a few hundred Gauss 
to a few kiloGauss. 

To understand the implication of the contact configuration and evolutionary state of the studied systems
on the magnetic field strengths, we considered for each system the spectral classification of the components,
their mass ratio, and the fillout factor.
The strongest longitudinal magnetic fields of 4 to 5\,kG were discovered in the massive O-type
triple system MY\,Ser in both components of the contact binary with an orbital period of  3.3216\,d,
a mass ratio $q = 0.95$, and a fillout factor of 1.99.
For the more massive O7.5\,III component in this binary, we obtained a definite field detection of about 4\,kG
at one epoch  and a marginal detection $\left< B_{\rm z} \right>=8619\pm662$\,G at another epoch.
Obviously, spectropolarimetric monitoring over the rotation periods of the components is urgently needed to study 
the structure of such extremely strong magnetic fields in the components of the contact system.
As such a strong longitudinal magnetic field has never been detected
in contact binaries in the past, our measurements strongly suggest that interaction between the components in
contact systems is of great importance for the generation of magnetic fields.
Notably,  because the amplitudes of the Zeeman signatures are lower in multiple systems in comparison with the size of
these features in single stars (e.g.\ \citealt{Hubrig2023}), their magnetic fields can even be stronger.
According to the spectral classification of this system, O7.5\,III$+$O9.5\,III$+$O9.5\,I, it is already
evolved from the main sequence.
Importantly, the fillout factor for this system is the largest among all studied systems.

Apart from MY\,Ser, only two other studied systems have a mass ratio close to 1.
These two systems with almost identical component masses are the contact eclipsing system
O9.5\,V$+$O9.5\,V in the quadruple system V402\,Pup with an orbital period of 1.019\,d
and a fillout factor of 1.31, and the contact eclipsing system B1-1.5\,V$+$B1-1.5\,V
with an orbital period of 0.762\,d and a fillout factor of 1.55 belonging to the system V701\,Sco.
While the measurement of the magnetic field in the quadruple system is very demanding, the magnetic field measured in the system  
V701\,Sco is rather strong with $\left< B_{\rm z} \right>=678\pm150$\,G (${\rm FAP}=10^{-7}$) detected at one epoch and
$\left< B_{\rm z} \right>=-876\pm105$\,G (${\rm FAP}=4\times10^{-4}$) measured at another epoch.
For this system, we concluded that both B-type components are magnetic.

For a number of systems, in particular systems with B-type components, the detected Zeeman signatures do not extend
over the full LSD Stokes~$I$ profile.
As has already been discussed by \citet{Hubrig2025}, this can be explained by the fact that magnetic B-type stars
are characterised by a chemical abundance distribution
that is non-uniform and non-symmetric with respect to the rotation axis, but shows some symmetry between the topology
of the magnetic field and the chemical spot distribution. A rotation modulation of intensities and radial velocities of He and metal lines has
also been discovered in magnetic O-type stars (e.g.\ \citealt{Grunhut2012}). The authors suggested that the observed line
profile variability can be a result of variations in the flattened distribution of magnetospheric plasma around the magnetic star.

While MY\,Ser appears to be a record holder among the studied systems,
kiloGauss-order magnetic fields were also detected in two other
systems. We measured  $\left< B_{\rm z} \right>=1559\pm131$\,G with ${\rm FAP}<10^{-10}$ in the detached O9.5IV(n)$+$B0V
system V1294\,Sco with an orbital period of 5.6\,d and a mass ratio of 0.75.
A mean longitudinal magnetic field  $\left< B_{\rm z} \right>=1327\pm128$\,G with ${\rm FAP}<10^{-10}$
was measured in the triple system V606\,Cen hosting a contact B0.5\,V$+$B3\,V system with an orbital period of 1.495\,d,
a mass ratio of 0.55, and a very low fillout factor of 2\%.

Admittedly, the number of systems studied is still too low for us to be able
to test possible correlations between the occurrence of
strong magnetic fields and system characteristics.
On the other hand, some characteristics, such as the presence of additional bodies
in the system, mass ratios
between the components, and probably large fillout factors corresponding
to deeper contacts seem to be favourable for the detection
of stronger magnetic fields in interacting systems.
In any case, the fact that the presence of magnetic fields is detected in all but one of the studied systems
strongly suggests that interaction between the system components plays a definite role in the generation of
magnetic fields in massive stars.

In contrast, previous surveys of magnetic fields in massive stars reported only a very low occurrence of
magnetic fields and usually concentrated on presumably single stars or members of wide binaries.
These surveys 
avoided complex systems with interacting binaries and frequently used inadequate measurement strategies.
The best example for a lack of success in the description of the magnetic companion in the interacting system
Plaskett's star was demonstrated in the works of \citet{Grunhut2013,Grunhut2022}.
Only a little more than a dozen bona fide
contact systems have been reported in the literature  (e.g.\ \citealt{Abdul-Masih2025}, and references therein).
Although our results on magnetic field measurements in interacting binaries present the first assessment of the occurrence
rate of magnetic fields in a sample of such systems,
additional spectropolarimetric observations are necessary to better understand  which
evolutionary channels dominate in the magnetic field generation and which end products of such systems hosting magnetic fields are produced.

Knowledge about the presence of magnetic components in contact systems is especially important in view of
theoretical modelling that considers them as potential progenitors of gravitational wave sources
detectable by facilities such as the Laser Interferometer Gravitational-Wave Observatory (LIGO; 
\citealt{Abramovici1992}) and the Virgo Interferometer (Virgo; \citealt{Abbott2016}).
As was already mentioned in Sect.~\ref{sec:intro}, among the well-studied contact systems,
almost 70\% have at least one confirmed companion and about 25\% have more than one
additional companion \citep{Abdul-Masih2025}. The author reviewed the current state of the field
of massive contact binary observations and suggested that the high multiplicity statistics is not
entirely unexpected due to the compactness of the inner orbits and that it is possible
that the companion may play a role in the formation of the contact systems.
Indeed, multiple systems are important in various astrophysical phenomena, significantly contributing 
to the genesis of black holes, neutron stars, and progenitors of stellar explosions and mergers. 
Several studies of the evolutionary pathways of triple systems with non-magnetic components
demonstrated their importance for our understanding of the formation of compact object binaries, the production
of explosive transients, and gravitational wave sources (e.g.\ \citealt{Toonen2016}).

Especially intriguing are studies exploring the role of the third body for producing binary black holes (BBHs) that have the potential to
merge within the age of the Universe. According to \citet{Vigna2025}, for triples in low metallicities and in which the inner
binary does not undergo a stellar merger, a BBH may form and the merger of BBHs can be
facilitated by a tertiary companion with an orbital period of up to a few thousand days. In favourable
orbital configurations, the time-to-coalescence may be as short as the von Zeipel-Lidov-Kozai (ZLK; 
\citealt{Zeipel1910,Lidov1962,Kozai1962}) mechanism timescale. \citet{Dorozsmai2024} focussed on the evolution of
hierarchical triples in which the stars of the inner binaries are chemically homogeneously
evolving and demonstrated that triple systems such as these can experience tertiary mass
transfer onto a BBH. The evolution of multiple systems is, however, especially complex and is
affected by several bodies dynamics and tidal and mass transfer processes.
Obviously, a particularly careful analysis of a representative sample of contact systems is necessary to test
theoretical predictions. Our results showing a high occurrence rate of magnetic fields in interacting systems, frequently
appearing as members  of multiple systems, additionally increase the complexity
of evolutionary stellar models of massive systems.

\begin{acknowledgements}

We would like to thank the referee Gautier~Mathys for his useful comments.
Our work is based on observations made with ESO telescopes at the La~Silla Paranal Observatory
under programmes 087.D-0800(A), 0113.D-0359(B), 0114.D-0183(A), 0115.D-2108(A), and 
0115.D-2150(A).
One observation was obtained with the Canada-France-Hawaii Telescope, which is operated by the
National Research Council of Canada, the Institut National des Sciences de l'Univers of the Centre National de la
Recherche Scientifique of France, and the University of Hawaii.

\end{acknowledgements}

%
   \bibliographystyle{aa} 
   \bibliography{aa58337-25} 
%

\end{document}